\documentclass[12pt]{article}
\usepackage[latin1]{inputenc}
\usepackage[twoside,left=2cm,right=2cm,top=2.5cm,bottom=3.5cm]{geometry}
\usepackage{amsmath}
\usepackage{amsfonts}
\usepackage{amssymb}
\usepackage{indentfirst}
\usepackage{fancyhdr}
\usepackage[dvips]{graphicx}
\usepackage{amssymb}
\usepackage{amsmath}
\usepackage{latexsym}
\usepackage{amsthm}
\usepackage{eucal}
\usepackage{natbib}
\usepackage[T1]{fontenc}
\usepackage{hyperref}
\usepackage{amsxtra}
\usepackage{amstext}
\usepackage{epsf,epsfig}
\usepackage{rotating}
\usepackage{multirow} 
\usepackage{slashbox}
\usepackage{pict2e}
\usepackage[table]{xcolor}

\title{Mixture of Latent Trait Analyzers for Model-Based Clustering of Categorical Data}
\author{
        {\Large Isabella Gollini$^\dagger$  \& Thomas Brendan Murphy$^\star$}\\[.5cm]
{$^\dagger$National Centre for Geocomputation, National University of Ireland Maynooth, Ireland}\\
{$^\star$School of Mathematical Sciences, University College Dublin, Ireland}
}
\date{}
\begin{document}

\maketitle

\begin{abstract}
\noindent Model-based clustering methods for continuous data are well established and commonly used in a wide range of applications. However, model-based clustering methods for categorical data are less standard. Latent class analysis is a commonly used method for model-based clustering of binary data and/or categorical data, but due to an assumed local independence structure there may not be a correspondence between the estimated latent classes and groups in the population of interest. The mixture of latent trait analyzers model extends latent class analysis by assuming a model for the categorical response variables that depends on both a categorical latent class and a continuous latent trait variable; the discrete latent class accommodates group structure and the continuous latent trait accommodates dependence within these groups. Fitting the mixture of latent trait analyzers model is potentially difficult because the likelihood function involves an integral that cannot be evaluated analytically. We  develop a variational approach for fitting the mixture of latent trait models and this provides an efficient model fitting strategy. The mixture of latent trait analyzers model is demonstrated on the analysis of data from the National Long Term Care Survey (NLTCS) and voting in the U.S. Congress. The model is shown to yield intuitive clustering results and it gives a much better fit than either latent class analysis or latent trait analysis alone. 
\end{abstract}

\section{Introduction}
\label{se:introduction}

\noindent Model-based clustering methods are widely used because they offer a coherent strategy for clustering data where uncertainties can be appropriately quantified using probabilities. Many model-based clustering methods are based on the finite Gaussian mixture model \citep[eg.][]{CelGov95,fraley02,mcnicholas08} and these methods have been successfully applied to the clustering of multivariate continuous measurement data; more recent extensions of model-based clustering for continuous data use finite mixtures of non-Gaussian models \citep[eg.][]{lin07,karlis08,lin10,andrews11}.

Categorical data arise in a wide range of applications including the social sciences, health sciences and marketing. Latent class analysis and latent trait analysis are two commonly used latent variable models for categorical data \citep[eg.][]{Bar11}. In the latent class analysis model, the dependence in the data is explained by a categorical latent variable that identifies groups (or classes) within which the response variables are independent (also known as the local independence assumption). Latent class analysis is used widely for model-based clustering of categorical data, however,  if the condition of independence within the groups is violated, then the latent class model will suggest more than the true number of groups and so the results are difficult to interpret and can be potentially misleading. The latent trait analysis model uses a continuous univariate or multivariate latent variable, called a latent trait, to model the dependence in categorical response variables.
If data come from a heterogeneous source or multiple groups, then a single multi-dimensional latent trait may not be sufficient to model the data. For these reasons, the proposed mixture of latent trait analyzers (MLTA) model is developed for model-based clustering of categorical data where a categorical latent variable identifies groups within the data and a latent trait is being used to accommodate the dependence between outcome variables within each cluster. 

In this paper, we focus on the particular case of binary data and we propose two different mixture of latent trait analyzers models for model-based clustering of binary data: a general model that supposes that the latent trait has a different effect in each group and the other more restrictive, but more parsimonious, model that supposes that the latent trait has the same effect in all groups. Thus, the proposed family of models (MLTA) is a categorical analogue of some parsimonious model families used for clustering continuous data \citep[eg.][]{fraley02, mcnicholas08,andrews11}. The MLTA family of models is most like the PGMM family \citep{mcnicholas08} which is based on the mixture of factor analyzers model \citep{ghahramani97} however the MLTA model accommodates binary response variables instead of continuous variables. In addition, the model can be easily applied to nominal categorical data by coding the data in binary form. 
The MLTA model is a generalization of the mixture of Rasch models \citep{Rost90}, and is a special case of multilevel mixture item response models \citep{vermunt07}; these connections are explored in more detail in Section~\ref{se:related}, after the MLTA model has been fully introduced. 

Fitting the mixture of latent trait analyzers model is potentially difficult because the likelihood function involves an integral that cannot be evaluated analytically. However, we propose using a variational approximation of the likelihood as proposed by \cite{Tip99} for model fitting purposes because it is extremely easy to implement and it converges quickly, so it can easily deal with high dimensional data and high dimensional latent traits.

The mixture of latent trait analyzers model is demonstrated through the analysis of two data sets: the National Long Term Care Survey (NLTCS) data set \citep{EFJ07} and a U.S. Congressional Voting data set \citep{UCI}.
The latent class analysis and the latent trait analysis models are not sufficient to summarize these two data sets. However,  the mixture of latent trait analyzers model detects the presence of several classes within these data and the dependence structure within groups is explained by a latent trait.

The paper is organized as follows.
Section \ref{se:LCA} provides an introduction to latent class analysis.
Section \ref{se:LTA} provides an introduction to latent trait analysis with a description of three of the most common techniques to evaluate the likelihood in this model: the Gauss-Hermite quadrature approach (Section \ref{sec:GH}), Monte Carlo sampling (Section \ref{sec:MC}), and the variational approach (Section \ref{sec:ltaV}). 
In Section \ref{se:MLTA} we introduce the mixture of latent trait analyzers model and a parsimonious version of the model (Section \ref{sec:parsimonious}). An interpretation of the model parameters is outlined in Section \ref{sec:interpret}, a discussion of models related to the MLTA model is given in Section~\ref{se:related}, and a variational approach for estimating the model parameters is proposed in Section~\ref{sec:fitting}. The issue of model identifiability is discussed in Section~\ref{sec:identifiability}. 
In Section \ref{sec:modelsel} an adjustment to the BIC \citep{Sch78} is proposed as a model selection criterion and Pearson's $\chi^2$ test and the truncated sum of squares Pearson residual criterion \citep{EFJ07} are used to assess model fit. Computational aspects of fitting the MLTA model are outlined in Section~\ref{sec:computational}.
The two applications are presented in Section~\ref{se:NLTCS} and Section~\ref{se:voting}. We conclude in Section~\ref{se:conclusions} by discussing the model and the results of its application.

\section{Latent Class \& Trait Analysis}

In this section, we give an overview of latent class and latent trait analysis which are the basis of the mixture of latent trait analyzers model. We also outline how the models are fitted in practice. 

\subsection{Latent Class Analysis}
\label{se:LCA}

\noindent The latent class analysis (LCA) can be used to model a set of $N$ observations $\mathbf{x}_1,\ldots,\mathbf{x}_N$ each involving $M$ binary (categorical) variables. Each observation ${\mathbf x}_n=(x_{n1},x_{n2},\ldots,x_{nM})$, where $x_{nm}$ records the value of binary variable $m$ for observation $n$ where $x_{nm}\in\{0,1\}$. 

LCA assumes that there is a latent categorical variable $\mathbf{z}_n=(z_{n1},z_{n2},\ldots,z_{nG})$ for which $z_{ng}=1$ if observation $n$ belongs to class $g$ and $z_{ng}=0$ otherwise. We assume that  $\mathbf{z}_{n}\sim \mbox{Multinomial}(1,(\eta_1,\eta_2,\ldots,\eta_G))$, where $\eta_{g}$ is the prior probability that a randomly chosen individual is in the $g$th  class ($\sum_{g=1}^{G}\eta_{g}=1$ and $\eta_{g}\geq 0$ for all $\, g=1,\ldots,G$). Furthermore, conditional on the latent class variable $z_{ng}=1$, the response variables $\mathbf{x}_n=(x_{n1},\ldots,x_{nM})$ are distributed as independent Bernoulli random variables with parameters $\pi_{g1},\pi_{g2},\ldots,\pi_{gM}$; therefore, $\pi_{gm}$ is the conditional probability of observing $x_{nm}=1$ if the observation is from class $g$ (that is, $z_{ng}=1$). Hence, the dependence between variables is explained by the latent class variable. LCA can be seen as a finite mixture model with mixing proportions $\eta_g$ in which the component distributions are multivariate independent Bernoulli distributions. Consequently the log-likelihood function is given as,
\begin{equation} \label{lca.ll}
\ell = \sum_{n=1}^N \log \left( \sum_{g=1}^G\eta_{g}\prod_{m=1}^M \pi_{gm}^{x_{nm}}(1-\pi_{gm})^{1-x_{nm}} \right).
\end{equation}
\cite{Bar11} describe how to estimate the parameters by maximum likelihood estimation via the expectation-maximisation (EM) algorithm \citep{dempster77}.

When LCA is used in a model-based clustering context, the inferred latent classes are assumed to correspond to clusters in the data; this may be a valid conclusion when the model assumptions are correct but may not be otherwise. 

\subsection{Latent Trait Analysis}
\label{se:LTA}
\noindent Latent trait analysis (LTA) can also be used to model a set of $N$ multivariate binary (categorical) observations. LTA assumes that there is a $D$ dimensional continuous latent variable $\mathbf{y}_n$ ($n=1,\ldots,N$) underlying the behavior of the $M$ categorical response variables within each observation $\mathbf{x}_n$ \citep{Bar11}. The LTA model assumes that 
\begin{equation}\label{lta.px}
p(\mathbf{x}_n)=\int p(\mathbf{x}_n|\mathbf{y}_n) p(\mathbf{y}_n)\,d\mathbf{y}_n
\end{equation}
where the conditional distribution of $\mathbf{x}_n$ given $\mathbf{y}_n$ is
\begin{equation}\label{lta.pxy}
p(\mathbf{x}_n|\mathbf{y}_n)=\prod_{m=1}^Mp\left( x_{nm}|\mathbf{y}_n\right)=\prod_{m=1}^M\left(\pi_{m}(\mathbf{y}_n)\right)^{x_{nm}}\left(1-\pi_{m}(\mathbf{y}_n)\right)^{1-x_{nm}},
\end{equation}
and the response function is a logistic function
\begin{equation*} \label{lta.pi}
\pi_{m}(\mathbf{y}_n)=p\left( x_{nm}=1|\mathbf{y}_n\right)= \dfrac{1}{1+\exp [-(b_{m}+\mathbf{w}_{m}^T\mathbf{y}_n)]} ,\quad 0 \leq \pi_{m}(\mathbf{y}_n) \leq 1 
\end{equation*}
where $b_m$ is the intercept parameter and ${\mathbf w}_m$ are the slope parameters in the logistic function. Thus, the conditional probability that $x_{nm}=1$ given $\mathbf{y}_{n}$ is an increasing function of $b_{m}+\mathbf{w}_m^T\mathbf{y}_n$ and if $\mathbf{w}_m={\mathbf 0}$ we get a probability equal to $1/(1+\exp(-b_m))$ independently of the value of $\mathbf{y}_n$. Finally, it is assumed that $\mathbf{y}_n\sim \mathcal{N}(\mathbf{0},\mathbf{I})$. Thus, although the variables within $\mathbf{x}_n$ are conditionally independent given $\mathbf{y}_n$, the marginal distribution of $\mathbf{x}_n$ accommodates dependence between the variables. 

Therefore, the log-likelihood is
\begin{equation} \label{lta.ll}
\ell =\sum_{n=1}^N \log\left(\int p({\mathbf x}_{n}|\mathbf{y}_n) p(\mathbf{y}_n)\,d\mathbf{y}_n\right)=\sum_{n=1}^N \log\left(\int \prod_{m=1}^M p(x_{nm}|\mathbf{y}_n) p(\mathbf{y}_n)\,d\mathbf{y}_n\right)
\end{equation}
The integral on (\ref{lta.ll}) cannot be evaluated analytically, so it is necessary to use other methods to evaluate it. These are briefly reviewed in Section~\ref{sec:GH}-Section~\ref{sec:ltaV}.

\subsubsection{Gauss-Hermite Quadrature} \label{sec:GH}
\noindent \cite{BA81} proposed a method to evaluate the integral in (\ref{lta.ll}) by using the Gauss-Hermite quadrature \citep{AS64}:
\begin{equation*} \label{lta.GH.ll}
\ell \approx \sum_{n=1}^N\log\left( \sum_{q=1}^Q p(\mathbf{x}_n|\mathbf{y}_q)h(\mathbf{y}_q)\right)=\sum_{n=1}^N \log\left(\sum_{q=1}^{Q} \prod_{m=1}^M p(x_{nm}|\mathbf{y}_q) h(\mathbf{y}_q)\right),  
\end{equation*}
where $Q$ is the number of set of integration points and $h(\mathbf{y}_{1}),\ldots,h(\mathbf{y}_{Q})$ ($\sum_{q=1}^Q h(\mathbf{y}_q)=1$) are the weights for the sets of points $\mathbf{y}_{1},\ldots,\mathbf{y}_{Q}$. In practice this method treats the latent variables as discrete taking values $\mathbf{y}_{1},\ldots,\mathbf{y}_{Q}$ with probabilities $h(\mathbf{y}_{1}),\ldots,h(\mathbf{y}_{Q})$. Thus, the model density can be approximated by a finite mixture model with $Q$ component densities, where $p(\mathbf{x}_n|\mathbf{y}_1),\ldots,p(\mathbf{x}_n|\mathbf{y}_Q)$ are the component distributions and $h(\mathbf{y}_1)\ldots,h(\mathbf{y}_Q)$ are fixed mixing proportions. \cite{Bar11} explain this approach in detail and outline its drawbacks. The number of component densities, $Q$, required increases exponentially in $D$, so the Gauss-Hermite quadrature can be hard to implement and can be quite slow. Furthermore when the $\mathbf{w}_m$ parameters are very unequal their estimates can diverge to infinity.

\subsubsection{Monte Carlo Sampling} \label{sec:MC}
\noindent An alternative approach is the Monte Carlo method \citep{SRL97}, that samples $\mathbf{y}_l \; (l=1,\ldots,L)$ from the $D$-dimensional latent distribution, and approximates the log-likelihood as
\begin{equation*} \label{lta.mc}
\ell \approx \sum_{n=1}^N \log  \left( \dfrac{1}{L} \sum_{l=1}^L p(\mathbf{x}_{n}|\mathbf{y}_l)\right)=\sum_{n=1}^N \log  \left( \dfrac{1}{L} \sum_{l=1}^L \prod_{m=1}^{M}p({x}_{nm}|\mathbf{y}_l)\right).
\end{equation*}
This approximation of the model density can be seen as a finite mixture model with component densities $p(\mathbf{x}_n|\mathbf{y}_1),\ldots,p(\mathbf{x}_n|\mathbf{y}_L)$ and the mixing proportions are equal to $\frac{1}{L}$, but usually $L$ is quite large, making the implementation of this method quite difficult.

\subsubsection{Variational Approximation} \label{sec:ltaV}

\noindent \cite{Tip99} proposed to use a variational approximation \citep{JJ96} to fit the latent trait analysis model in a manner that is fast in convergence and easy to implement. The main aim of this approach is to maximize a lower bound that approximates the likelihood function. In latent trait analysis the log-likelihood function (\ref{lta.ll}) is governed by the logistic sigmoid function, that can be approximated by the exponential of a quadratic form involving variational parameters $\boldsymbol{\xi}_{n}=(\xi_{n1},\ldots,\xi_{nM})$ where $\xi_{nm}\neq 0$ for all $m=1,\ldots,M$. This allows for the computation of an approximate log-likelihood in closed form. In this case the lower bound of each term in the log-likelihood is given by,
\begin{equation*}	\label{lta.v.lb}
\begin{split}
\mathcal{L}(\boldsymbol{\xi}_{n})&= \log\left(  \tilde{p}\left(\mathbf{x}_{n}|\boldsymbol{\xi}_{n}\right) \right)\\
&= \log\left(  \int \prod_{m=1}^M \tilde{p}(x_{nm}|\mathbf{y}_n,\xi_{nm}) p(\mathbf{y}_n)\,d\mathbf{y}_n\right),
\end{split}
\end{equation*}
where 
\begin{equation*}\label{lta.v.pxyXi}
\tilde{p}\left( x_{nm}|\mathbf{y}_n,\xi_{nm}\right)=\sigma(\xi_{nm})\exp\left( \dfrac{A_{nm}-\xi_{nm}}{2}+\lambda(\xi_{nm})\left( A_{nm}^2-\xi_{nm}^2\right) \right),  
\end{equation*}
$\sigma(\xi_{nm})=(1+\exp(-\xi_{nm}))^{-1}$, $A_{nm}=\left( 2x_{nm}-1\right)  \left( b_{m}+\mathbf{w}_{m}^T\mathbf{y}_n\right) $ and $\lambda(\xi_{nm})=(\frac{1}{2}-\sigma(\xi_{nm}))/2\xi_{nm}$. This approximation has the property that $\tilde{p}\left( x_{nm}|\mathbf{y}_n,\xi_{nm}\right)\leq p\left( x_{nm}|\mathbf{y}_n\right)$ with equality when $|\xi_{nm}|=A_{nm}$ and thus it follows $ \tilde{p}\left(\mathbf{x}_{n}|\boldsymbol{\xi}_{n}\right)\leq p\left(\mathbf{x}_{n}\right)$.
\cite{Tip99} and \cite{Bis06} outline a variational EM algorithm that maximizes this lower bound and thus fit a latent trait analysis model. 

However, the approximation of the log-likelihood obtained by using the variational approach is always less or equal than the true log-likelihood, so it may be advantageous to get a more accurate estimate the log-likelihood at the last step of the algorithm using Gauss-Hermite quadrature (Section~\ref{sec:GH}); this is discussed further in Section~\ref{sec:computational}. 

\subsubsection{Comparison of estimation methods}
\label{sec:comparison}

There are some advantages when using the variational approach to approximate the integral in the LTA likelihood. Firstly, the method involved avoids the need for user specified quantities like the number of quadrature points or the number of Monte Carlo samples. In addition, the variational approach involves iterating a series of closed-form updates that are easy to compute. Secondly, the algorithm usually converges considerably more quickly than in Gauss-Hermite quadrature and Monte Carlo sampling case, particularly so for large data sets and the approximation of the likelihood is more accurate as dimensionality increases, as the likelihood becomes more Gaussian in form.
But, since the variational approach estimates the true likelihood function by using a lower bound, the approximation of the likelihood given by this method is always less or equal than the true value, so for low dimensional latent traits it may be better to use the Gauss-Hermite quadrature. Additionally, there are also the usual drawbacks of the EM algorithm: the results can vary because of the initialization of this algorithm and there is the risk of converging to a local maximum instead of the global maximum of the lower bound of the log-likelihood. 

\begin{figure}[]
		\begin{center}
		\includegraphics[scale=0.7]{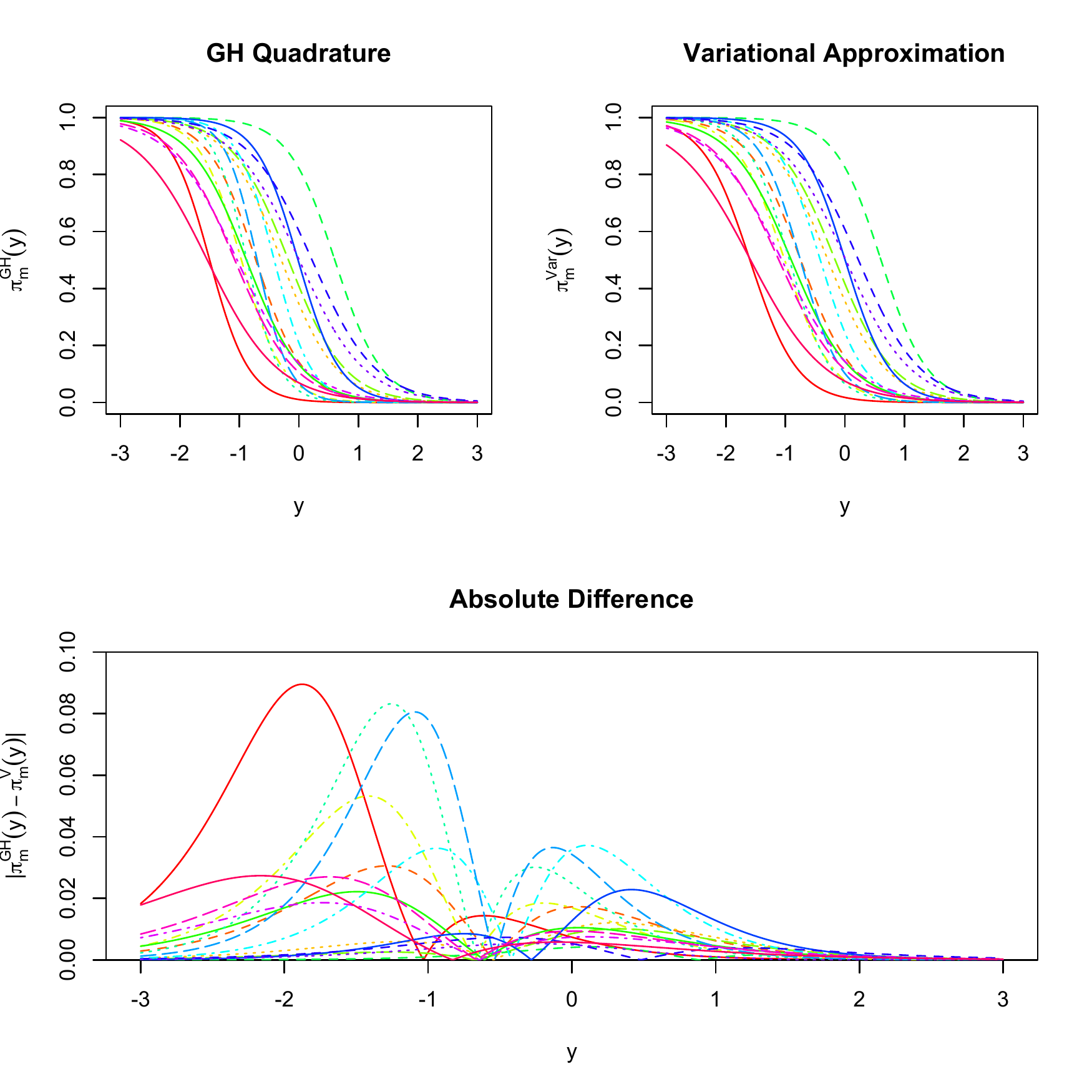}
		\end{center}
		\caption{Response function for the Gauss-Hermite Quadrature and Variational Approach. The absolute difference of the response functions is also shown.}

		\label{fig:ghv}
\end{figure}

In Figure~\ref{fig:ghv} we compare the response function estimated by the Gauss-Hermite quadrature with 6 quadrature points and the variational approach for a unidimensional latent trait model. The data consists of  $M=16$ binary variables and $N=21574$ observations, and are fully described in Section~\ref{se:NLTCS}. In this example the two methods bring to very similar results, as shown by the plot of the two response functions, and their absolute differences that are always $<0.1$.

Furthermore, \cite{Tip99} investigates the accuracy of the variational approach in a high dimensional context comparing the estimates given by the variational and the Monte Carlo sample approaches in terms of both computing time and errors.

\section{Mixture of Latent Trait Analyzers}
\label{se:MLTA}
\noindent The mixture of latent trait analyzers (MLTA) model generalizes the latent class analysis and latent trait analysis by assuming that a set of $N$ observations $\mathbf{x}_1,\ldots,\mathbf{x}_N$ comes from $G$ different groups, and the behavior of the $M$ categorical response variables given by each observation $\mathbf{x}_n$ $(n=1,\ldots,N)$ depends on both the group and the $D$ dimensional continuous latent variable $\mathbf{y}_n$.  Thus, the MLTA model is a mixture model for binary data but where observations are not necessarily conditionally independent given the group memberships. In fact, the observations within groups are modeled using a latent trait analysis model and thus dependence is accommodated.

Suppose that each observation comes from one of $G$ groups and we have $\mathbf{z}_{n}=(z_{n1},z_{n2},\ldots,z_{nG})$ which is an indicator of the group membership. We assume that $\mathbf{z}_n\sim\mbox{Multinomial}(1,(\eta_1,\eta_2,\ldots,\eta_G))$ where $\eta_g$ is the prior probability of a randomly chosen observation coming from the $g$th group ($\sum_{g'=1}^{G}\eta_{g'}=1$ and $\eta_{g}\geq 0$ $\forall \, g=1,\ldots,G$). Further, we assume that the conditional distribution of $\mathbf{x}_n$ given that the observation is from group $g$ (ie. $z_{ng}=1$) is a latent trait analysis model with parameters $b_{mg}$ and $\mathbf{w}_{mg}$; this yields the mixture of latent trait analyzers (MLTA) model. 

Thus, the MLTA model is of the form,
\begin{equation*}
\begin{split}
p( \mathbf{x}_n) 
&= \sum_{g=1}^{G}\eta_g p\left({\mathbf x}_n|z_{ng}=1\right)=\sum_{g=1}^G  \eta_{g} \int p\left( \mathbf{x}_n|\mathbf{y}_n,z_{ng}=1\right) p(\mathbf{y}_n)\,d\mathbf{y}_n
\end{split}
\end{equation*}
where the conditional distribution given $\mathbf{y}_n$ and $z_{ng}=1$ is
\begin{equation} \label{mlta.pxy}
\begin{split}
p\left( \mathbf{x}_n|\mathbf{y}_n,z_{ng}=1\right)&=\prod_{m=1}^M p\left( x_{nm}|\mathbf{y}_n,z_{ng}=1\right)\\
&=\prod_{m=1}^M\left(\pi_{mg}(\mathbf{y}_n)\right)^{x_{nm}}\left(1-\pi_{mg}(\mathbf{y}_n)\right)^{1-x_{nm}},
\end{split}
\end{equation}
and the response function for each group is given by
\begin{equation} \label{mix.pig}
\pi_{mg}(\mathbf{y}_n)= p\left( x_{nm}=1|\mathbf{y}_n,z_{ng}=1\right) = \dfrac{1}{1+\exp\left[-( b_{mg}+\mathbf{w}_{mg}^T\mathbf{y}_n) \right]} ,  
\end{equation}
where $b_{mg}$ and $\mathbf{w}_{mg}$ are the model parameters. In addition, it is assumed that the latent variable $\mathbf{y}_n\sim \mathcal{N}(\mathbf{0},\mathbf{I}) $.

The log-likelihood can be written as
\begin{equation} \label{mix.ll}
\ell = \sum_{n=1}^N \log\left( \sum_{g=1}^G  \eta_{g} \int \prod_{m=1}^M p\left( x_{nm}|\mathbf{y}_n,z_{ng}=1\right) p(\mathbf{y}_n)\,d\mathbf{y}_n \right),
\end{equation}
so, the model is a finite mixture model in which the component distributions are latent trait analysis models and the mixing proportions are $\eta_1,\eta_2,\ldots,\eta_G$. 

\subsection{Parsimonious Model}
\label{sec:parsimonious}
\noindent It is sometimes useful to use a more parsimonious response function than the one presented in (\ref{mix.pig}); this is especially important when the data set is high dimensional, comes from several different groups and the continuous latent variable is high dimensional. 

Similarly to the factor analysis setting \citep{Bar11}, $\frac{D\times(D-1)}{2}$ parameters are constrained because of the
indeterminacy arising from all the possible rotations of $\mathbf{w}_g=(\mathbf{w}_{1g},\ldots,\mathbf{w}_{Mg})$, so the model in (\ref{mix.pig}) involves $(G-1)+(G \times M)+G \times \left(M \times D - {D\times(D-1)}/{2}\right)$ free parameters, of which $G \times \left(M \times D - {D\times(D-1)}/{2}\right)$ are the  parameters $\mathbf{w}_1,\mathbf{w}_2,\ldots,\mathbf{w}_G$. 

If the $\mathbf{w}_{g}$ parameters are constrained to be the same in each group, a more parsimonious model would be: 
\begin{equation} \label{mix.w.pig}
\pi_{mg}(\mathbf{y}_n)= \dfrac{1}{1+\exp\left[-( b_{mg}+\mathbf{w}_{m}^T\mathbf{y}_n) \right]} ,\quad 0\leq \pi_{gm}(\mathbf{y}_n)\leq 1,
\end{equation} 
that supposes that the $\mathbf{w}_{mg}\equiv\mathbf{w}_m$ for each group, so that it involves $(G-1)+(G \times M)+\left( M \times D - {D\times(D-1)}/{2}\right)$ free parameters.

\subsection{Interpretation of Model Parameters}
\label{sec:interpret}
\noindent In the mixture of latent trait analyzers model $\eta_g$ is the mixing proportion for the group $g$  $(g=1,\ldots,G)$, that corresponds to the prior probability that a randomly chosen individual is in the $g$-th group. 
The behavior of the individuals within the group $g$ is characterized by the parameters $b_{mg}$ and $\mathbf{w}_{mg}$. In particular $b_{mg}$ has a direct effect on the probability of a positive response to the variable $m$ given by an individual in group $g$, through the relationship
\begin{equation} \label{mix.pi0}
\pi_{mg}(\mathbf{0})=p(x_{nm}=1|\mathbf{y}_n=\mathbf{0},z_{ng}=1)=\dfrac{1}{1+\exp(-b_{mg})}.
\end{equation} 
The value $\pi_{mg}(\mathbf{0})$ is the probability that the median individual in group $g$ has a positive response for the variable $m$, since the continuous latent variable is distributed as a $\mathcal{N}(\mathbf{0},\mathbf{I})$. 
A measure of the heterogeneity of the values of variable $m$ within group $g$ is given by the slope value $\mathbf{w}_{mg}$; the larger the value of $\mathbf{w}_{mg}$ the greater the differences in the probabilities of positive response in the variable $m$ for observations from group $g$. The value of the slope parameters also account for the dependence between observed data variables. For example, if two variables $m$ and $m'$ yield   a positive (negative) value for $\mathbf{w}_{mg}^T\mathbf{w}_{m'g}$ then they will both simultaneously have a probability of a positive outcome greater (lesser) than the median probability for group $g$ more often than expected under local independence. 

The quantity $w_{dmg}$ can be used to calculate the correlation coefficient, within each group $g$, between the observed variables $\mathbf{x}_m$ and the latent variable $\mathbf{y}_d$ $(d=1,\ldots,D)$ which is given by its standardized value \citep{BSMG02},
\begin{equation} \label{mix.sw}
w_{dmg}^*=\dfrac{w_{dmg}}{\sqrt{\sum_{d'=1}^Dw_{d'mg}^2+1}}. 
\end{equation}

Another useful quantity for analyzing the dependence within groups is a version of $\mathit{lift}$ \citep{BMUT97}. We use $\mathit{lift}$ within each group to quantify  the effect of the dependence on the probability of two positive responses compared to the probability of two positive responses under an independence model. 
The $\mathit{lift}$ is defined as,
\begin{equation} \label{mix.lift}
\mathit{lift}=\dfrac{\mathbb{P}\left(x_{nm}=1,x_{nk}=1|z_{ng}=1 \right)}{\mathbb{P}\left(x_{nm}=1|z_{ng}=1\right)\mathbb{P}\left(x_{nk}=1|z_{ng}=1 \right)} \qquad
\end{equation}
where $m=1,2,\ldots,M$ and $m\neq k$.
Two independent positive responses have $\mathit{lift}=1$: the more the variables are dependent, the further the value of the $\mathit{lift}$ is from 1. It is possible to estimate it by using the Gauss-Hermite quadrature as,
\begin{equation} \label{mix.liftgh}
\mathit{lift}\approx \dfrac{\sum_{q=1}^Q \pi_{mg}(\mathbf{y}_q) \pi_{kg}(\mathbf{y}_q) h(\mathbf{y}_q)}{\left( \sum_{q=1}^Q \pi_{mg}(\mathbf{y}_q) h(\mathbf{y}_q)\right) \left( \sum_{q=1}^Q \pi_{kg}(\mathbf{y}_q) h(\mathbf{y}_q)\right) },
\end{equation}
where $h(\mathbf{y}_1),\ldots, h(\mathbf{y}_Q)$ are the appropriate weights associated with the quadrature set of points $\mathbf{y}_1,\ldots, \mathbf{y}_Q$ and the parameters are estimated using the variational method (Section~\ref{sec:fitting}). Lift values that are much less than 1 are evidence of negative dependence within groups and lift values that are greater than 1 are evidence of positive dependence within groups. 

Finally, the posterior distribution of the latent variable $\mathbf{y}_n$ conditional on the observation belonging to a particular group can be obtained from the model outputs (see Section~\ref{sec:fitting}) 
and the posterior mean estimates of these $\mathbf{y}_n$ scores can be used to interpret the latent variables within each group. 

\subsection{Related Models}
\label{se:related}

The MLTA model has a lot of common characteristics with a number of models in the statistical and wider scientific literature. 

The MLTA model can be seen as a discrete response variable analogue of the continuous response variable mixture of factor analyzers (MFA) model \citep{ghahramani97, mclachlan03} and more recently developed parsimonious analogues of this model \citep{mcnicholas08,mcnicholas10,baek10}. In these models, a discrete latent class variable accounts for grouping structure and a latent factor score accounts for correlation between response variables within groups. The component means in the MFA model are analogous to the intercepts in the MLTA model, the loading matrix is analogous to the slope parameters, the mixing proportions take an identical role in both models. Additionally, \cite{muthen01} describes a general latent variable model for continuous outcome variables as used by the Mplus software and this model also uses both continuous and discrete latent variables in the model framework. 

The Rasch model \citep{rasch60} is a commonly used model for analyzing binary (and categorical data) and it has been widely used in educational statistics. The Rasch model has a very similar structure to the LTA model. In the Rasch model, the probability of a positive outcome is given as $${\mathbb P}(x_{nm}=1|\beta_{n},\delta_m)=\frac{\exp(\beta_n-\delta_m)}{1+\exp(\beta_n-\delta_m)}, $$ where the $\beta_{n}$ values are called {\em ability} parameters and the $\delta_m$ values are called {\em difficulty} parameters and these correspond to the latent trait and the intercept parameters in the LTA model. Furthermore, the model makes a local independence assumption when constructing the probability $${\mathbb P}(\mathbf{x}_n|\beta_n,\boldsymbol\delta)=\prod_{m=1}^{M}{\mathbb P}(x_{nm}|\beta_n,\delta_m),$$ where $\boldsymbol\delta=(\delta_1,\delta_2,\ldots,\delta_m)$. In some formulations of the Rasch model, the ability parameters are treated as fixed unknown parameters and in other formulations they are treated as random effects thus making the model equivalent to the univariate LTA model. 

Consequently, the mixture of latent trait analyzers (MLTA) model has a similar structure to the mixture Rasch model \citep{Rost90,Rost95,VDY07}, where the response function is: $${\mathbb P}(x_{nm}=1|\beta_{n},\delta_{mg},z_{ng}=1)=\frac{\exp(\beta_{n}-\delta_{mg})}{1+\exp(\beta_{n}-\delta_{mg})}. $$
However, as with the Rasch model, the mixture Rasch model usually includes a univariate ability parameter $\beta_{n}$ that is the same across all the variables and groups, that would be equivalent to assuming that MLTA model $w_{1g}\equiv w_{2g}\equiv\ldots\equiv w_{Mg}$ and $w_{mg}\equiv w_m$ for all $g$. 

In \cite{vDY07a} they extend the mixture Rasch model in order to allow multivariate ability parameters:
$${\mathbb P}(x_{nm}=1|\mathbf{q}_{m},\boldsymbol{\beta}_{n},\delta_m,z_{ng}=1)=\frac{\exp(\mathbf{q}_{m}^T\boldsymbol{\beta}_{n}-\delta_{mg})}{1+\exp(\mathbf{q}_{m}^T\boldsymbol{\beta}_{n}-\delta_{mg})}, $$
where $\mathbf{q}_{m}$ are variable-specific parameters, and $\boldsymbol{\beta}_{n}$ is the $D$-dimensional ability parameter.  This is equivalent to a MLTA model with $w_{mg}\equiv w_m$ for all $g$, as in our parsimonious model.

Thus, many versions of the mixture Rasch model are included as special cases of the MLTA model. Again, the latent ability parameter is either assumed to be a fixed or random effect that varies across observations to accommodate different outcome probabilities; the latent ability parameter is the primary quantity of interest in many Rasch modeling applications.

Another model that has a very similar structure to the mixture Rasch model is the mixed latent trait (MLT) model described in  \cite{Ueb99}, this model also has a univariate latent trait, uses a probit structure and numerical integration are required for computing the likelihood, so they can be difficult to apply to large heterogeneous data sets. \cite{qu96} and \cite{hadgu98} use a two component model with a similar form to the MLT model in medical diagnosis. 

Additionally, \cite{Ueb99} also describes a probit latent class model (PLCA) which has a similar structure to the MLTA model, but this model uses a multivariate latent trait which has the same dimensionality as outcome variable. \cite{VM05} developed a latent class factor analysis (LCFA) model which uses discrete latent variables to develop a model with similar modeling aims to latent trait analysis, but which can be fitted in a much more computationally efficient manner. 

Perhaps, the MLTA model proposed has closest connections to the family of models identified as multilevel mixture item response models \citep{vermunt07}.
Key differences between our model and the multilevel mixture item response model are that we focus on a multivariate trait parameter (which is optional only in the model of Vermunt) and that we further also introduce a constrained parsimonious version of the model. 
In addition, we offer a computationally efficient alternative algorithm for fitting this model without the need to resort to numerical quadrature methods. 
The MLTA and multilevel mixture item response models can be estimated using the software Latent GOLD \citep{LatentGold08}, but the fact that this software uses quadrature methods for the numerical integration of continuous latent variables probably makes it less suitable for analyzing large datasets with underlying highdimensional latent trait structures.

\subsection{Model Fitting}
\label{sec:fitting}

When fitting the MLTA model, the integral in the log-likelihood (\ref{mix.ll}) cannot be solved analytically since it is exactly the same as in latent trait analysis (\ref{lta.ll}). To obtain the approximation of the likelihood it is possible to use an EM algorithm:
\begin{enumerate}
\item \label{em.e}  \textit{E-step:} Estimate $z_{ng}$ as the approximate posterior probability that an individual with response vector $\mathbf{x}_n$ belongs to group $g$.
\item \textit{M-step:} Estimate $\eta_{g}$ as the proportion of observations in group $g$.

Estimate the integral in the complete-data log-likelihood by using the variational approach.

Estimate the parameters $b_{mg}$ and $\mathbf{w}_{mg}$ for $m=1,2,\ldots,M$ and $g=1,2,\ldots,G$. 

Also, estimate the approximate log-likelihood value. 
\item Return to \textit{step} \ref{em.e} until convergence is attained.
\end{enumerate}
The advantages and the drawbacks of the different estimate methods for the likelihood are the same as in latent trait analysis (Section~\ref{sec:comparison}).

We further focus on the variational approach to approximate the log-likelihood because it can be efficiently implemented.
Similarly to the latent trait analysis case, we introduce the variational parameters $\boldsymbol{\xi}_{ng}=(\xi_{n1g},\ldots,\xi_{nMg})$ where $\xi_{nmg}\neq 0$ for all $m=1,\ldots,M$ to approximate the logarithm of the component densities with a lower bound:
\begin{equation*}
\begin{split}
\mathcal{L}(\boldsymbol{\xi}_{ng})&=\log\left( \tilde{p}(\mathbf{x}_n|z_{ng}=1,\boldsymbol{\xi}_{ng}) \right)\\
&=\log\left(\int \prod_{m=1}^M \tilde{p}\left( x_{nm}|\mathbf{y}_n,z_{ng}=1,\xi_{nmg}\right) p(\mathbf{y}_n)\,d\mathbf{y}_n \right) \\
\end{split}
\end{equation*}
where the conditional distribution $p(x_{mn}|,{\mathbf y}_{n},z_{ng}=1)$ is approximated by,
\begin{equation*}	\label{mix.pxyXi}
\tilde{p}\left( x_{nm}|\mathbf{y}_n,z_{ng}=1,\xi_{nmg}\right)=\sigma(\xi_{nmg})\exp\left( \dfrac{A_{nmg}-\xi_{nmg}}{2}+\lambda(\xi_{nmg})\left( A_{nmg}^2-\xi_{nmg}^2\right) \right), 
\end{equation*}
where $\sigma(\xi_{nmg})=(1+\exp(-\xi_{nmg}))^{-1}$, $A_{nmg}=(2x_{nm}-1)(b_{mg}+\mathbf{w}_{mg}^T\mathbf{y}_n)$ and $\lambda(\xi_{nmg})=(\frac{1}{2}-\sigma(\xi_{nmg}))/2\xi_{nmg}$. 

To obtain the approximation of the log-likelihood it is necessary to use a double EM algorithm on the $(i+1)$th iteration:
\begin{enumerate}
\item \label{var.em.e} \textit{E-step:} Estimate $ z^{(i+1)}_{ng} $ with:
\begin{equation*}	\label{mix.zg.hat}
z^{(i+1)}_{ng} =\dfrac{\eta^{(i)}_{g} \exp\left( \mathcal{L}(\boldsymbol{\xi}^{(i)}_{ng})\right) }{\sum_{g'=1}^G\eta^{(i)}_{g'}\exp\left( \mathcal{L}(\boldsymbol{\xi}^{(i)}_{ng'})\right) }.
\end{equation*}
\item  \textit{M-step:} Estimate $ \eta^{(i+1)}_{g} $ using:
\begin{equation*} \label{mix.etag.hat}
\eta^{(i+1)}_{g}=\dfrac{\sum_{n=1}^N z^{(i+1)}_{ng}}{N}.
\end{equation*}
\item Estimate the likelihood:
\begin{enumerate}  
\item \textit{E-step:} Compute the latent posterior statistics for $\tilde{p}\left( \mathbf{y}_{n}|\mathbf{x}_{n},z^{(i+1)}_{ng}=1, \boldsymbol{\xi}^{(i)}_{ng}\right)$ which is a $\mathcal{N}(\boldsymbol{\mu}^{(i+1)}_{ng},\mathbf{C}^{(i+1)}_{ng})$ density:
\begin{equation*} \label{mix.cng.mung}
\mathbf{C}^{(i+1)}_{ng}=\left[  \mathbf{I} -2\sum_{m=1}^M\lambda(\xi^{(i)}_{nmg})\mathbf{w}^{(i)}_{mg}\mathbf{w}^{(i)T}_{mg}\right] ^{-1}
\end{equation*}
and
\begin{equation*} \label{mix.mung}
\boldsymbol{\mu}^{(i+1)}_{ng}=\mathbf{C}^{(i+1)}_{ng}\left[\sum_{m=1}^M\left(x_{nm}-\dfrac{1}{2}+2\lambda(\xi^{(i)}_{nmg})b^{(i)}_{mg}\right)\mathbf{w}^{(i)}_{mg}\right]. 
\end{equation*}
\item  \textit{M-step:} Optimize the variational parameter $\xi^{(i+1)}_{nmg}$ in order to make the approximation $\tilde{p}(\mathbf{x}_{n}|z^{(i+1)}_{ng}=1,\boldsymbol{\xi}^{(i+1)}_{ng})$ as close as possible to $p(\mathbf{x}_{n}|z^{(i+1)}_{ng}=1)$ for all $\mathbf{x}_n$ using:
\begin{equation*} \label{mix.xi2g}
\xi^{(i+1)2}_{nmg}=\mathbf{w}^{(i)T}_{mg}\left( \mathbf{C}^{(i+1)}_{ng}+\boldsymbol{\mu}^{(i+1)}_{ng}\boldsymbol{\mu}^{(i+1)T}_{ng}\right) \mathbf{w}^{(i)}_{mg} + 2 b^{(i)}_{mg} \mathbf{w}^{(i)T}_{mg}\boldsymbol{\mu}^{(i+1)}_{ng}+b^{(i)2}_{mg},
\end{equation*}
since $\tilde{p}\left( x_{nm}|\mathbf{y}_n,z_{ng}=1,\xi_{nmg}\right)$ is symmetric in $\xi_{nmg}$ choosing either the positive or the negative root of $\xi_{nmg}^2$ yields to the same results.
\item \label{var.em.w}  Optimise the model parameters $\mathbf{w}^{(i+1)}_{mg}$ and $b^{(i+1)}_{mg}$
 in order to increase the approximate likelihood $\tilde{p}\left(\mathbf{x}_1,\ldots,\mathbf{x}_N|\boldsymbol{\xi}^{(i+1)}_{11},\ldots,\boldsymbol{\xi}^{(i+1)}_{NG}\right)$
  using:
\begin{equation*} \label{mix.what}
\hat{\mathbf{w}}^{(i+1)}_{mg}=-\left[2\sum_{n=1}^N z^{(i+1)}_{ng}\lambda(\xi_{nmg}^{(i+1)}) \mathbb{E}[\hat{\mathbf{y}}_{n}\hat{\mathbf{y}}_n^T]^{(i+1)}_g \right]^{-1} \left[ \sum_{n=1}^N z^{(i+1)}_{ng}\left( x_{nm} -\dfrac{1}{2}\right)\hat{\boldsymbol{\mu}}^{(i+1)}_{ng} \right], 
\end{equation*}
where $\hat{\mathbf{w}}^{(i+1)}_{mg}=(\mathbf{w}^{(i+1)T}_{mg},b^{(i+1)}_{mg})^T,\; \hat{\boldsymbol{\mu}}^{(i+1)}_{ng}=(\boldsymbol{\mu}^{(i+1)T}_{ng},1)^T,$ and\\ $ \mathbb{E}[\hat{\mathbf{y}}_{n}\hat{\mathbf{y}}_n^T]^{(i+1)}_g=
\begin{bmatrix}
\mathbf{C}^{(i+1)}_{ng}+\boldsymbol{\mu}^{(i+1)}_{ng}\boldsymbol{\mu}^{(i+1)T}_{ng} & \boldsymbol{\mu}^{(i+1)}_{ng}\\
\boldsymbol{\mu}^{(i+1)T}_{ng} & 1
\end{bmatrix} $.
\item  Estimate the lower bound:
\begin{equation*}	\label{mix.lb}
\begin{split}
\mathcal{L}(\boldsymbol{\xi}^{(i+1)}_{ng})
&=\sum_{m=1}^M\left[ \log\left( \sigma(\xi^{(i+1)}_{nmg})\right)-\dfrac{\xi^{(i+1)}_{nmg}}{2}-\lambda(\xi^{(i+1)}_{nmg})\xi^{(i+1)2}_{nmg}+\left(x_{nm}-\dfrac{1}{2}\right)b^{(i+1)}_{mg}+\lambda(\xi^{(i+1)}_{nmg})b^{(i+1)2}_{mg} \right] \\
&\quad +\dfrac{\log|\mathbf{C}^{(i+1)}_{ng}|}{2}+\dfrac{\boldsymbol{\mu}^{(i+1)T}_{ng}[\mathbf{C}^{(i+1)}_{ng}]^{-1}\boldsymbol{\mu}^{(i+1)}_{ng}}{2},\\
\end{split}
\end{equation*}
and the log-likelihood:
\begin{equation*}	\label{mix.var.applogl}
\ell^{(i)} \approx \sum_{n=1}^N\log\left( \sum_{g=1}^G\eta^{(i+1)}_g \exp(\mathcal{L}(\boldsymbol{\xi}^{(i+1)}_{ng})) \right).\\
\end{equation*}
\end{enumerate}
\item Return to \textit{step} \ref{var.em.e} until convergence is attained.
\item Estimate the log-likelihood by using the Gauss-Hermite quadrature:
\begin{equation} \label{mix.GHll}
\ell \approx \sum_{n=1}^N\log\left( \sum_{g=1}^G\eta_g \sum_{q=1}^Q p(\mathbf{x}_n|\mathbf{y}_q,z_{ng}=1)h(\mathbf{y}_q) \right),
\end{equation}
where $h(\mathbf{y}_1),\ldots, h(\mathbf{y}_Q)$ are the appropriate weights associated to the quadrature set of points $\mathbf{y}_1,\ldots, \mathbf{y}_Q$ and $p(\mathbf{x}_n|\mathbf{y}_q,z_{ng}=1)$ is given by (\ref{mlta.pxy}) and (\ref{mix.pig}). 
\end{enumerate}


To fit the parsimonious model outlined in Section~\ref{sec:parsimonious} it is possible to use the variational approach using the EM algorithm described above, except for the estimate of the model parameters at \textit{step} \ref{var.em.w}:
\begin{equation*} \label{mix.w.what}
\hat{\mathbf{w}}^{(i+1)}_{m}=-\left[2\mathbf{K}^{(i+1)}_m\right]^{-1} \boldsymbol{\gamma}^{(i+1)}_m,
\end{equation*}
where 
\begin{equation*}	\label{mix.w.wh.v}
\hat{\mathbf{w}}^{(i+1)}_{m}=(\mathbf{w}^{(i+1)T}_{m},b^{(i+1)}_{m1},\ldots, b^{(i+1)}_{mG})^T, \quad \boldsymbol{\gamma}^{(i+1)}_m=
\begin{bmatrix}
\sum_{g=1}^G \sum_{n=1}^N z^{(i+1)}_{ng}\left(x_{nm}-\dfrac{1}{2}\right) \boldsymbol{\mu}^{(i+1)}_{ng}\\
\sum_{n=1}^N z^{(i+1)}_{n1} \left(x_{nm}-\dfrac{1}{2}\right)\\
\vdots\\
\sum_{n=1}^N z^{(i+1)}_{nG} \left(x_{nm}-\dfrac{1}{2}\right)
\end{bmatrix}
\end{equation*}
and,
{\small
\begin{equation*}	\label{mix.w.K}
\mathbf{K}^{(i+1)}_{m}=
\begin{bmatrix}
\sum_{g=1}^G \sum_{n=1}^N z^{(i+1)}_{ng}\lambda(\xi^{(i+1)}_{nmg})\mathbb{E}\left[\mathbf{y}_{n}\mathbf{y}_{n}^T\right]^{(i+1)}_g 
& \sum_{n=1}^N z^{(i+1)}_{n1}\lambda(\xi^{(i+1)}_{nm1})\boldsymbol{\mu}^{(i+1)}_{n1} 
& \ldots & 
\sum_{n=1}^N z^{(i+1)}_{nG}\lambda(\xi^{(i+1)}_{nmG})\boldsymbol{\mu}^{(i+1)}_{nG}\\
\sum_{n=1}^N z^{(i+1)}_{n1}\lambda(\xi^{(i+1)}_{nm1})\boldsymbol{\mu}^{(i+1)T}_{n1} & \sum_{n=1}^N z^{(i+1)}_{n1}\lambda(\xi^{(i+1)}_{nm1}) & 0 & 0\\
\vdots & 0 & \ddots & 0\\
\sum_{n=1}^N z^{(i+1)}_{nG}\lambda(\xi^{(i+1)}_{nmG})\boldsymbol{\mu}^{(i+1)T}_{nG} & 0 & 0 & \sum_{n=1}^N z^{(i+1)}_{nG}\lambda(\xi^{(i+1)}_{nmG})\\
\end{bmatrix}
\end{equation*}}

The derivation of these parameter estimates is given in Section 1 of the supplementary material. 

\subsection{Model Identifiability}
\label{sec:identifiability}
Model identifiability is an important issue for a model involving many parameters.
\cite{Goo74}, \cite{Bar11} and \cite{DR10} give a detailed explanation of model identifiability in latent class analysis and
\cite{Bar80} introduces this issue in the latent trait analysis context. 
In \cite{Allman2009} they argue that the classical definition of identifiability is too strong for a lot of latent variable models, and they introduce the concept of ``generic identifiability'' that implies that the set of points for which identfiability does not hold has measure zero. They explore ``generic identifiability'' for different models, including LCA.

A necessary condition for model identifiability is that the number of the free estimated parameters not exceed the number of possible data patterns.
Nevertheless this condition is not sufficient as the actual information in a dataset can be less depending of the size or the frequency of pattern occurrences within the dataset.

As with the loading matrices in mixture models with a factor analytic structure \citep[eg.][]{ghahramani97,mcnicholas08} the slope parameters $\mathbf{w}_g$ are only identifiable up to a rotation of the factors. The rotational freedom of the factor scores is important when determining the number of parameters in the model (Section~\ref{sec:parsimonious} and Section~\ref{sec:modelsel}).

In addition, model identifiability holds if the observed information matrix is full rank. As a result possible non-identifiability can manifest itself through high standard errors of the parameter estimates. Another empirical method that can be used to assess non-identifiability consists of checking whether the same maximized likelihood value is reached with different estimates of the model parameter values when starting the EM algorithm from different values. 

We found that these checks for identifiability were all satisfied in the empirical examples discussed in Section~\ref{sec:applications}.

\section{Model Selection}
\label{sec:modelsel}
\noindent The Bayesian Information Criterion ($\mathrm{BIC}$) (\cite{Sch78}) can be used to select the model,
\begin{equation*}	\label{mix.bic}
\mathrm{BIC}=-2\ell+ k \log(N),
\end{equation*}
where $k$ is the number of free parameters in the model and $N$ is the number of observations. The model with the lower value of $\mathrm{BIC}$ is preferable. It is important to remember that its value could be overestimated if the log-likelihood is approximated by using the variational approach; hence, the proposal to use Gauss Hermite quadrature to evaluate the maximized log-likelihood for model selection purposes.

In the context of MLTA, the values of $G$, $D$ and whether the $\mathbf{w}_g$ values are constrained to be equal across groups need to be determined.

In the mixture of latent trait analyzers context (and more widely within finite mixture models) the $\mathrm{BIC}$ penalizes too much the models with high $D$ and/or high $G$. The $\mathrm{BIC}$ as defined by \cite{Sch78} implicitly assumes that all the parameter estimates depend on the entire set of $N$ observations, but in MLTA the estimates of $\mathbf{b}_{g}$ and $\mathbf{w}_g$ depend just on the observations that belong to group $g$. So, an alternative penalized $\mathrm{BIC}$ is given as
\begin{equation*}	\label{mix.bicp}
\begin{split}
\mathrm{BIC^*}&=-2\ell+ (G-1)\log(N)+ k^* \sum_{g=1}^G \log(\eta_g N)  \\
&=\mathrm{BIC} + k^* \sum_{g=1}^G \log(\eta_g),
\end{split}
\end{equation*}
where $k^*$ is the number of free parameters depending on each group of observations. This penalty penalizes parameters to a lesser extent than BIC because only the estimated number of observations involved in the estimation of each parameter is used in the penalty. This version of BIC has previously been proposed by \cite{Ste02} and similar criteria have been proposed by \cite{Pau98} and \cite{RNSK07}. It is worth noting that $\mathrm{BIC^*}$ will behave in a similar way to $\mathrm{BIC}$ for large sample sizes. Hence, it can be seen as a small sample adjustment to $\mathrm{BIC}$.

The Pearson's $\chi^2$-test can be used to check the goodness of the model fit. The $\chi^2$ statistic  is calculated as
\begin{equation*} \label{mix.chisq}
\begin{split}
\chi^2&=\sum_{p=1}^P\dfrac{(\mathrm{Observed}_{p}-\mathrm{Expected}_p)^2}{\mathrm{Expected}_p}\\
&=\sum_{\underline{n}=1}^{\underline{N}}\dfrac{(\mathrm{Observed}_{\underline{n}}-\mathrm{Expected}_{\underline{n}})^2}{\mathrm{Expected}_{\underline{n}}}+ \left(N - \sum_{\underline{n}=1}^{\underline{N}} \mathrm{Expected}_{\underline{n}}\right) ,\\
\end{split}
\end{equation*}
where $P=2^M$ is the total number of possible patterns, $\underline{N}$ is the number of observed patterns and $\mathrm{Observed}_{\underline{n}}$ and $\mathrm{Expected}_{\underline{n}} = N p(\mathbf{x}_{\underline{n}})$ represent the observed and expected frequencies for the $\underline{n}$-th response pattern, respectively. Under the null hypothesis the $\chi^2$ statistic is asymptotically distributed as a $\chi^2_{P-k-1}$ as $N\rightarrow\infty$. If $M$ is large it is common to have a large number of very small counts and the Pearson's $\chi^2$-test is not applicable. In this case, \cite{EFJ07} suggest the truncated SSPR criterion to examine deviations between expected and observed frequencies via the sum of squared Pearson residual only for the patterns with large observed frequencies.

\section{Computational Aspects}
\label{sec:computational}

\noindent The estimates of the parameters in the latent class analysis models $(D=0)$ are the exact maximum likelihood estimates. Since the dimensionality of the data in the two data sets is large, the parameters for the model with $D \geq 1$  are estimated by using the variational approach and the log-likelihood has been calculated at the last step of the algorithm by using the Gauss-Hermite quadrature, with 5 quadrature points per dimension; the results were not sensitive to the number of quadrature points but the computational time was very heavily dependent on this number. 

The categorical latent variables $\mathbf{z}_n$ $(n=1,\ldots,N)$ have been initialized by randomly assigning each observation to one of the $G$ possible groups.
The variational parameters $\xi_{nmg}$ $(n=1,\ldots,N, \; m=1,\ldots,M, \; g=1,\ldots, G)$ are initialized to be equal to $20$, this implies the initial approximation of the conditional distribution to be very close to $0$ and it reduces the dependence of the final estimates on the initializing values. The model parameters $b_{mg}$ and $\mathbf{w}_{mg}$ have been initialized by random generated numbers from a $\mathcal{N}(0,1)$. Ten random starts of the algorithm were used and the solution with the maximum likelihood value was selected. 

The standard errors of the model parameter have been calculated using the jackknife method \citep{Efr81}. It is worth noting that when employing this method the estimates of the parameter without the $n$-th observation $(n=1,\ldots,N)$ can be obtained in just a few iterations.

Since the EM algorithm is linearly convergent, a criterion based on the Aitken acceleration \citep{MP00} has been used to determine the convergence of the algorithm. The EM algorithm has been stopped when
\begin{equation*} \label{mix.Astop}
\| \ell_{A}^{(i+1)}-\ell_{A}^{(i)}\|<\mathrm{tol}
\end{equation*}
where $i$ is the iteration, $\mathrm{tol}=10^{-2}$ is the desired tolerance, 
\begin{equation*} \label{mix.lA}
\ell_{A}^{(i+1)}=\ell^{(i)}+\dfrac{1}{1-a^{(i)}}\left( \ell^{(i+1)}-\ell^{(i)} \right) \qquad \mathrm{and} \qquad a^{(i)}=\dfrac{\ell^{(i+1)}-\ell^{(i)}}{\ell^{(i)}-\ell^{(i-1)}}.
\end{equation*}

\section{Applications}
\label{sec:applications}

\subsection{National Long Term Care Survey (NLTCS)}
\label{se:NLTCS}

\noindent \cite{E02,E03, E04} and \cite{EFJ07} report on 16 binary outcome variables recorded for 21574 elderly (age 65 and above) people who took part in the National Long-Term Care Survey in the years 1982, 1984, 1989 and 1994. The outcome variables record the level of disability and can be divided in two subsets: the "activities of daily living" (ADLs) and "instrumental activities of daily living" (IADLs). The first subset is composed of the first six variables which include basic activities of hygiene and personal care: \textit{(1)} eating, \textit{(2)} getting in/out of bed, \textit{(3)} getting around inside, \textit{(4)} dressing, \textit{(5)} bathing, \textit{(6)} using the toilet. The second subset concern the basic activities necessary to reside in the community: \textit{(7)} doing heavy house work, \textit{(8)} doing light house work, \textit{(9)} doing laundry, \textit{(10)} cooking, \textit{(11)} grocery shopping, \textit{(12)} getting about outside, \textit{(13)} travelling, \textit{(14)} managing money, \textit{(15)} taking medicine, \textit{(16)} telephoning. The responses are coded as 1 = disabled and 0 = able.

The MLTA model is fitted to these data for $D=0,1,2,3$ and $G=1,2,\ldots,11$; the $D=0$ case corresponds to the LCA model and the $G=1 \; \& \; D>0$ case corresponds to the LTA model.

Table \ref{tab:nltcs.ll} records the log-likelihood evaluated for the parameter estimates found using the algorithm outlined in Section~\ref{sec:ltaV}.
Estimates of the log-likelihood and the lower bound are reported to emphasize the importance of estimating the log-likelihood at the last step of the variational approach by using the Gauss-Hermite quadrature instead of the lower bound values.

\begin{table}[ht]
\caption{Approximated log-likelihood and lower bound (in parentheses).}
\label{tab:nltcs.ll}
\smallskip
\centering
\footnotesize
\begin{tabular}{l|c|cc|cc|cc}
  \hline
\multicolumn{1}{c}{} & \multicolumn{1}{c}{$D=0$} & \multicolumn{2}{c}{$D=1$} & \multicolumn{2}{c}{$D=2$} & \multicolumn{2}{c}{$D=3$} \\ 
  \hline
$G=1$ & -200085.10 & -140318.06 & (-145827.00) & -136169.53 & (-144483.30) & -136075.66 & (-144483.30)  \\ 
  $G=2$ & -152527.30 & -135301.29 & (-139930.20) & -134273.79 & (-139283.00) & -134275.46 & (-139091.30) \\ 
  $G=3$ & -141277.10 & -134362.61 & (-136349.10) & -133025.27 & (-136080.80) & -133008.17 & (-136066.50) \\ 
  $G=4$ & -137464.20 & -133120.36 & (-134540.70) & -131839.77 & (-134253.60) & -132116.82 & (-134145.00) \\ 
  $G=5$ & -135216.20 & -131813.29 & (-133203.40) & -131505.23 & (-133061.10) & -131393.42 & (-132919.80) \\ 
  $G=6$ & -133643.80 & -131396.59 & (-132261.50) & -131154.94 & (-132161.60) & -130992.52 & (-132123.00) \\ 
  $G=7$ & -132659.70 & -131120.79 & (-131840.30) & -130729.39 & (-131785.80) & -130607.37 & (-131731.10) \\ 
  $G=8$ & -132202.90 & -130708.20 & (-131565.90) & -130450.55 & (-131146.60) & -130403.20 & (-131106.60) \\ 
  $G=9$ & -131367.70 & -130342.81 & (-130866.20) & -130164.32 & (-130855.20) & -130155.19 & (-130798.00) \\ 
  $G=10$ & -131155.90 & -130135.91 & (-130806.60) & -130049.64 & (-130681.20) & -129936.33 & (-130544.50) \\ 
  $G=11$ & -130922.60 & -130110.22 & (-130574.40) & -129860.74 & (-130475.10) & -129881.83 & (-130404.50) \\ 
   \hline
\end{tabular}

\end{table}
The estimated $\mathrm{BIC}$ and $\mathrm{BIC}^{*}$ have been used to select the best model (Table \ref{tab:nltcs.bic}), both criteria indicate that the model with 10 groups and a one-dimensional latent variable as the best model. \cite{FHRZ09} show that the LCA model that minimizes the BIC has nineteen latent classes, so the mixture of latent trait analyzers suggests that there are much fewer groups. The nineteen latent class model has a lower BIC (BIC=262165.07) but they argue that this number of classes is not sensible in the context of the application. This suggests that the LCA is using multiple groups to account for dependence and that the latent classes do not necessarily correspond to data clusters.
 
\begin{table}[ht]
\caption{The estimated $\mathrm{BIC}$ (left) and $\mathrm{BIC}^{*}$ (right) for the models with $D=0,1,2,3$ and $G=1,2,\ldots,11$.}
\label{tab:nltcs.bic}
\smallskip
\centering
\scriptsize
\begin{tabular}{@{\extracolsep{-4pt}}l|c|c|c|c}
  \hline
\multicolumn{1}{c}{} & \multicolumn{1}{c}{$D=0$} & \multicolumn{1}{c}{$D=1$} & \multicolumn{1}{c}{$D=2$} & \multicolumn{1}{c}{$D=3$}\\ 
  \hline
$G=1$ & 400329.84 & 280955.46 & 272808.09 & 272760.05 \\ 
$G=2$ & 305383.99 & 271251.24 & 269495.61 & 269778.37 \\ 
$G=3$ & 283053.38 & 269703.18 & 267477.56 & 267862.51 \\ 
$G=4$ & 275597.01 & 267548.00 & 265585.58 & 266698.51 \\ 
$G=5$ & 271270.65 & 265263.18 & 265395.51 & 265870.43 \\ 
$G=6$ & 268295.46 & 264759.09 & 265173.93 & 265687.34 \\ 
$G=7$ & 266496.97 & 264536.80 & 264801.82 & 265535.74 \\ 
$G=8$ & 264882.64 & 264040.94 & 264723.15 & 265746.12 \\ 
$G=9$ & 264252.28 & 263639.47 & 264629.69 & 265868.82 \\ 
$G=10$ & 263998.39 & \textbf{263554.99} & 264879.33 & 266049.81 \\ 
$G=11$ & 263701.27 & 263832.93 & 264980.55 & 266559.53 \\ 
   \hline
\end{tabular}
\quad
\begin{tabular}{@{\extracolsep{-3pt}}l|c|c|c|c}
  \hline
\multicolumn{1}{c}{} & \multicolumn{1}{c}{$D=0$} & \multicolumn{1}{c}{$D=1$} & \multicolumn{1}{c}{$D=2$} & \multicolumn{1}{c}{$D=3$}\\ 
  \hline
$G=1$ & 400329.84 & 280955.46 & 272808.09 & 272760.05 \\ 
$G=2$ & 305360.21 & 271204.68 & 269427.88 & 269690.47 \\ 
$G=3$ & 282997.58 & 269596.55 & 267322.64 & 267661.41 \\ 
$G=4$ & 275504.40 & 267367.43 & 265322.50 & 266345.65 \\ 
$G=5$ & 271136.22 & 264993.58 & 265006.92 & 265366.66 \\ 
$G=6$ & 268113.98 & 264403.40 & 264614.73 & 264973.09 \\ 
$G=7$ & 266267.39 & 264069.53 & 264102.10 & 264643.31 \\ 
$G=8$ & 264591.64 & 263464.76 & 263877.58 & 264649.68 \\ 
$G=9$ & 263910.91 & 262952.98 & 263618.30 & 264544.87 \\ 
$G=10$ & 263600.50 & \textbf{262766.36} & 263684.61 & 264524.20 \\ 
$G=11$ & 263242.58 & 262921.59 & 263581.68 & 264754.70 \\ 
 \hline
\end{tabular}

\end{table}

Since there is a large number of response patterns with a very small number of observations (of all the $2^{16}=65536$ possible response patterns, 62384 (95.2\%) contain zero counts and only 481 (0.7\%) contain more than 5 counts), the Pearson's $\chi^2$ test is not applicable and the truncated SSPR criterion has been used to test the goodness of fit of the model. Three different levels of truncation are shown in Table \ref{tab:nltcs.ss}. 

The best model selected by the $\mathrm{BIC}$ and the $\mathrm{BIC}^*$ is one of the best fits as indicated by the SSPR over all the three levels of truncation. Table 1 of the supplementary material shows a comparison of  the observed and the expected frequencies for the response patterns with more than 100 observations under the best model selected. The table shows that there is a close match between the observed and expected frequencies under this model. 

From Table \ref{tab:nltcs.bic} and \ref{tab:nltcs.ss} it is also possible to see how the mixture of latent trait analyzers model is more appropriate than the latent class analysis and the latent trait analysis, in terms both of BIC and goodness of fit.

\begin{table}[ht]
\caption{The sum of squared of Pearson residuals for different levels of truncation ($\geq100$, $\geq 25$, $\geq 10$).}
\label{tab:nltcs.ss}
\smallskip
\centering
\footnotesize
\begin{tabular}{@{\extracolsep{-3pt}}l|rrrr|rrrr|rrrr}
\multicolumn{1}{c}{} & \multicolumn{4}{c}{\textit{observed frequencies} $\geq 100$} & \multicolumn{4}{c}{\textit{observed frequencies} $\geq 25$} & \multicolumn{4}{c}{\textit{observed frequencies} $\geq 10$} \\ 
  \hline
 & $D=0$ & $D=1$ & $D=2$ & $D=3$ & $D=0$ & $D=1$ & $D=2$ & $D=3$ & $D=0$ & $D=1$ & $D=2$ & $D=3$ \\ 
  \hline
$G=1$ & 6.1e+09 & 2441 & 1924 & 4082 & 6.3e+09 & 20524 & 21513 & 25795 & 6.3e+09 & 38927 & 25130 & 30009 \\ 
  $G=2$ & 80199 & 1600 & 1318 & 1718 & 129642 & 4614 & 3590 & 4822 & 182265 & 8126 & 6614 & 9654 \\ 
  $G=3$ & 5304 & 1470 & 1410 & 1597 & 16588 & 4610 & 2974 & 3318 & 61331 & 7610 & 4898 & 5814 \\ 
  $G=4$ & 4875 & 1439 & 638 & 806 & 10762 & 3060 & 1874 & 2530 & 17405 & 5498 & 2993 & 4609 \\ 
  $G=5$ & 2717 & 598 & 490 & 545 & 6321 & 1788 & 1498 & 1683 & 10886 & 3414 & 3037 & 3255 \\ 
  $G=6$ & 1434 & 323 & 533 & 462 & 3901 & 1423 & 1787 & 1530 & 6652 & 2498 & 2758 & 2817 \\ 
  $G=7$ & 561 & 561 & 561 & 418 & 2412 & 2412 & 2412 & 1450 & 4808 & 4808 & 4808 & 2468 \\ 
  $G=8$ & 413 & 413 & 413 & 234 & 1837 & 1837 & 1837 & 1228 & 3792 & 3792 & 3792 & 2115 \\ 
  $G=9$ & 417 & 253 & 183 & 229 & 1647 & 987 & 938 & 1054 & 3010 & 1725 & 1534 & 1761 \\ 
  $G=10$ & 348 & \textbf{160} & 292 & \textbf{150} & 1348 & \textbf{723} & 827 & 946 & 2695 & \textbf{1367} & 1495 & 1558 \\ 
  $G=11$ & 280 & 196 & 171 & \textbf{118} & 1336 & 872 & 760 & \textbf{700} & 2246 & 1518 & \textbf{1348} & 1481 \\ 
   \hline
\end{tabular}

\end{table}

The parameter estimates and standard errors for the selected model are reported in Table~\ref{tab:nltcs.bwsp}. The standardized values $w_{mg}^{*}$ and median probabilities $\pi_{mg}(0)$ are also reported to aid interpretation of the groups found by this model. 

\begin{table}
\caption{Parameter estimates (\textit{standard errors}) for the selected model ($G=10$, $D=1$)}
\label{tab:nltcs.bwsp}
\smallskip
\centering
\tiny
\begin{tabular}{@{\extracolsep{-5.5pt}}l|rrrc|rrrc|rrrc|}
\multicolumn{13}{c}{}
\vspace{-15pt}\\
\multicolumn{1}{c}{} & \multicolumn{4}{c}{$g=1$} & \multicolumn{4}{c}{$g=2$} &\multicolumn{4}{c}{$g=3$}\\
  \hline
 & \multicolumn{1}{c}{$b_{mg}$} & \multicolumn{1}{c}{$w_{mg}$} & \multicolumn{1}{c}{$w_{mg}^*$} & \multicolumn{1}{c}{$\pi_{mg}(0)$} \vline  & \multicolumn{1}{c}{$b_{mg}$} & \multicolumn{1}{c}{$w_{mg}$} & \multicolumn{1}{c}{$w_{mg}^*$} & \multicolumn{1}{c}{$\pi_{mg}(0)$} \vline  & \multicolumn{1}{c}{$b_{mg}$} & \multicolumn{1}{c}{$w_{mg}$} & \multicolumn{1}{c}{$w_{mg}^*$} & \multicolumn{1}{c}{$\pi_{mg}(0)$} \vline \\  
  \hline
$m=1$ & -3.16 \thinspace      (\textit{0.11}) & 0.02 \thinspace        (\textit{0.04}) & 0.02 & 0.04 & -2.89 \thinspace         (\textit{0.12}) & 0.96 \thinspace        (\textit{0.12}) & 0.69 & 0.07 & -5.68 \thinspace         (\textit{0.06}) & 0.00 \thinspace   (\textit{0.00}) & 0.00 & 0 \\ 
  $m=2$ & -0.97 \thinspace    (\textit{0.08}) & 1.11 \thinspace        (\textit{0.06}) & 0.74 & 0.31 & -4.53 \thinspace         (\textit{0.12}) & 0.05 \thinspace        (\textit{0.06}) & 0.05 & 0.01 & -0.26 \thinspace         (\textit{0.05}) & 0.13 \thinspace        (\textit{0.04}) & 0.13 & 0.44 \\ 
  $m=3$ & 0.38 \thinspace     (\textit{0.09}) & 1.36 \thinspace        (\textit{0.05}) & 0.81 & 0.57 & -2.81 \thinspace         (\textit{0.15}) & -0.01 \thinspace       (\textit{0.07}) & -0.01 & 0.06 & 2.37 \thinspace         (\textit{0.07}) & -0.08 \thinspace       (\textit{0.02}) & -0.08 & 0.91 \\ 
  $m=4$ & -1.46 \thinspace    (\textit{0.09}) & 0.38 \thinspace        (\textit{0.07}) & 0.36 & 0.19 & -1.90 \thinspace  (\textit{0.12}) & 1.50 \thinspace         (\textit{0.11}) & 0.83 & 0.20 & -2.72 \thinspace  (\textit{0.08}) & 0.07 \thinspace        (\textit{0.02}) & 0.07 & 0.06 \\ 
  $m=5$ & 0.64 \thinspace     (\textit{0.08}) & 0.64 \thinspace        (\textit{0.07}) & 0.54 & 0.65 & -0.49 \thinspace         (\textit{0.10}) & 1.42 \thinspace         (\textit{0.09}) & 0.82 & 0.41 & 0.38 \thinspace  (\textit{0.05}) & 0.83 \thinspace        (\textit{0.02}) & 0.64 & 0.59 \\ 
  $m=6$ & -1.19 \thinspace    (\textit{0.08}) & 0.73 \thinspace        (\textit{0.07}) & 0.59 & 0.25 & -2.74 \thinspace         (\textit{0.12}) & 1.05 \thinspace        (\textit{0.12}) & 0.72 & 0.09 & -0.92 \thinspace         (\textit{0.06}) & 0.67 \thinspace        (\textit{0.04}) & 0.56 & 0.29 \\ 
  $m=7$ & 5.95 \thinspace     (\textit{0.08}) & -0.05 \thinspace       (\textit{0.02}) & -0.05 & 1 & 1.78 \thinspace    (\textit{0.11}) & 1.70 \thinspace         (\textit{0.09}) & 0.86 & 0.77 & 2.51 \thinspace  (\textit{0.08}) & 0.11 \thinspace        (\textit{0.03}) & 0.11 & 0.92 \\ 
  $m=8$ & 0.69 \thinspace     (\textit{0.08}) & -0.14 \thinspace       (\textit{0.07}) & -0.14 & 0.67 & -1.54 \thinspace        (\textit{0.11}) & 1.77 \thinspace        (\textit{0.08}) & 0.87 & 0.26 & -3.44 \thinspace         (\textit{0.07}) & -0.01 \thinspace       (\textit{0.01}) & -0.01 & 0.03 \\ 
  $m=9$ & 2.84 \thinspace     (\textit{0.11}) & -0.12 \thinspace       (\textit{0.05}) & -0.12 & 0.94 & 0.61 \thinspace         (\textit{0.11}) & 1.80 \thinspace         (\textit{0.08}) & 0.87 & 0.60 & -0.37 \thinspace  (\textit{0.05}) & 0.23 \thinspace        (\textit{0.05}) & 0.22 & 0.41 \\ 
  $m=10$ & 1.54 \thinspace    (\textit{0.10}) & 0.31 \thinspace         (\textit{0.07}) & 0.30 & 0.82 & 0.69 \thinspace   (\textit{0.11}) & 1.73 \thinspace        (\textit{0.08}) & 0.87 & 0.61 & -2.21 \thinspace         (\textit{0.07}) & 0.01 \thinspace        (\textit{0.02}) & 0.01 & 0.10 \\ 
  $m=11$ & 3.10 \thinspace     (\textit{0.12}) & 0.29 \thinspace        (\textit{0.07}) & 0.28 & 0.96 & 2.30 \thinspace   (\textit{0.13}) & 0.82 \thinspace        (\textit{0.12}) & 0.63 & 0.89 & 1.68 \thinspace  (\textit{0.07}) & 0.05 \thinspace        (\textit{0.03}) & 0.05 & 0.84 \\ 
  $m=12$ & 1.91 \thinspace    (\textit{0.10}) & 1.00 \thinspace    (\textit{0.08}) & 0.71 & 0.84 & -0.28 \thinspace         (\textit{0.09}) & 0.67 \thinspace        (\textit{0.09}) & 0.56 & 0.44 & 4.39 \thinspace  (\textit{0.06}) & 0.00 \thinspace   (\textit{0.01}) & 0.00 & 0.99 \\ 
  $m=13$ & 1.62 \thinspace    (\textit{0.09}) & 0.66 \thinspace        (\textit{0.08}) & 0.55 & 0.82 & 1.72 \thinspace  (\textit{0.11}) & 0.08 \thinspace        (\textit{0.10}) & 0.08 & 0.85 & 1.60 \thinspace    (\textit{0.07}) & 0.00 \thinspace   (\textit{0.03}) & 0.00 & 0.83 \\ 
  $m=14$ & 0.11 \thinspace    (\textit{0.09}) & 1.14 \thinspace        (\textit{0.06}) & 0.75 & 0.52 & 2.02 \thinspace  (\textit{0.14}) & 0.25 \thinspace        (\textit{0.07}) & 0.25 & 0.88 & -1.59 \thinspace         (\textit{0.06}) & -0.03 \thinspace       (\textit{0.03}) & -0.03 & 0.17 \\ 
  $m=15$ & -0.38 \thinspace   (\textit{0.08}) & 0.89 \thinspace        (\textit{0.06}) & 0.67 & 0.42 & 0.52 \thinspace  (\textit{0.09}) & 0.83 \thinspace        (\textit{0.09}) & 0.64 & 0.61 & -2.29 \thinspace         (\textit{0.07}) & 0.01 \thinspace        (\textit{0.02}) & 0.01 & 0.09 \\ 
  $m=16$ & -0.98 \thinspace   (\textit{0.08}) & 0.93 \thinspace        (\textit{0.06}) & 0.68 & 0.30 & 0.31 \thinspace   (\textit{0.09}) & 0.47 \thinspace        (\textit{0.09}) & 0.42 & 0.57 & -3.09 \thinspace         (\textit{0.08}) & -0.07 \thinspace       (\textit{0.02}) & -0.07 & 0.04 \\ 
\hline
& \multicolumn{4}{c}{$\eta_g=0.070$ \thinspace (\textit{0.002})} \vrule & \multicolumn{4}{c}{$\eta_g=0.048$ \thinspace (\textit{0.002})} \vrule & \multicolumn{4}{c}{$\eta_g=0.108$ \thinspace (\textit{0.003})} \vrule \\
   \hline
\multicolumn{13}{c}{}
\vspace{-9pt}\\
   \multicolumn{1}{c}{} &\multicolumn{4}{c}{$g=4$} &\multicolumn{4}{c}{$g=5$} & \multicolumn{4}{c}{$g=6$} \\
  \hline
 & \multicolumn{1}{c}{$b_{mg}$} & \multicolumn{1}{c}{$w_{mg}$} & \multicolumn{1}{c}{$w_{mg}^*$} & \multicolumn{1}{c}{$\pi_{mg}(0)$} \vline  & \multicolumn{1}{c}{$b_{mg}$} & \multicolumn{1}{c}{$w_{mg}$} & \multicolumn{1}{c}{$w_{mg}^*$} & \multicolumn{1}{c}{$\pi_{mg}(0)$} \vline  & \multicolumn{1}{c}{$b_{mg}$} & \multicolumn{1}{c}{$w_{mg}$} & \multicolumn{1}{c}{$w_{mg}^*$} & \multicolumn{1}{c}{$\pi_{mg}(0)$} \vline \\
  \hline
$m=1$ & 0.27 \thinspace       (\textit{0.13}) & -0.84 \thinspace       (\textit{0.13}) & -0.64 & 0.56 & -4.91 \thinspace        (\textit{0.05}) & 0.02 \thinspace        (\textit{0.01}) & 0.02 & 0.01 & 0.79 \thinspace  (\textit{0.05}) & -0.69 \thinspace       (\textit{0.04}) & -0.57 & 0.68 \\ 
  $m=2$ & 2.51 \thinspace     (\textit{0.25}) & -0.82 \thinspace       (\textit{0.16}) & -0.63 & 0.91 & -4.07 \thinspace        (\textit{0.05}) & 0.02 \thinspace        (\textit{0.01}) & 0.02 & 0.02 & 3.27 \thinspace  (\textit{0.08}) & -0.05 \thinspace       (\textit{0.02}) & -0.05 & 0.96 \\ 
  $m=3$ & 1.76 \thinspace     (\textit{0.20}) & -0.23 \thinspace        (\textit{0.18}) & -0.22 & 0.85 & -4.61 \thinspace        (\textit{0.02}) & 0.00 \thinspace   (\textit{0.01}) & 0.00 & 0.01 & 3.88 \thinspace     (\textit{0.08}) & 0.03 \thinspace        (\textit{0.02}) & 0.03 & 0.98 \\ 
  $m=4$ & 2.32 \thinspace     (\textit{0.21}) & -0.87 \thinspace       (\textit{0.19}) & -0.66 & 0.89 & -3.23 \thinspace        (\textit{0.06}) & -0.01 \thinspace       (\textit{0.02}) & -0.01 & 0.04 & 3.34 \thinspace         (\textit{0.06}) & -0.08 \thinspace       (\textit{0.02}) & -0.08 & 0.97 \\ 
  $m=5$ & 4.01 \thinspace     (\textit{0.23}) & -0.29 \thinspace       (\textit{0.13}) & -0.27 & 0.98 & -0.77 \thinspace        (\textit{0.05}) & -0.36 \thinspace       (\textit{0.04}) & -0.34 & 0.32 & 4.90 \thinspace  (\textit{0.07}) & 0.00 \thinspace   (\textit{0.01}) & 0.00 & 0.99 \\ 
  $m=6$ & 1.84 \thinspace     (\textit{0.20}) & -0.30 \thinspace         (\textit{0.17}) & -0.29 & 0.86 & -2.78 \thinspace        (\textit{0.07}) & -0.07 \thinspace       (\textit{0.03}) & -0.07 & 0.06 & 2.11 \thinspace         (\textit{0.07}) & 0.09 \thinspace        (\textit{0.04}) & 0.09 & 0.89 \\ 
  $m=7$ & 1.81 \thinspace     (\textit{0.16}) & -1.73 \thinspace       (\textit{0.17}) & -0.87 & 0.77 & 1.85 \thinspace         (\textit{0.05}) & 0.22 \thinspace        (\textit{0.03}) & 0.21 & 0.86 & 6.41 \thinspace  (\textit{0.06}) & 0.01 \thinspace        (\textit{0.01}) & 0.01 & 1 \\ 
  $m=8$ & -0.95 \thinspace    (\textit{0.17}) & -2.37 \thinspace       (\textit{0.11}) & -0.92 & 0.38 & -2.92 \thinspace        (\textit{0.07}) & 0.22 \thinspace        (\textit{0.03}) & 0.22 & 0.05 & 2.90 \thinspace   (\textit{0.07}) & -0.02 \thinspace       (\textit{0.02}) & -0.02 & 0.95 \\ 
  $m=9$ & -0.18 \thinspace    (\textit{0.17}) & -2.63 \thinspace       (\textit{0.10}) & -0.93 & 0.47 & -0.83 \thinspace         (\textit{0.05}) & 0.41 \thinspace        (\textit{0.04}) & 0.38 & 0.31 & 4.60 \thinspace   (\textit{0.07}) & 0.00 \thinspace   (\textit{0.01}) & 0.00 & 0.99 \\ 
  $m=10$ & -0.25 \thinspace   (\textit{0.16}) & -2.47 \thinspace       (\textit{0.10}) & -0.93 & 0.46 & -2.42 \thinspace         (\textit{0.06}) & 0.13 \thinspace        (\textit{0.03}) & 0.13 & 0.08 & 4.25 \thinspace  (\textit{0.07}) & -0.03 \thinspace       (\textit{0.01}) & -0.03 & 0.99 \\ 
  $m=11$ & 0.22 \thinspace    (\textit{0.16}) & -1.79 \thinspace       (\textit{0.15}) & -0.87 & 0.53 & 0.67 \thinspace         (\textit{0.05}) & -0.72 \thinspace       (\textit{0.03}) & -0.59 & 0.65 & 4.48 \thinspace         (\textit{0.08}) & -0.03 \thinspace       (\textit{0.02}) & -0.03 & 0.99 \\ 
  $m=12$ & -1.00 \thinspace      (\textit{0.22}) & 0.69 \thinspace        (\textit{0.14}) & 0.57 & 0.29 & -0.24 \thinspace         (\textit{0.05}) & -0.79 \thinspace       (\textit{0.03}) & -0.62 & 0.45 & 4.34 \thinspace         (\textit{0.07}) & 0.03 \thinspace        (\textit{0.02}) & 0.03 & 0.99 \\ 
  $m=13$ & -0.80 \thinspace    (\textit{0.17}) & 0.13 \thinspace        (\textit{0.14}) & 0.13 & 0.31 & 0.82 \thinspace  (\textit{0.05}) & -0.95 \thinspace       (\textit{0.03}) & -0.69 & 0.68 & 2.52 \thinspace         (\textit{0.07}) & -0.03 \thinspace       (\textit{0.03}) & -0.03 & 0.93 \\ 
  $m=14$ & -0.20 \thinspace    (\textit{0.17}) & -1.27 \thinspace       (\textit{0.14}) & -0.79 & 0.46 & -1.66 \thinspace        (\textit{0.06}) & -0.25 \thinspace       (\textit{0.04}) & -0.24 & 0.16 & 1.32 \thinspace         (\textit{0.06}) & -1.04 \thinspace       (\textit{0.04}) & -0.72 & 0.76 \\ 
  $m=15$ & 0.32 \thinspace    (\textit{0.14}) & -0.90 \thinspace        (\textit{0.13}) & -0.67 & 0.57 & -2.25 \thinspace        (\textit{0.06}) & 0.06 \thinspace        (\textit{0.03}) & 0.06 & 0.10 & 1.56 \thinspace   (\textit{0.06}) & -0.68 \thinspace       (\textit{0.05}) & -0.56 & 0.81 \\ 
  $m=16$ & -0.66 \thinspace   (\textit{0.17}) & -1.59 \thinspace       (\textit{0.14}) & -0.85 & 0.39 & -3.08 \thinspace        (\textit{0.07}) & 0.06 \thinspace        (\textit{0.02}) & 0.06 & 0.04 & 0.47 \thinspace  (\textit{0.06}) & -1.23 \thinspace       (\textit{0.03}) & -0.78 & 0.60 \\ 
   \hline
& \multicolumn{4}{c}{$\eta_g=0.022$ \thinspace (\textit{0.002})} \vrule & \multicolumn{4}{c}{$\eta_g=0.149$ \thinspace (\textit{0.003})} \vrule & \multicolumn{4}{c}{$\eta_g=0.112$ \thinspace (\textit{0.002})} \vrule\\
      \hline
      \multicolumn{13}{c}{}
      \vspace{-9pt}
   \\
   \multicolumn{1}{c}{} & \multicolumn{4}{c}{$g=7$} &\multicolumn{4}{c}{$g=8$} &\multicolumn{4}{c}{$g=9$}\\
  \hline
 & \multicolumn{1}{c}{$b_{mg}$} & \multicolumn{1}{c}{$w_{mg}$} & \multicolumn{1}{c}{$w_{mg}^*$} & \multicolumn{1}{c}{$\pi_{mg}(0)$} \vline  & \multicolumn{1}{c}{$b_{mg}$} & \multicolumn{1}{c}{$w_{mg}$} & \multicolumn{1}{c}{$w_{mg}^*$} & \multicolumn{1}{c}{$\pi_{mg}(0)$} \vline  & \multicolumn{1}{c}{$b_{mg}$} & \multicolumn{1}{c}{$w_{mg}$} & \multicolumn{1}{c}{$w_{mg}^*$} & \multicolumn{1}{c}{$\pi_{mg}(0)$} \vline \\ 
  \hline
$m=1$ & -7.61 \thinspace      (\textit{0.19}) & 0.00 \thinspace   (\textit{0.00}) & 0.00 & 0 & -5.87 \thinspace  (\textit{0.05}) & 0.00 \thinspace   (\textit{0.00}) & 0.00 & 0 & -4.90 \thinspace   (\textit{0.07}) & -0.02 \thinspace       (\textit{0.01}) & -0.02 & 0.01 \\ 
  $m=2$ & -8.79 \thinspace    (\textit{0.26}) & 0.00 \thinspace   (\textit{0.00}) & 0.00 & 0 & -4.43 \thinspace  (\textit{0.03}) & 0.00 \thinspace   (\textit{0.00}) & 0.00 & 0.01 & -0.90 \thinspace        (\textit{0.06}) & -0.35 \thinspace       (\textit{0.04}) & -0.33 & 0.29 \\ 
  $m=3$ & -10.13 \thinspace   (\textit{0.00}) & 0.00 \thinspace      (\textit{0.00}) & 0.00 & 0 & -4.81 \thinspace  (\textit{0.02}) & 0.00 \thinspace   (\textit{0.00}) & 0.00 & 0.01 & 1.36 \thinspace        (\textit{0.06}) & 0.18 \thinspace        (\textit{0.03}) & 0.17 & 0.79 \\ 
  $m=4$ & -6.99 \thinspace    (\textit{0.10}) & 0.00 \thinspace    (\textit{0.00}) & 0.00 & 0 & -3.55 \thinspace  (\textit{0.05}) & 0.00 \thinspace   (\textit{0.00}) & 0.00 & 0.03 & -2.65 \thinspace       (\textit{0.10}) & -0.19 \thinspace        (\textit{0.04}) & -0.19 & 0.07 \\ 
  $m=5$ & -6.50 \thinspace     (\textit{0.26}) & 0.00 \thinspace   (\textit{0.00}) & 0.00 & 0 & -1.55 \thinspace  (\textit{0.04}) & 0.00 \thinspace   (\textit{0.00}) & 0.00 & 0.17 & -0.28 \thinspace       (\textit{0.06}) & -0.91 \thinspace       (\textit{0.03}) & -0.67 & 0.44 \\ 
  $m=6$ & -8.72 \thinspace    (\textit{0.33}) & 0.00 \thinspace   (\textit{0.00}) & 0.00 & 0 & -3.29 \thinspace  (\textit{0.06}) & 0.00 \thinspace   (\textit{0.00}) & 0.00 & 0.04 & -1.43 \thinspace       (\textit{0.07}) & -0.60 \thinspace        (\textit{0.05}) & -0.51 & 0.20 \\ 
  $m=7$ & -7.35 \thinspace    (\textit{0.19}) & 0.00 \thinspace   (\textit{0.00}) & 0.00 & 0 & -0.07 \thinspace  (\textit{0.03}) & 0.00 \thinspace   (\textit{0.00}) & 0.00 & 0.48 & 0.49 \thinspace        (\textit{0.06}) & -0.60 \thinspace        (\textit{0.04}) & -0.52 & 0.62 \\ 
  $m=8$ & -11.07 \thinspace   (\textit{0.10}) & 0.00 \thinspace    (\textit{0.00}) & 0.00 & 0 & -8.32 \thinspace  (\textit{0.25}) & 0.00 \thinspace   (\textit{0.00}) & 0.00 & 0 & -4.11 \thinspace  (\textit{0.08}) & -0.05 \thinspace       (\textit{0.01}) & -0.05 & 0.02 \\ 
  $m=9$ & -8.88 \thinspace    (\textit{0.04}) & 0.00 \thinspace   (\textit{0.00}) & 0.00 & 0 & -5.96 \thinspace  (\textit{0.03}) & 0.00 \thinspace   (\textit{0.00}) & 0.00 & 0 & -2.77 \thinspace  (\textit{0.08}) & -0.09 \thinspace       (\textit{0.02}) & -0.09 & 0.06 \\ 
  $m=10$ & -8.16 \thinspace   (\textit{0.04}) & 0.00 \thinspace   (\textit{0.00}) & 0.00 & 0 & -5.48 \thinspace  (\textit{0.05}) & 0.00 \thinspace   (\textit{0.00}) & 0.00 & 0 & -8.48 \thinspace  (\textit{0.21}) & 0.00 \thinspace   (\textit{0.00}) & 0.00 & 0 \\ 
  $m=11$ & -9.60 \thinspace    (\textit{0.04}) & 0.00 \thinspace   (\textit{0.00}) & 0.00 & 0 & -3.23 \thinspace  (\textit{0.05}) & 0.00 \thinspace   (\textit{0.00}) & 0.00 & 0.04 & -2.25 \thinspace       (\textit{0.08}) & 0.01 \thinspace        (\textit{0.02}) & 0.01 & 0.10 \\ 
  $m=12$ & -3.91 \thinspace   (\textit{0.07}) & 0.00 \thinspace   (\textit{0.00}) & 0.00 & 0.02 & -1.38 \thinspace       (\textit{0.04}) & 0.00 \thinspace   (\textit{0.00}) & 0.00 & 0.20 & 2.51 \thinspace         (\textit{0.08}) & 0.04 \thinspace        (\textit{0.03}) & 0.04 & 0.92 \\ 
  $m=13$ & -4.28 \thinspace   (\textit{0.03}) & 0.00 \thinspace   (\textit{0.00}) & 0.00 & 0.01 & -1.94 \thinspace       (\textit{0.05}) & 0.00 \thinspace   (\textit{0.00}) & 0.00 & 0.13 & -1.26 \thinspace       (\textit{0.07}) & 0.03 \thinspace        (\textit{0.04}) & 0.03 & 0.22 \\ 
  $m=14$ & -6.11 \thinspace   (\textit{0.09}) & 0.00 \thinspace   (\textit{0.00}) & 0.00 & 0 & -3.70 \thinspace   (\textit{0.06}) & 0.00 \thinspace   (\textit{0.00}) & 0.00 & 0.02 & -6.28 \thinspace       (\textit{0.04}) & 0.00 \thinspace   (\textit{0.00}) & 0.00 & 0 \\ 
  $m=15$ & -7.95 \thinspace   (\textit{0.14}) & 0.00 \thinspace   (\textit{0.00}) & 0.00 & 0 & -2.92 \thinspace  (\textit{0.05}) & 0.00 \thinspace   (\textit{0.00}) & 0.00 & 0.05 & -3.60 \thinspace        (\textit{0.08}) & -0.05 \thinspace       (\textit{0.02}) & -0.05 & 0.03 \\ 
  $m=16$ & -4.03 \thinspace   (\textit{0.04}) & 0.00 \thinspace   (\textit{0.00}) & 0.00 & 0.02 & -3.55 \thinspace       (\textit{0.06}) & 0.00 \thinspace   (\textit{0.00}) & 0.00 & 0.03 & -4.55 \thinspace       (\textit{0.06}) & 0.03 \thinspace        (\textit{0.01}) & 0.03 & 0.01 \\ 
   \hline
& \multicolumn{4}{c}{$\eta_g=0.138$ \thinspace (\textit{0.003})} \vrule& \multicolumn{4}{c}{$\eta_g=0.201$ \thinspace (\textit{0.003})} \vrule& \multicolumn{4}{c}{$\eta_g=0.094$ \thinspace (\textit{0.003})} \vrule\\
\hline
\multicolumn{13}{c}{}
\vspace{-9pt}
\\
\multicolumn{1}{c}{} & \multicolumn{4}{c}{$g=10$} &\multicolumn{8}{c}{}\\
\cline{1-5}
 & \multicolumn{1}{c}{$b_{mg}$} & \multicolumn{1}{c}{$w_{mg}$} & \multicolumn{1}{c}{$w_{mg}^*$} & \multicolumn{1}{c}{$\pi_{mg}(0)$} \vline & \multicolumn{8}{c}{} \\ 
\cline{1-5}
$m=1$ & -1.83 \thinspace      (\textit{0.09}) & 0.74 \thinspace        (\textit{0.08}) & 0.59 & 0.16 & \multicolumn{8}{c}{}\\ 
  $m=2$ & 1.52 \thinspace     (\textit{0.09}) & 0.29 \thinspace        (\textit{0.07}) & 0.28 & 0.82 & \multicolumn{8}{c}{}\\
  $m=3$ & 3.70 \thinspace      (\textit{0.10}) & -0.06 \thinspace        (\textit{0.04}) & -0.06 & 0.98 & \multicolumn{8}{c}{}\\
  $m=4$ & 0.36 \thinspace     (\textit{0.09}) & 0.97 \thinspace        (\textit{0.06}) & 0.70 & 0.58 & \multicolumn{8}{c}{}\\ 
  $m=5$ & 2.95 \thinspace     (\textit{0.12}) & 0.35 \thinspace        (\textit{0.06}) & 0.33 & 0.95 & \multicolumn{8}{c}{}\\ 
  $m=6$ & 1.05 \thinspace     (\textit{0.09}) & 0.94 \thinspace        (\textit{0.06}) & 0.68 & 0.71 & \multicolumn{8}{c}{}\\ 
  $m=7$ & 5.44 \thinspace     (\textit{0.09}) & -0.06 \thinspace       (\textit{0.03}) & -0.06 & 1 & \multicolumn{8}{c}{}\\ 
  $m=8$ & 0.16 \thinspace     (\textit{0.10}) & -1.40 \thinspace         (\textit{0.05}) & -0.81 & 0.53 & \multicolumn{8}{c}{}\\ 
  $m=9$ & 1.35 \thinspace     (\textit{0.10}) & -1.29 \thinspace        (\textit{0.07}) & -0.79 & 0.74 & \multicolumn{8}{c}{}\\ 
  $m=10$ & -0.03 \thinspace   (\textit{0.10}) & -1.11 \thinspace        (\textit{0.06}) & -0.74 & 0.50 & \multicolumn{8}{c}{}\\ 
  $m=11$ & 1.86 \thinspace    (\textit{0.12}) & -0.36 \thinspace       (\textit{0.09}) & -0.34 & 0.86 & \multicolumn{8}{c}{}\\
  $m=12$ & 3.27 \thinspace    (\textit{0.15}) & -0.02 \thinspace       (\textit{0.06}) & -0.02 & 0.96 & \multicolumn{8}{c}{}\\ 
  $m=13$ & 1.44 \thinspace    (\textit{0.11}) & 0.19 \thinspace        (\textit{0.09}) & 0.19 & 0.81 & \multicolumn{8}{c}{}\\
  $m=14$ & -1.88 \thinspace   (\textit{0.11}) & 0.30 \thinspace         (\textit{0.08}) & 0.28 & 0.14 & \multicolumn{8}{c}{}\\ 
  $m=15$ & -1.33 \thinspace   (\textit{0.10}) & 0.18 \thinspace         (\textit{0.08}) & 0.18 & 0.21 & \multicolumn{8}{c}{}\\
  $m=16$ & -3.66 \thinspace   (\textit{0.12}) & 0.10 \thinspace         (\textit{0.06}) & 0.10 & 0.03 & \multicolumn{8}{c}{}\\
\cline{1-5}
 & \multicolumn{4}{c}{$\eta_g=0.057$ \thinspace (\textit{0.002})} \vline & \multicolumn{8}{c}{} \\  
 \cline{1-5}    
\end{tabular}
\end{table}

\subsubsection{Analysis of the Groups in the Selected Model}

From Table~\ref{tab:nltcs.bwsp} it is possible to observe that Groups 7 and 8 consist mainly of the people that are able to do all the activities examined by the NLTCS survey. However, Group 8 has higher probability of being unable to do the activities.  In particular, the probability of not being able to do heavy house work (activity 7) is much higher in Group 8 than in Group 7. In both groups, the estimated $\mathbf{w}_{g}$ values are zero, indicating that the local independence assumption is adequate in these groups. The fact that $\mathbf{w}_g$ is zero also indicates that there is no more variability in each outcome than can be described by a Bernoulli distribution.

In contrast, Group 6 consists mainly of disabled people. In addition, the $\mathbf{w}_g$ value is non-zero and the slope parameters are very negative for the activities of managing money (activity 14), taking medicine (activity 15) and telephoning (activity 16). The slope parameters for these activities indicate that there is considerable variability in the outcomes in these activities and that the outcomes are positively dependent (that is, those able/disabled in one activity will tend to be able/disabled in the others). It is worth noting that these activities are the least physically taxing of the instrumental activities of daily living, as recorded in these data. 

Group 9 is characterized by people able to do the activities 1, 4, 8, 9, 10, 11, 13, 14, 15, 16 that correspond to
eating, dressing, doing light house work, doing laundry, cooking, grocery shopping, travelling, managing money, taking medicine, telephoning, but these individuals are unable to get around inside and outside (activities 3 and 12). For the other activities (number 2, 5, 6, 7) there is quite a big variability in responses explained by the within group latent trait structure; the slope parameters for these activities are all negatively signed which again indicates that these activities exhibit strong positive dependence. These activities getting in/out of bed (activity 2), bathing (activity 5), using the toilet (activity 6) and doing heavy housework (activity 7) require mobility which may explain the positive dependence between the outcomes in these. 

Group 1 consists mainly of people able to eat  (activity 1), but unable to do heavy house work (activity 7), for the other activities there is quite a big variability explained by the within group latent trait structure, but the median individual in this group has a probability in excess of 80\% to be disabled in activities 9, 10, 11, 12, 13. Activities 2, 3, 4, 5, 6, 10, 11, 12, 13, 14, 15, 16 have highly positive slope parameters, whereas activities 7, 8, 9 have moderately negative slopes. Thus there is positive dependence within these sets of activities but negative dependence across these groups. 

Group 2 consists mainly of people able to get in/out of bed, and get around outside (activities 2,3), but unable to travel (activity 13), for the other activities there is quite a big variability explained by the within group latent trait structure, but the probability of the median individual of this group being disabled is large (over the 88\%) for the activities 11, 14 (grocery shopping and managing money), and small (less than the 10\%) for the activities 1 and 6 (eating and using the toilet). Interestingly, the median individual tends to have lower probability of being disabled in the activities of daily living (ADLs) than in the instrumental activities of daily living (IADLs). The slopes for most activities have equal sign, so there is evidence of strong positive dependence between activities in this group. Thus people are either able or unable to do multiple activities in this group to a greater degree than independence would suggest.

Group 3 consists mainly of people able to do the activities 1, 4, 8, 10, 14, 15, 16, but unable to do the activities 3, 7, 11, 12, 13, for the other activities there is quite a big variability explained by the within group latent trait structure. The activities with large slope parameters involve physical movement and these again exhibit positive dependence in a similar manner to Group 9. 

Group 4 is characterized mainly by people unable to do the activities of daily living 2, 3, 4, 5, 6, 7, but there is large variability in the responses explained by the within group latent trait structure. In contrast to Group 2, for this group the median individual  tends to have a higher probability of being disabled in the activities of daily living (ADLs) than in the instrumental activities of daily living (IADLs). The signs of the large magnitude slope parameters are equal, so there is strong positive dependence within this group also. 

Group 5 consists mainly of people able to do the activities 1, 2, 3, 4, 6, 8, 10, 15, 16, but unable to do heavy house work (activity 7), for the other activities there is quite a big variability explained by the within group latent trait structure. The large magnitude slope parameters are equal in sign, so there is strong positive dependence within these activities within this group also.

Group 10 consists mainly of people able to telephone (activity 16), but unable to do heavy house work and get around inside and outside (activities 7, 3 and 12), for the other activities there is quite a big variability explained by the within group latent trait structure, but the probability of being disabled for the median individual of this group is large (over the 80\%) for the variables 2, 5, 11, 13, and quite small (between the 14\% and the 21\%) for the variables 1, 14, 15. In this group, activities 1, 2, 4, 5, 6, 14, 15 have strongly positive slope and activities 8, 9, 10, 11 have negative slopes. Thus, there is positive dependence within these activity groups and negative dependence across the activity groups. The second group of activities require a large amount of movement whereas the first group require less so. 

The groups found in this analysis suggest that there is a wide range of experiences in terms of activities of daily living in the individuals included in the NLTCS data. The groups have a similar structure to the one found by \cite{EFJ07} but their model allows individuals to have mixed membership across groups, so they suggest that slightly fewer groups (between 7 and 9) are needed.

\subsection{U.S. Congressional Voting}
\label{se:voting}

\noindent The voting records of the U.S. House of Representatives Congressmen in the second session of the 98th congress (January 23rd, 1984--October 12th, 1984) are recorded in \cite{CQA}. During this session the House of Representatives recorded votes on 408 issues. Sixteen of the key issues voted on by the House are described in this publication \cite[][Pages 7-C to 11-C]{CQA}. The voting data for these sixteen issues was collated in \cite{Schl87} and is publicly available in the UCI Machine Learning data repository \citep{UCI}.  

The data set contains the votes of all 435 House of Representatives congressmen on the sixteen issues: each vote has a value of \textit{yes} (voted for, paired for, and announced for), \textit{no} (voted against, paired against, and announced against) or \textit{undecided} (voted present, voted present to avoid conflict of interest, and did not vote or otherwise make a position known). It is also known if the voter is a \textit{Democrat} or a \textit{Republican} congressman. The issues voted on were: \textit{(Q1)} Handicapped Infants, \textit{(Q2)} Water Project Cost-Sharing, \textit{(Q3)} Adoption of the Budget Resolution, \textit{(Q4)} Physician Fee Freeze, \textit{(Q5)} El Salvador Aid, \textit{(Q6)} Religious Groups in Schools, \textit{(Q7)} Anti-Satellite Test Ban, \textit{(Q8)} Aid to Nicaraguan `Contras', \textit{(Q9)} MX Missile, \textit{(Q10)} Immigration, \textit{(Q11)} Synfuels Corporation Cutback, \textit{(Q12)} Education Spending, \textit{(Q13)} Superfund Right to Sue, \textit{(Q14)} Crime, \textit{(Q15)} Duty-Free Exports, \textit{(Q16)} Export Administration Act/South Africa.

Each question has been coded using two binary variables $a$ and $b$: the responses for the $a$ variables are coded as 1 = \textit{yes}/\textit{no} and 0 =  \textit{undecided} , and $b$ variable are 1 = \textit{yes}, 0 = \textit{no}/\textit{undecided}. 
The MLTA model was fitted to these data for $D=0,1,2,3$ and $G=1,2,\ldots, 5$. The log-likelihood for the fitted models are reported in Table~\ref{tab:cong.ll.GH}. 

\begin{table}[ht]
\caption{Approximated log-likelihood for the models with $D=0,1,2,3$ and $G=1,2,\ldots,5$}
\label{tab:cong.ll.GH}
\smallskip
\centering
\footnotesize
\begin{tabular}{@{\extracolsep{-4pt}}l|c|cc|cc|cc}
  \hline
\multicolumn{1}{c}{} & \multicolumn{1}{c}{$D=0$} & \multicolumn{2}{c}{$D=1$} & \multicolumn{2}{c}{$D=2$} & \multicolumn{2}{c}{$D=3$} \\ 
  \hline
  \multicolumn{1}{c}{} & \multicolumn{1}{c}{} & \multicolumn{1}{c}{} & \multicolumn{1}{c}{$\mathbf{w}$ \textit{fixed}} & \multicolumn{1}{c}{} & \multicolumn{1}{c}{$\mathbf{w}$ \textit{fixed}} &  \multicolumn{1}{c}{} & \multicolumn{1}{c}{$\mathbf{w}$ \textit{fixed}}\\ 
  \hline 
$G=1$ & -6109.61 & -4789.10 & & -4565.47 &  &-4468.45 & \\ 
$G=2$ & -4888.64 & -4533.89 & -4741.79 & -4364.65 & -4492.78 & -4317.43 & -4383.82\\  
$G=3$ & -4699.47 & -4332.34 & -4580.95 & -4245.82 & -4417.75 & -4174.52 & -4340.71 \\ 
$G=4$ & -4613.10 & -4212.98 & -4453.24 & -4141.93 & -4260.51 & -4074.31 & -4378.15\\ 
$G=5$ & -4533.50 & -4149.19 & -4378.84 & -4037.96 & -4241.45 & -3956.23 & -4236.31 \\ 
   \hline
\end{tabular}
\label{tab:cvot.ll}
\end{table}

The $\mathrm{BIC}$ and the $\mathrm{BIC}^*$ (Table \ref{tab:cvot.BIC} and \ref{tab:cvot.BICs}) agree in selecting the model with 4 groups, a bivariate latent trait and with slope parameters, $\mathbf{w}$, equal across groups as the best model.

\begin{table}[ht]
\caption{The estimated BIC for the models with $D=0,1,2,3$ and $G=1,2,\ldots,5$}
\smallskip
\centering 
\footnotesize
\begin{tabular}{@{\extracolsep{-4pt}}l|r|rr|rc|rr}
  \hline
\multicolumn{1}{c}{} & \multicolumn{1}{c}{$D=0$} & \multicolumn{2}{c}{$D=1$} & \multicolumn{2}{c}{$D=2$} & \multicolumn{2}{c}{$D=3$} \\ 
  \hline
  \multicolumn{1}{c}{} & \multicolumn{1}{c}{} & \multicolumn{1}{c}{} & \multicolumn{1}{c}{$\mathbf{w}$ \textit{fixed}} & \multicolumn{1}{c}{} & \multicolumn{1}{c}{$\mathbf{w}$ \textit{fixed}} &  \multicolumn{1}{c}{} & \multicolumn{1}{c}{$\mathbf{w}$ \textit{fixed}}\\ 
  \hline 
$G=1$ & 12413.64 & 9967.02 & & 9708.10 & & 9696.32 & \\
$G=2$ & 10172.18 & 9851.49 & 10072.90 & 9889.69 & 9763.21 & 10159.77 &  9727.54\\
$G=3$ & 9994.32  & 9843.29 & 9951.70 & 10235.27 & 9813.63 & 10639.44 & 9841.81\\ 
$G=4$ & 10022.07 & 9999.47 & 9896.76 & 10610.72 & \textbf{9699.65} & 11204.53 & 10117.18\\ 
$G=5$ & 10063.36 & 10266.79 & 9948.44 & 10986.01 & 9862.00 & 11733.84 & 10033.99\\ 
   \hline
\end{tabular}
\label{tab:cvot.BIC}
\end{table}

\begin{table}[ht]
\caption{The estimated $\mathrm{BIC}^{*}$ for the models with $D=0,1,2,3$ and $G=1,2,\ldots,5$}
\smallskip
\centering
\footnotesize
\begin{tabular}{@{\extracolsep{-4pt}}l|r|rr|rc|rr}
  \hline
\multicolumn{1}{c}{} & \multicolumn{1}{c}{$D=0$} & \multicolumn{2}{c}{$D=1$} & \multicolumn{2}{c}{$D=2$} & \multicolumn{2}{c}{$D=3$} \\ 
  \hline
  \multicolumn{1}{c}{} & \multicolumn{1}{c}{} & \multicolumn{1}{c}{} & \multicolumn{1}{c}{$\mathbf{w}$ \textit{fixed}} & \multicolumn{1}{c}{} & \multicolumn{1}{c}{$\mathbf{w}$ \textit{fixed}} &  \multicolumn{1}{c}{} & \multicolumn{1}{c}{$\mathbf{w}$ \textit{fixed}}\\ 
  \hline 
$G=1$ & 12413.64 & 9967.02 & & 9708.10 & & 9696.32 & \\ 
$G=2$ & 10127.64 & 9761.98 & 10028.51 & 9746.68 & 9718.82 & 9986.49 & 9681.51 \\ 
$G=3$ & 9883.15 & 9614.18 & 9801.00 & 9871.50 & 9698.52 & 10204.99 & 9729.45\\ 
$G=4$ & 9834.05 & 9600.57 & 9666.16 & 10017.47 & \textbf{9464.28} & 10404.27 & 9920.84 \\ 
$G=5$ & 9787.05 & 9711.72 &  9615.51 & 10128.07 & 9572.70 & 10531.05 & 9741.57\\ 
  \hline
\end{tabular}
\label{tab:cvot.BICs}
\end{table}

\subsubsection{Analysis of the Groups in the Selected Model}

Table \ref{tab:cvot.vot4} shows a cross tabulation of group membership with party membership (\textit{Republican} or \textit{Democrat}) for the 435 congressmen. From this table it is clear that Group 4 consists mainly of \textit{Republican} congressmen, Group 1 and 3 consist mainly of \textit{Democrat} congressmen and Group 2 is a small group consisting of members of both parties. Interestingly, even though the Democrat congressmen are divided into two voting blocs, these do not correspond to the Northern Democrats and Southern Democrats which were perceived to be voting blocs within the party (Table~\ref{tab:cvot.state}). However, of the eleven \textit{Democrat} congressmen who are in Group 4, nine are Southern Democrats, suggesting that they vote in a similar way to the \textit{Republican} congressmen. It is worth emphasizing that the data under consideration only records voting on sixteen key issues and that more voting blocs may be revealed if voting on all issues were examined. 

\begin{table}[ht]
\caption{Classification table for party in the selected model ($G=4$, $D=2$, $\mathbf{w}$ \textit{fixed})}
\smallskip
\centering
\footnotesize
\begin{tabular}{l|rrrr}
  \hline
 & $g=1$ & $g=2$ & $g=3$ & $g=4$ \\ 
  \hline
\textit{Democrat} &  65 &   8 & 183 &  11 \\ 
\textit{Republican} &  2 &   3 &  18 & 145 \\ 
   \hline
\end{tabular}
\label{tab:cvot.vot4}
\end{table}

\begin{table}[ht]
\caption{Classification table for location and party in the selected model}
\smallskip
\centering
\footnotesize
\begin{tabular}{l|rrrr}
  \hline
 & $g=1$ & $g=2$ & $g=3$ & $g=4$ \\ 
  \hline
\textit{Northern Democrat} &  48 &   7 & 121 &  2 \\ 
\textit{Southern Democrat} &  17 &   1 &  62 & 9 \\ 
\textit{Northern Republican} &  2 &   2 & 13 &  111 \\ 
\textit{Southern Republican} &  0 &   1 &  5 & 34 \\ 
   \hline
\end{tabular}
\label{tab:cvot.state}
\end{table}

Table~\ref{tab:cvot.par} gives the parameter estimate $b_{mg}$ and median probability $\pi_{mg}(\mathbf{0})$ for each of the groups. This table reveals that Group 2 consists of congressmen who have a high probability of being {\em undecided} on many of the issues.

The probabilities of a positive response for the $a$ variables for the median individuals in Groups 1, 3 and 4 are always very large (over the 0.90) with two exceptions in Group 1, for the variable number 3 where $\pi_{31}(\mathbf{0})= 0.82$ and for variable number 31 where $\pi_{311}(\mathbf{0})= 0$. So, the majority of congressmen in the Groups 1, 3 and 4 voted on most issues, but with a slightly lower voting rate in Group 1 on the Water Project Cost-Sharing and a very low voting rate on the Export Administration Act/South Africa. As a result of these high voting rates, most of the probabilities given for $b$ variables (\textit{yes} versus \textit{no/undecided}) in these groups can be interpreted as a voting \textit{yes} versus \textit{no} probability.

\begin{figure}[htp]
		
		\begin{center}
		\includegraphics[scale=0.8]{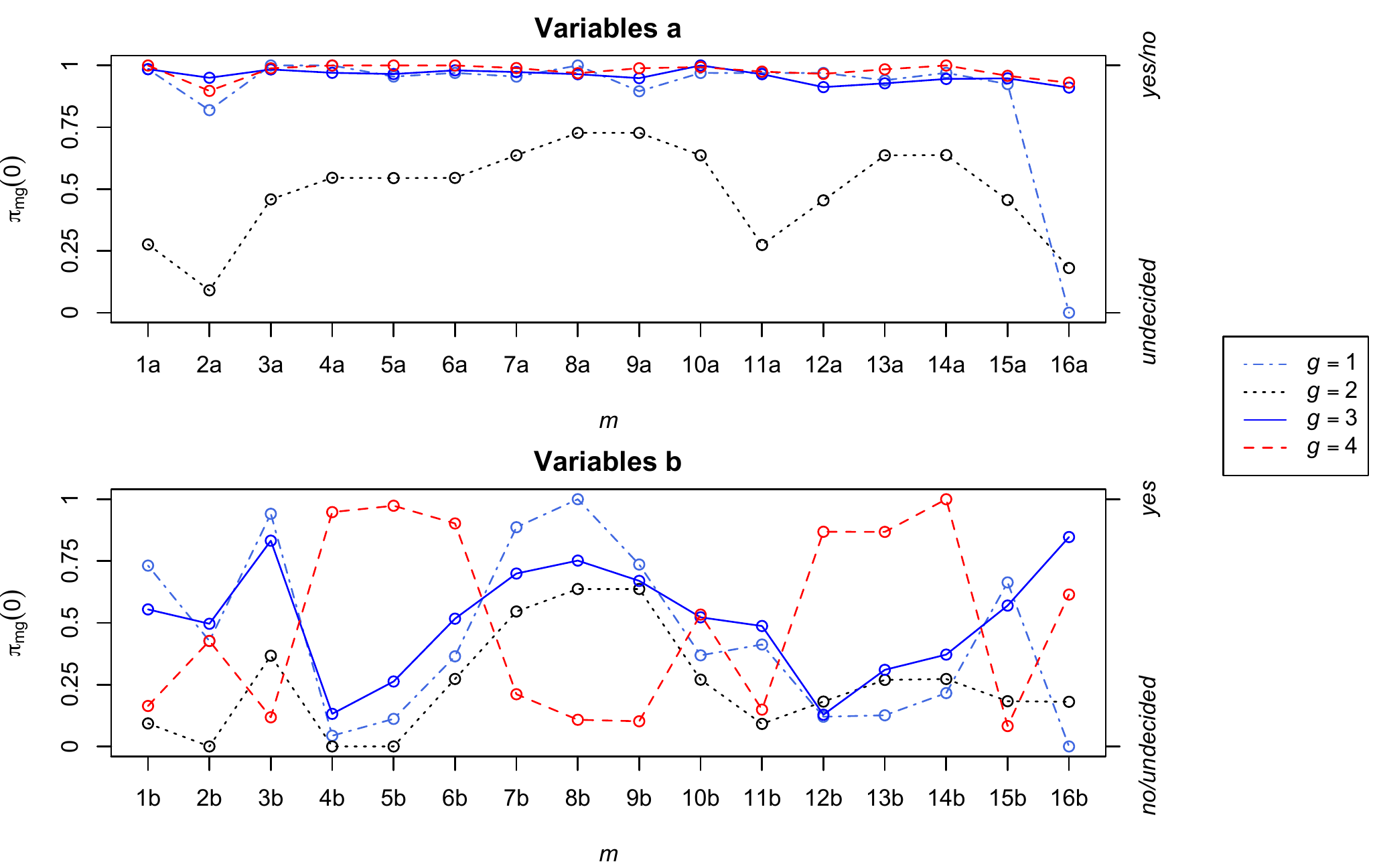}
		\end{center}
		\caption{The median probabilities, $\pi_{mg}(\mathbf{0})$, for the selected model ($G=4$, $D=2$, $\mathbf{w}$ \textit{fixed}). The variable coded as $a$ records the congress person voting {\em yes/no} versus {\em undecided} and the vote coded as $b$ records the congress person voting {\em yes} versus {\em no/undecided}. In both cases the median probability gives the probability of the first outcome.}
		\label{fig:congplot}
\end{figure}

From Table~\ref{tab:cvot.par} and Figure~\ref{fig:congplot} it is possible to observe that the responses for the median individual in the mainly \textit{Republican} group (Group 4) are the opposite to the ones given by the median individuals in the mainly \textit{Democrat} groups (Groups 1 and 3) for the majority of the issues.  

In fact the median individual in the \textit{Republican} group (Group 4) has high probabilities to give positive responses for the variables $4b,5b,6b,12b,13b,14b$, which are concerned with the Physician Fee Freeze, the El Salvador Aid, the Religious Groups in Schools, the Education Spending, the Superfund Right to Sue and the Crime; the \textit{Democrat} groups (Groups 1 and 3) have low probabilities of giving these issues a positive response. Additionally, the \textit{Democrat} groups have high probability of positive responses for the variables $3b,7b,8b,9b$, which are concerned with the Adoption of the Budget Resolution, the Anti-Satellite Test Ban, the Aid to the Nicaraguan `Contras' and the MX Missile, whereas the \textit{Republican} groups have low probabilities for these issues. 

The choice of a model with the $\mathbf{w}$ common in all groups suggests the latent trait has the same effect in all groups.
The estimated posterior mean of the latent variable $\mathbf{y}_n$ conditional on $z_{ng}=1$ was estimated for each voter and for each group and these are shown in Figures 1--4 of the supplementary material. The posterior mean is plotted a number of times in these figures to show the impact of the latent variable on outcome variables. The variation in the positions of the congressmen in these plots shows how the voting is influenced by the latent trait in addition to the baseline median probability of voting as characterized by the intercept parameters $\mathbf{b}$. It is clear from the plots that the latent trait can introduce significant variation in the outcome variable within a group if the slope parameter $\mathbf{w}_g$ has large magnitude.

\begin{sidewaystable}
\caption{Parameter estimates (\textit{standard errors}) for the best model selected ($G=4$, $D=2$, $\mathbf{w}$ \textit{fixed}). The variable coded as $a$ records the congress person voting {\em yes/no} versus {\em undecided} and the vote coded as $b$ records the congress person voting {\em yes} versus {\em no/undecided}. In both cases the median probability gives the probability of the first outcome. }
\label{tab:cvot.par}
\smallskip
\centering
\scriptsize
\rowcolors*{2}{gray!30}{}
\begin{tabular}{@{\extracolsep{-4pt}}clrcrcrcrcrrrr}
\multicolumn{2}{c}{} & \multicolumn{2}{c}{$g=1$} & \multicolumn{2}{c}{$g=2$} &\multicolumn{2}{c}{$g=3$}&\multicolumn{2}{c}{$g=4$}&\multicolumn{2}{c}{$d=1$}&\multicolumn{2}{c}{$d=2$}\\
  \hline
\multicolumn{2}{c}{} \vrule &  \multicolumn{1}{c}{$b_{mg}$} & \multicolumn{1}{c}{$\pi_{mg}(\mathbf{0})$} \vrule & \multicolumn{1}{c}{$b_{mg}$} & \multicolumn{1}{c}{$\pi_{mg}(\mathbf{0})$} \vrule & \multicolumn{1}{c}{$b_{mg}$} & \multicolumn{1}{c}{$\pi_{mg}(\mathbf{0})$} \vrule & \multicolumn{1}{c}{$b_{mg}$} & \multicolumn{1}{c}{$\pi_{mg}(\mathbf{0})$} \vrule & \multicolumn{1}{c}{$w_{dm}$} & \multicolumn{1}{c}{$w_{dm}^*$} \vrule & \multicolumn{1}{c}{$w_{dm}$} & \multicolumn{1}{c}{$w_{dm}^*$} \\ 
  \hline
\cellcolor[gray]{1} \multirow{2}{*}{Q1} & $a$ & 4.23 \;       (\textit{1.99}) & 0.98 & -0.98 \;        (\textit{0.89}) & 0.28 & 4.23 \;         (\textit{0.56}) & 0.99 & 9.67 \;         (\textit{0.25}) & 1 & -0.24 \;   (\textit{0.23}) & -0.23 & 0.14 \;        (\textit{0.08}) & 0.14 \\ 
&  $b$ & 1.10 \;      (\textit{0.30}) & 0.73 & -2.49 \;         (\textit{2.22}) & 0.09 & 0.24 \;         (\textit{0.16}) & 0.55 & -1.74 \;        (\textit{0.24}) & 0.16 & -0.54 \;        (\textit{0.13}) & -0.44 & 0.45 \;        (\textit{0.14}) & 0.37 \\ 
\cellcolor[gray]{1} \multirow{2}{*}{Q2} &  $a$ & 1.55 \;     (\textit{0.33}) & 0.82 & -2.45 \;        (\textit{2.26}) & 0.09 & 3.04 \;         (\textit{0.32}) & 0.95 & 2.23 \;         (\textit{0.28}) & 0.90 & 0.35 \;  (\textit{0.19}) & 0.31 & 0.30 \;  (\textit{0.12}) & 0.27 \\ 
&  $b$ & -0.35 \;    (\textit{0.29}) & 0.43 & -24.80 \;        (\textit{1.81}) & 0 & 0.02 \;    (\textit{0.18}) & 0.50 & -0.38 \;         (\textit{0.20}) & 0.43 & 0.90 \;   (\textit{0.16}) & 0.57 & 0.81 \;         (\textit{0.16}) & 0.52 \\ 
\cellcolor[gray]{1} \multirow{2}{*}{Q3} &  $a$ & 10.39 \;    (\textit{0.31}) & 1 & -0.17 \;   (\textit{0.88}) & 0.46 & 4.21 \;         (\textit{0.63}) & 0.98 & 4.52 \;         (\textit{0.81}) & 0.99 & -0.39 \;        (\textit{0.22}) & -0.36 & 0.02 \;        (\textit{0.07}) & 0.02 \\ 
&  $b$ & 3.35 \;     (\textit{0.60}) & 0.94 & -0.72 \;         (\textit{1.16}) & 0.37 & 2.08 \;         (\textit{0.28}) & 0.83 & -2.43 \;        (\textit{0.32}) & 0.12 & -1.21 \;        (\textit{0.19}) & -0.77 & 0.06 \;        (\textit{0.12}) & 0.04 \\ 
\cellcolor[gray]{1} \multirow{2}{*}{Q4} &  $a$ & 11.72 \;    (\textit{0.41}) & 1 & 0.21 \;    (\textit{0.62}) & 0.55 & 3.67 \;         (\textit{0.36}) & 0.97 & 11.17 \;        (\textit{0.36}) & 1 & -0.56 \;   (\textit{0.16}) & -0.49 & -0.08 \;       (\textit{0.10}) & -0.07 \\ 
&  $b$ & -3.84 \;    (\textit{0.74}) & 0.04 & -15.91 \;       (\textit{1.12}) & 0 & -2.55 \;   (\textit{0.25}) & 0.13 & 3.58 \;         (\textit{0.53}) & 0.95 & 1.33 \;         (\textit{0.19}) & 0.80 & -0.11 \;         (\textit{0.12}) & -0.07 \\ 
\cellcolor[gray]{1} \multirow{2}{*}{Q5} &  $a$ & 3.11 \;     (\textit{0.69}) & 0.95 & 0.19 \;         (\textit{0.84}) & 0.54 & 3.40 \;  (\textit{0.37}) & 0.97 & 9.30 \;  (\textit{0.35}) & 1 & -0.36 \;   (\textit{0.22}) & -0.34 & -0.09 \;       (\textit{0.10}) & -0.09 \\ 
&  $b$ & -3.69 \;   (\textit{0.64}) & 0.11 & -26.22 \;       (\textit{2.14}) & 0 & -1.93 \;   (\textit{0.25}) & 0.26 & 5.97 \;         (\textit{0.78}) & 0.97 & 2.41 \;         (\textit{0.20}) & 0.92 & 0.04 \;  (\textit{0.08}) & 0.01 \\ 
\cellcolor[gray]{1} \multirow{2}{*}{Q6} &  $a$ & 3.50 \;     (\textit{0.84}) & 0.97 & 0.18 \;         (\textit{0.86}) & 0.55 & 3.94 \;         (\textit{0.49}) & 0.98 & 10.26 \;        (\textit{0.39}) & 1 & 0.01 \;    (\textit{0.24}) & 0.01 & -0.24 \;        (\textit{0.11}) & -0.23 \\ 
&  $b$ & -0.88 \;   (\textit{0.38}) & 0.36 & -1.98 \;        (\textit{1.36}) & 0.27 & 0.30 \;  (\textit{0.22}) & 0.52 & 3.44 \;         (\textit{0.43}) & 0.90 & 2.06 \;  (\textit{0.20}) & 0.88 & -0.48 \;         (\textit{0.14}) & -0.21 \\ 
\cellcolor[gray]{1} \multirow{2}{*}{Q7} &  $a$ & 3.46 \;    (\textit{0.74}) & 0.95 & 0.76 \;         (\textit{0.88}) & 0.64 & 4.20 \;  (\textit{0.38}) & 0.97 & 4.99 \;         (\textit{2.53}) & 0.99 & -1.00 \;   (\textit{0.23}) & -0.71 & -0.11 \;       (\textit{0.12}) & -0.08 \\ 
& $b$ & 3.44 \;    (\textit{0.59}) & 0.89 & 0.61 \;         (\textit{1.15}) & 0.55 & 1.43 \;         (\textit{0.26}) & 0.70 & -2.07 \;         (\textit{0.27}) & 0.21 & -2.22 \;        (\textit{0.19}) & -0.91 & -0.09 \;       (\textit{0.10}) & -0.04 \\ 
\cellcolor[gray]{1} \multirow{2}{*}{Q8} &  $a$ & 16.20 \;    (\textit{1.09}) & 1 & 1.33 \;    (\textit{0.93}) & 0.73 & 3.95 \;         (\textit{0.42}) & 0.96 & 3.84 \;         (\textit{0.55}) & 0.97 & -1.04 \;        (\textit{0.30}) & -0.71 & -0.25 \;        (\textit{0.15}) & -0.17 \\ 
&  $b$ & 30.61 \;   (\textit{2.79}) & 1 & 1.57 \;    (\textit{1.23}) & 0.64 & 2.19 \;         (\textit{0.30}) & 0.75 & -3.73 \;         (\textit{0.38}) & 0.11 & -2.55 \;        (\textit{0.21}) & -0.93 & -0.14 \;       (\textit{0.11}) & -0.05 \\ 
\cellcolor[gray]{1} \multirow{2}{*}{Q9} &  $a$ & 2.25 \;    (\textit{0.41}) & 0.90 & 1.09 \;  (\textit{0.90}) & 0.73 & 3.09 \;  (\textit{0.33}) & 0.95 & 4.64 \;         (\textit{1.83}) & 0.99 & -0.55 \;        (\textit{0.14}) & -0.47 & -0.19 \;       (\textit{0.11}) & -0.16 \\ 
&  $b$ & 1.59 \;    (\textit{0.41}) & 0.74 & 1.21 \;         (\textit{1.28}) & 0.64 & 1.07 \;         (\textit{0.23}) & 0.67 & -3.26 \;        (\textit{0.37}) & 0.10 & -1.94 \;         (\textit{0.18}) & -0.88 & -0.28 \;       (\textit{0.13}) & -0.13 \\ 
\cellcolor[gray]{1} \multirow{2}{*}{Q10} &  $a$ & 3.51 \;    (\textit{0.95}) & 0.97 & 0.58 \;         (\textit{0.83}) & 0.64 & 10.46 \;        (\textit{0.28}) & 1 & 5.11 \;    (\textit{2.29}) & 0.99 & -0.29 \;        (\textit{0.23}) & -0.27 & -0.16 \;       (\textit{0.12}) & -0.15 \\ 
&  $b$ & -0.58 \;   (\textit{0.28}) & 0.37 & -1.17 \;        (\textit{1.36}) & 0.27 & 0.10 \;  (\textit{0.16}) & 0.52 & 0.14 \;         (\textit{0.18}) & 0.53 & -0.13 \;        (\textit{0.13}) & -0.11 & -0.74 \;       (\textit{0.16}) & -0.59 \\ 
\cellcolor[gray]{1} \multirow{2}{*}{Q11} &  $a$ & 3.54 \;    (\textit{0.75}) & 0.97 & -1.03 \;        (\textit{1.36}) & 0.27 & 3.41 \;         (\textit{0.34}) & 0.96 & 3.74 \;         (\textit{0.48}) & 0.98 & -0.40 \;         (\textit{0.25}) & -0.37 & -0.15 \;       (\textit{0.14}) & -0.13 \\ 
&  $b$ & -0.37 \;   (\textit{0.26}) & 0.41 & -2.43 \;        (\textit{3.37}) & 0.09 & -0.05 \;        (\textit{0.16}) & 0.49 & -1.80 \;         (\textit{0.29}) & 0.15 & 0.42 \;         (\textit{0.18}) & 0.38 & 0.20 \;  (\textit{0.22}) & 0.18 \\ 
\cellcolor[gray]{1} \multirow{2}{*}{Q12} &  $a$ & 3.56 \;    (\textit{0.91}) & 0.97 & -0.20 \;         (\textit{0.97}) & 0.45 & 2.44 \;         (\textit{0.29}) & 0.91 & 3.44 \;         (\textit{0.80}) & 0.97 & -0.38 \;         (\textit{0.23}) & -0.34 & -0.29 \;       (\textit{0.18}) & -0.26 \\ 
&  $b$ & -2.38 \;   (\textit{0.42}) & 0.12 & -1.86 \;        (\textit{1.17}) & 0.18 & -2.32 \;        (\textit{0.25}) & 0.13 & 2.18 \;         (\textit{0.39}) & 0.87 & 0.96 \;         (\textit{0.18}) & 0.66 & -0.45 \;        (\textit{0.16}) & -0.31 \\ 
\cellcolor[gray]{1} \multirow{2}{*}{Q13} &  $a$ & 2.87 \;    (\textit{0.53}) & 0.94 & 0.63 \;         (\textit{0.85}) & 0.64 & 2.71 \;         (\textit{0.27}) & 0.93 & 4.28 \;         (\textit{0.55}) & 0.98 & -0.55 \;        (\textit{0.20}) & -0.48 & -0.15 \;        (\textit{0.15}) & -0.13 \\ 
&  $b$ & -2.32 \;   (\textit{0.44}) & 0.13 & -1.32 \;        (\textit{2.06}) & 0.27 & -0.98 \;        (\textit{0.19}) & 0.31 & 2.20 \;  (\textit{0.30}) & 0.87 & 1.08 \;  (\textit{0.18}) & 0.73 & -0.18 \;        (\textit{0.20}) & -0.12 \\ 
\cellcolor[gray]{1} \multirow{2}{*}{Q14} &  $a$ & 3.71 \;    (\textit{0.67}) & 0.97 & 0.68 \;         (\textit{0.94}) & 0.64 & 3.13 \;         (\textit{0.28}) & 0.95 & 9.54 \;         (\textit{1.08}) & 1 & -0.69 \;   (\textit{0.24}) & -0.55 & -0.26 \;       (\textit{0.16}) & -0.21 \\ 
&  $b$ & -1.70 \;    (\textit{0.36}) & 0.22 & -1.38 \;        (\textit{0.97}) & 0.27 & -0.68 \;        (\textit{0.20}) & 0.37 & 11.16 \;         (\textit{3.25}) & 1 & 1.16 \;    (\textit{0.19}) & 0.70 & -0.65 \;         (\textit{0.15}) & -0.39 \\ 
\cellcolor[gray]{1} \multirow{2}{*}{Q15} &  $a$ & 2.56 \;    (\textit{0.47}) & 0.92 & -0.19 \;        (\textit{0.96}) & 0.46 & 2.98 \;         (\textit{0.33}) & 0.95 & 3.19 \;         (\textit{0.45}) & 0.96 & -0.37 \;        (\textit{0.24}) & -0.35 & -0.14 \;       (\textit{0.15}) & -0.13 \\ 
&  $b$ & 0.81 \;    (\textit{0.31}) & 0.66 & -1.94 \;        (\textit{1.50}) & 0.18 & 0.31 \;  (\textit{0.18}) & 0.57 & -2.80 \;         (\textit{0.35}) & 0.08 & -1.08 \;        (\textit{0.16}) & -0.73 & 0.07 \;        (\textit{0.15}) & 0.05 \\ 
\cellcolor[gray]{1} \multirow{2}{*}{Q16} &  $a$ & -26.79 \;  (\textit{2.38}) & 0 & -3.04 \;   (\textit{2.13}) & 0.18 & 4.17 \;         (\textit{0.35}) & 0.91 & 3.78 \;         (\textit{0.33}) & 0.93 & -2.18 \;        (\textit{0.17}) & -0.91 & 0.01 \;        (\textit{0.12}) & 0.00 \\ 
&  $b$ & -33.71 \;  (\textit{3.05}) & 0 & -3.69 \;   (\textit{2.54}) & 0.18 & 3.76 \;         (\textit{0.39}) & 0.85 & 0.99 \;         (\textit{0.28}) & 0.61 & -2.74 \;        (\textit{0.22}) & -0.94 & 0.03 \;        (\textit{0.14}) & 0.01 \\ 
   \hline
\multicolumn{2}{c}{} \vrule& \multicolumn{2}{c}{$\eta_1=0.152$ \; (\textit{0.017})} \vrule& \multicolumn{2}{c}{$\eta_2=0.025$ \; (\textit{0.009})} \vrule& \multicolumn{2}{c}{$\eta_3=0.465$ \; (\textit{0.029})} \vrule& \multicolumn{2}{c}{$\eta_4=0.358$ \; (\textit{0.027})} \vrule & \multicolumn{4}{c}{} \\
   \hline
\end{tabular}
\end{sidewaystable}

From the analysis of the $\mathit{lift}$ (Tables~2--5 of the supplementary material) it is possible to observe that: 
within Group 1 and Group 3 there is a strong positive dependence within the set of variables $(4b,5b,6b,12b,13b,14b)$; this is shown by $\mathit{lift}$ values in excess of one. 
Within Group 1 there is also a negative dependence between this set of variables and the set $(16a,16b)$, within which there is a positive dependence; this is evidenced by the $\mathit{lift}$ values well below one.
Within Group 4 there is a positive dependence within the set of variables $(3b,7b,8b,9b,15b)$, and a negative dependence between this set of variables and the variable $2b$.
In Group 2 there is a strong positive dependence within the sets of variables $(4b,5b,6b,12b,13b,14b)$,  $(2b,4b,5b,6b)$ and $(15b,16a,16b)$, and a negative dependence between the sets of variables $(2b,4b,5b,6b)$ and $(3b,7b,8b,9b,15b,16a,16b)$, and between the sets variables $(12b,13b,14b)$ and $(15b,16a,16b)$. These $\mathit{lift}$ values show that the model is accounting for dependence within the groups which may be missed in a latent class analysis of these data.

\section{Conclusions}
\label{se:conclusions}

\noindent The mixture of latent trait analyzers model provides a powerful methodology for model-based clustering for categorical data. It provides a suitable alternative to LCA for clustering binary data and it can be fitted in a computationally efficient manner using a variational EM algorithm. In addition, the model includes LCA as a special case so it is also applicable when the LCA modeling assumptions are appropriate. The continuous latent trait in model also facilitates investigating the dependence between variables, thus offering a combination of the properties offered by the Latent Class Analysis model and the Latent Trait Analysis model. The model parameters are interpretable and provide a characterization of the within cluster structure.

The variational EM algorithm proposed for model fitting gives an efficient mechanism for model fitting, thus making the method easily applicable. 

The model provides an extra model-based clustering framework for categorical data with a similar structure to  the continuous data methods of {\sf mclust} \citep{fraley02}, {\sf MIXMOD} \citep{biernacki06}, {\sf EMMIX} \citep{mclachlan99}, {\sf pgmm} \citep{mcnicholas08} and {\sf t-EIGEN}\citep{andrews11}.

The MLTA framework as proposed for the automatic selection using BIC (and a variant) of the number of clusters ($G$) and the latent trait dimensionality ($D$) required to appropriately model dependence within clusters. Alternative model selection techniques that are not based on a penalized likelihood criterion could offer a more computationally efficient manner for selecting $G$ and $D$. 

The excellent clustering behavior of this method has been shown by two applications on the National Long Term Care Survey (NLTCS) and U.S. Congressional Voting data sets. In both cases, the model found groups that were intuitive in their interpretation and fewer groups were found than the LCA model suggests. 

\section*{Acknowledgements}
This research was supported by a Science Foundation Ireland Research Frontiers Programme Grant (06/RFP/M040) and Strategic Research Cluster Grant (08/SRC/I1407).

\newpage
\setcounter{section}{0}
\setcounter{figure}{0}
\setcounter{table}{0}
\setcounter{page}{1}

\begin{center}
{\LARGE Supplementary Material for\\
 ``Mixture of Latent Trait Analyzers \\[.3cm]
 for Model-Based Clustering of Categorical Data''}\\[.3cm]

        {\Large Isabella Gollini$^\dagger \qquad \qquad \qquad $  Thomas Brendan Murphy$^\star$}\\[.3cm]
{$^\dagger$National Centre for Geocomputation, National University of Ireland Maynooth, Ireland}\\
{$^\star$School of Mathematical Sciences, University College Dublin, Ireland}

\end{center}

\section{Parameter Estimation in the Parsimonious Model}
\label{se:parsimonious}

It is necessary to use an EM algorithm to estimate the parameters in the parsimonious model presented in Section 3.1 of the main paper. The mathematical calculations required to derive the EM algorithm for the parsimonious model beyond those needed for the MLTA model are contained herein. \\
\vspace{-.3cm}
We define $\mathbf{X}=(\mathbf{x}_1,\ldots,\mathbf{x}_N)$, $\mathbf{Y}=(\mathbf{y}_1,\ldots,\mathbf{y}_N)$, $\mathbf{Z}=(\mathbf{z}_{11},\ldots,\mathbf{z}_{NG})$ and $\boldsymbol{\Xi}=(\boldsymbol{\xi}_{11},\ldots,\boldsymbol{\xi}_{NG})$.
\begin{equation*} \label{mix.lpxyxiw}
\begin{split}
&\log\left( \tilde{p}\left( \mathbf{X}|\mathbf{Y},\mathbf{Z},\boldsymbol{\Xi} \right) \right) = \sum_{g=1}^G \sum_{n=1}^N z_{ng}\log\left( \prod_{m=1}^M\tilde{p}\left( x_{nm}|\mathbf{y}_n,z_{ng}=1,\xi_{nmg}\right)\right) \\
&=\sum_{g=1}^G \sum_{n=1}^N z_{ng}\lambda(\xi_{nmg})\left[ \mathbf{w}_{m}^T\mathbf{y}_n\mathbf{y}_n^T\mathbf{w}_{m} +2b_{mg}\mathbf{w}_{m}^T\mathbf{y}_n+b_{mg}^2 \right] + \sum_{g=1}^G \sum_{n=1}^N z_{ng} \left(x_{nm}-\dfrac{1}{2}\right)\left(\mathbf{w}_{m}^T\mathbf{y}_n+b_{mg} \right) +\mathrm{const}_{\hat{\mathbf{w}}_m}\\
\end{split}
\end{equation*}
where $\mathrm{const}_{\hat{\mathbf{w}}_m}$ are all the terms not involving $\hat{\mathbf{w}}_m$.\\

\noindent {\bf \large E-step}
\begin{equation*}\label{mix.a.estepw}
\begin{split}
\mathbb{E}&\left[ \log\left( \tilde{p}\left( \mathbf{X}|\mathbf{Y},\mathbf{Z},\boldsymbol{\Xi} \right)\right)  \right] = 
\sum_{g=1}^G \sum_{n=1}^N z_{ng}\lambda(\xi_{nmg})\left[ \mathbf{w}_{m}^T \mathbb{E}\left[\mathbf{y}_{n}\mathbf{y}_{n}^T\right]_g \mathbf{w}_{m} +2b_{mg}\mathbf{w}_{m}^T\mathbb{E}[\mathbf{y}_{n}]_g+b_{mg}^2 \right]\\
& \quad + \sum_{g=1}^G \sum_{n=1}^N z_{ng} \left[ \left(x_{nm}-\dfrac{1}{2}\right)\left(\mathbf{w}_{m}^T\mathbb{E}[\mathbf{y}_{n}]_g+b_{mg} \right) \right] +\mathrm{const}_{\hat{\mathbf{w}}_m}\\
&= \mathbf{w}_{m}^T \left[ \sum_{g=1}^G \sum_{n=1}^N z_{ng}\lambda(\xi_{nmg}) \mathbb{E}\left[\mathbf{y}_{n}\mathbf{y}_{n}^T\right]_g\right]  \mathbf{w}_{m}+ 2 \mathbf{w}_{m}^T \left[ \sum_{g=1}^G \sum_{n=1}^N z_{ng}\lambda(\xi_{nmg}) \boldsymbol{\mu}_{ng} b_{mg}\right]+\sum_{g=1}^G \sum_{n=1}^N z_{ng}\lambda(\xi_{nmg})b_{mg}^2  \\
& \quad +  \mathbf{w}_{m}^T \left[ \sum_{g=1}^G \sum_{n=1}^N z_{ng} \left(x_{nm}-\dfrac{1}{2}\right)\boldsymbol{\mu}_{ng}\right] + \sum_{g=1}^G \sum_{n=1}^N z_{ng}\left(x_{nm}-\dfrac{1}{2}\right) b_{mg} +\mathrm{const}_{\hat{\mathbf{w}}_m}\\
&=\hat{\mathbf{w}}_{m}^T \mathbf{K}_m\hat{\mathbf{w}}_{m} + \hat{\mathbf{w}}_{m}^T \boldsymbol{\gamma}_m +\mathrm{const}_{\hat{\mathbf{w}}_m}\\
\end{split}
\end{equation*}
where 
\begin{equation*}	\label{mix.a.wh.v}
\hat{\mathbf{w}}_{m}=(\mathbf{w}_{m}^T,b_{m1},\ldots, b_{mG})^T, \quad \boldsymbol{\gamma}_m=
\begin{bmatrix}
\sum_{g=1}^G \sum_{n=1}^N z_{ng}\left(x_{nm}-\dfrac{1}{2}\right) \boldsymbol{\mu}_{ng}\\
\sum_{n=1}^N z_{n1} \left(x_{nm}-\dfrac{1}{2}\right)\\
\vdots\\
\sum_{n=1}^N z_{nG} \left(x_{nm}-\dfrac{1}{2}\right)
\end{bmatrix}
\end{equation*}
and,
\begin{equation*}	\label{mix.a.K}
\mathbf{K}_{m}=
\begin{bmatrix}
\sum_{g=1}^G \sum_{n=1}^N z_{ng}\lambda(\xi_{nmg})\mathbb{E}\left[\mathbf{y}_{n}\mathbf{y}_{n}^T\right]_g 
& \sum_{n=1}^N z_{n1}\lambda(\xi_{nm1})\boldsymbol{\mu}_{n1} 
& \ldots & 
\sum_{n=1}^N z_{nG}\lambda(\xi_{nmG})\boldsymbol{\mu}_{nG}\\
\sum_{n=1}^N z_{n1}\lambda(\xi_{nm1})\boldsymbol{\mu}_{n1}^{T} & \sum_{n=1}^N z_{n1}\lambda(\xi_{nm1}) & 0 & 0\\
\vdots & 0 & \ddots & 0\\
\sum_{n=1}^N z_{nG}\lambda(\xi_{nmG})\boldsymbol{\mu}_{nG}^T & 0 & 0 & \sum_{n=1}^N z_{nG}\lambda(\xi_{nmG})\\
\end{bmatrix}
\end{equation*}

\noindent {\bf \large M-step}

\begin{equation*}\label{mix.a.mstepw}
\hat{\mathbf{w}}_{m} = \arg\max_{\hat{\mathbf{w}}_{m}}\left(\mathbb{E}\left[ \log\left( \tilde{p}\left( \mathbf{X}|\mathbf{Y},\mathbf{Z},\boldsymbol{\Xi} \right)\right)  \right]\right) 
\end{equation*}
\begin{equation*}	\label{mix.a.derMw}
\dfrac{\partial \mathbb{E}\left[ \log\left( \tilde{p}\left( \mathbf{X}|\mathbf{Y},\mathbf{Z},\boldsymbol{\Xi} \right)\right)  \right]}{\partial \hat{\mathbf{w}}_{m}}=2\mathbf{K}_m\hat{\mathbf{w}}_{m} + \boldsymbol{\gamma}_m=0
\end{equation*}
therefore,
\begin{equation*} \label{mix.a.what}
\hat{\mathbf{w}}_{m}=-\left[2\mathbf{K}_m\right]^{-1} \boldsymbol{\gamma}_m
\end{equation*}

\section{Additional Results}

\subsection{National Long Term Care Survey (NLTCS) Example}

Table~\ref{tab:nltcs.ex100} presents the results mentioned in the example regarding the National Long Term Care Survey (NLTCS) (Section 6.1 of the main paper). It shows the comparison of the observed and the expected frequencies for the response patterns with more than 100 observations under the best model selected ($G=10$, $D=1$).

\newpage 
\begin{table}[ht]
\caption{Expected frequencies for the response patterns with more than 100 observations for the selected model}
\label{tab:nltcs.ex100}
\smallskip
\centering
\footnotesize
\begin{tabular}{crr}
  \hline
\textit{Response Pattern} & \textit{Observed} & \textit{Expected} \\ 
  \hline
0000000000000000 & 3853 & 3851 \\ 
0000100000000000 & 216 & 225 \\ 
0000001000000000 & 1107 & 1028 \\ 
0000101000000000 & 188 & 228 \\ 
0000001000100000 & 122 & 120 \\ 
0000000000010000 & 351 & 351 \\ 
0010000000010000 & 206 & 141 \\ 
0000001000010000 & 303 & 312 \\ 
0010001000010000 & 182 & 168 \\ 
0000101000010000 & 108 & 98 \\ 
0010101000010000 & 106 & 128 \\ 
0000000000001000 & 195 & 200 \\ 
0000001000001000 & 198 & 224 \\ 
0000001000101000 & 196 & 162 \\ 
0000001000011000 & 123 & 110 \\ 
0000001000111000 & 176 & 176 \\ 
0010001000111000 & 120 & 98 \\ 
0000101000111000 & 101 & 111 \\ 
0111111111111000 & 102 & 64 \\ 
1111111111111010 & 107 & 88 \\ 
0111111111111110 & 104 & 125 \\ 
1111111111111110 & 164 & 218 \\ 
0111111111111111 & 153 & 170 \\ 
1111111111111111 & 660 & 506 \\ 
   \hline
\end{tabular}
\end{table}

\subsection{U.S. Congressional Voting Example}
This section reports some details of the results for the best model selected ($G=4$, $D=2$, $\mathbf{w}$ \textit{fixed}) for the U.S. Congressional Voting Example mentioned in the paper (Section 6.2).

\begin{sidewaysfigure}
\begin{center}
\includegraphics[scale=0.73]{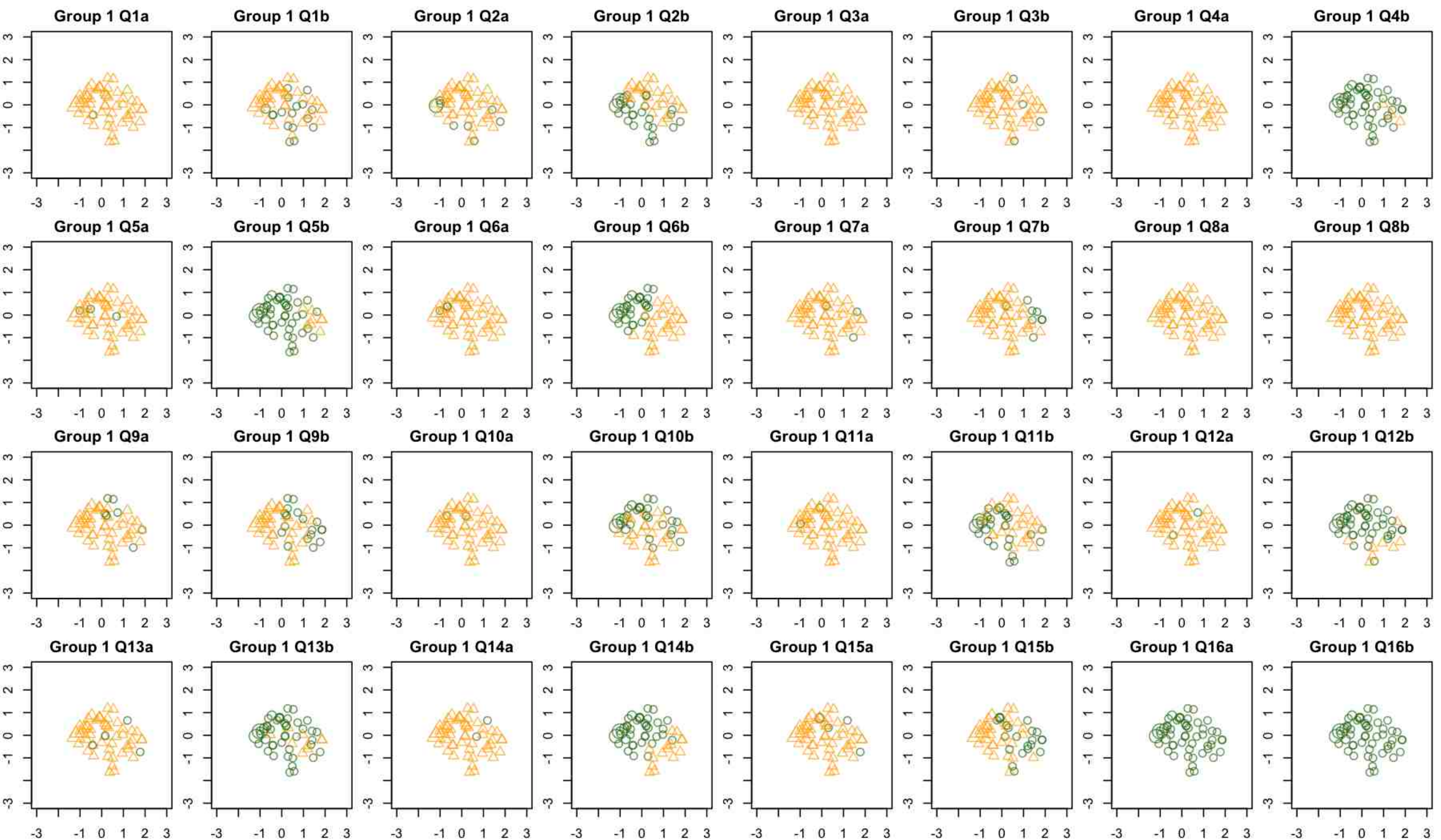}
\caption{Plot of the $\boldsymbol{\mu}_{n1}$ for the selected model $(G=4, D=2, \mathbf{w} fixed)$ in the U.S. congressional voting example.}\label{mlta.fig.mug1}
\end{center}
\end{sidewaysfigure}

\begin{sidewaysfigure}
\begin{center}
\includegraphics[scale=0.73]{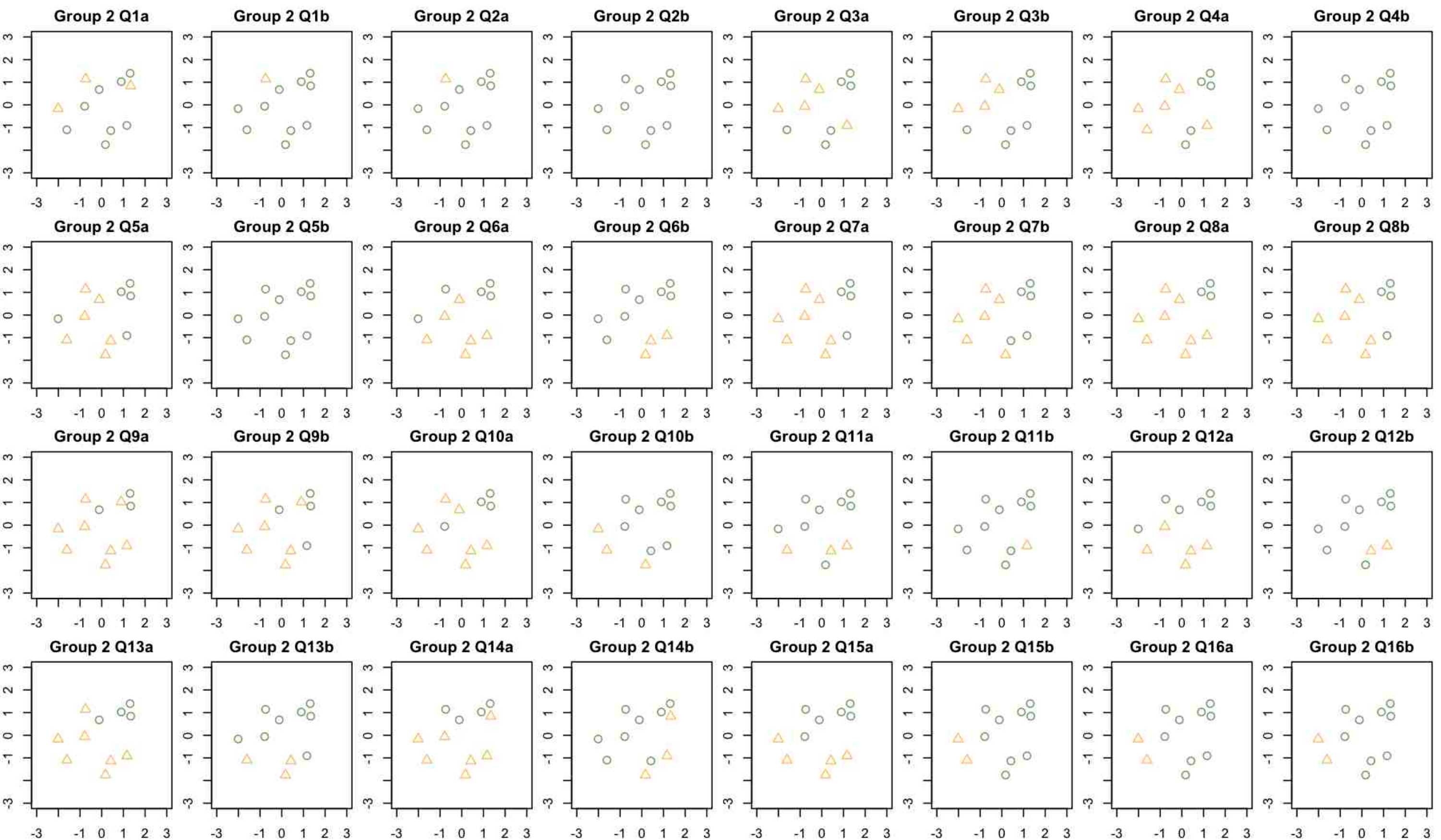}
\caption{Plot of the $\boldsymbol{\mu}_{n2}$ for the selected model $(G=4, D=2, \mathbf{w} fixed)$ in the U.S. congressional voting example.}\label{mlta.fig.mug2}
\end{center}
\end{sidewaysfigure}

\begin{sidewaysfigure}
\begin{center}
\includegraphics[scale=0.73]{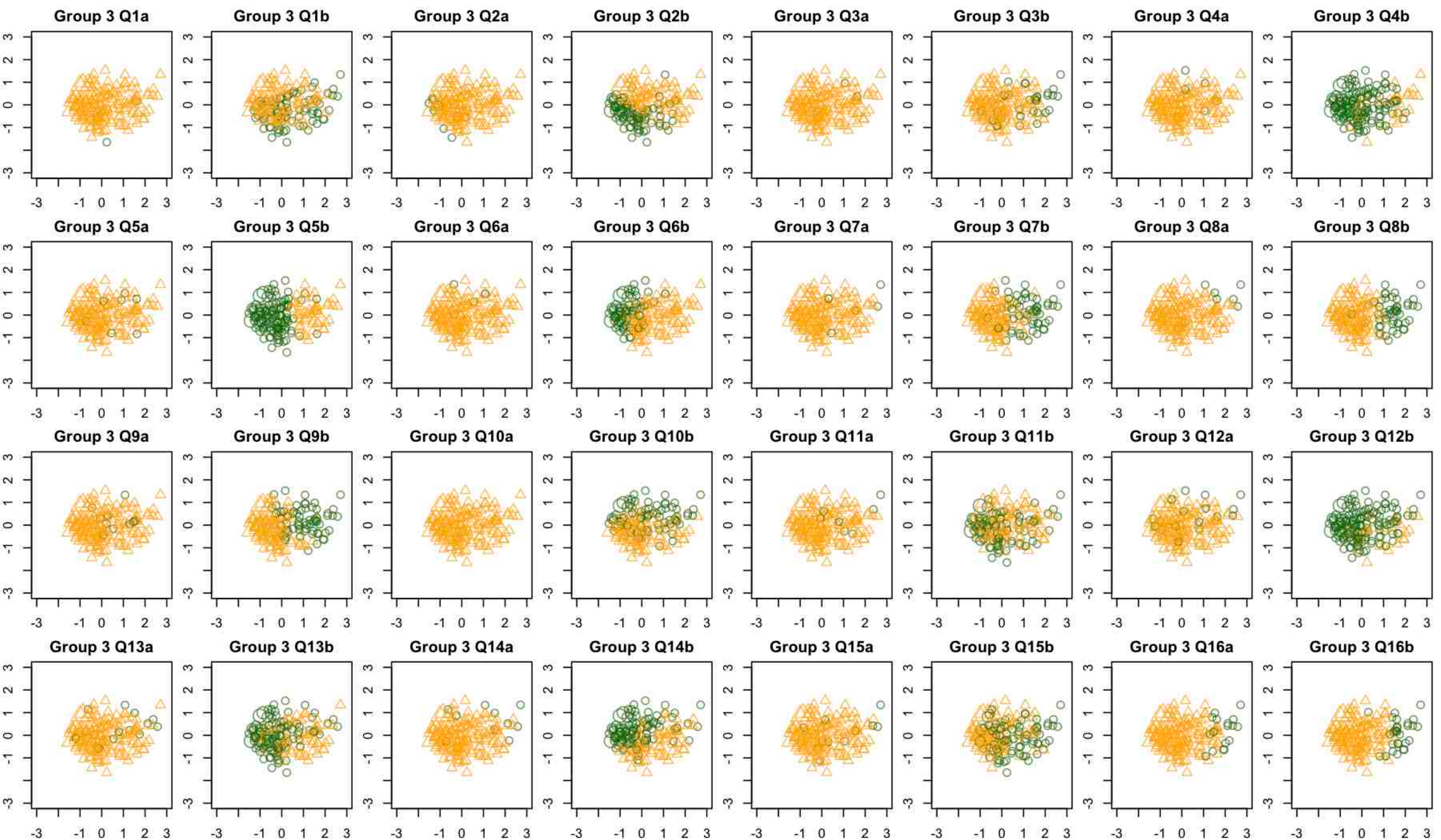}
\caption{Plot of the $\boldsymbol{\mu}_{n3}$ for the selected model $(G=4, D=2, \mathbf{w} fixed)$ in the U.S. congressional voting example.}\label{mlta.fig.mug3}
\end{center}
\end{sidewaysfigure}

\begin{sidewaysfigure}
\begin{center}
\includegraphics[scale=0.73]{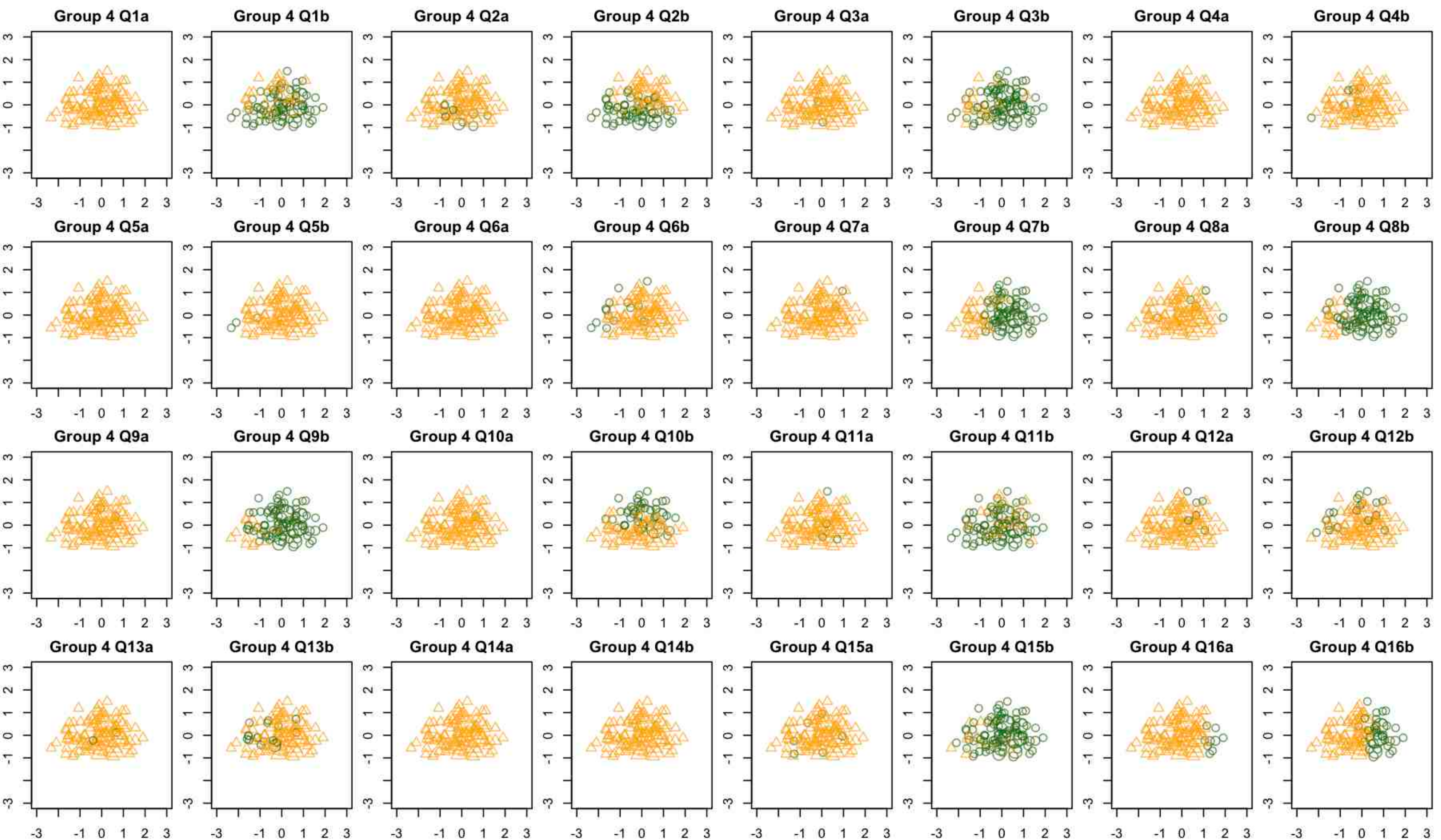}
\caption{Plot of the $\boldsymbol{\mu}_{n4}$ for the selected model $(G=4, D=2, \mathbf{w} fixed)$ in the U.S. congressional voting example.}\label{mlta.fig.mug4}
\end{center}
\end{sidewaysfigure}

We show the plots (Figures~\ref{mlta.fig.mug1}--\ref{mlta.fig.mug4}), mentioned in Section 6.2.1, of the estimated expectations $\boldsymbol{\mu}_{ng}$ of the latent variable $\mathbf{y}_n$. We have a plot for each group, and only the observations with the maximum a posteriori probability to belong to that group are shown.

In the variables coded as $a$ the green circles correspond to the congressmen voting $yes/no$ and the yellow triangles to the people $undecided$.
In the variables coded as $b$ the green circles correspond to the congressmen voting $yes$ and the yellow triangles to the people $no/undecided$.


In the next pages are also reported Tables \ref{tab:cvot.lift1}--\ref{tab:cvot.lift4} showing the estimates of the $\mathit{lift}$ mentioned in the paper (end of Section 6.2.1).

\begin{sidewaystable}
\caption{Estimate of the $\mathit{lift}$ for $g=1$ in the selected model} \label{tab:cvot.lift1}
\smallskip
\scriptsize
\begin{tabular}{@{\extracolsep{-5pt}}r|cccccccccccccccccccccccccccccccc}
  \multicolumn{16}{c}{}
    \vspace{-11pt}\\
  \hline
$m$ & 1b & 2a & 2b & 3a & 3b & 4a & 4b & 5a & 5b & 6a & 6b & 7a & 7b & 8a & 8b & 9a & 9b & 10a & 10b & 11a & 11b & 12a & 12b & 13a & 13b & 14a & 14b & 15a & 15b & 16a & 16b \\ 
  \hline
1a  & 1 & 1 & 1 & 1 & 1 & 1 & 1 & 1 & 0.99 & 1 & 1 & 1 & 1 & 1 & 1 & 1 & 1 & 1 & 1 & 1 & 1 & 1 & 1 & 1 & 1 & 1 & 1 & 1 & 1 & 1 & 1.01 \\ 
  1b  &  & 1 & 0.99 & 1 & 1.01 & 1 & 0.83 & 1 & 0.80 & 1 & 0.87 & 1.01 & 1.02 & 1 & 1 & 1.01 & 1.04 & 1 & 0.96 & 1 & 0.98 & 1 & 0.85 & 1 & 0.87 & 1 & 0.85 & 1 & 1.04 & 1.19 & 1.20 \\ 
  2a  &  &  & 1.04 & 1 & 1 & 1 & 1.06 & 1 & 1.07 & 1 & 1.04 & 1 & 0.99 & 1 & 1 & 1 & 0.98 & 1 & 0.97 & 1 & 1.02 & 1 & 1.03 & 1 & 1.04 & 1 & 1.02 & 1 & 0.98 & 0.89 & 0.88 \\ 
  2b &  &  &  & 1 & 0.98 & 1 & 1.41 & 0.99 & 1.56 & 1 & 1.24 & 0.98 & 0.93 & 1 & 1 & 0.97 & 0.85 & 0.99 & 0.83 & 0.99 & 1.13 & 0.99 & 1.17 & 0.98 & 1.28 & 0.99 & 1.14 & 0.99 & 0.89 & 0.41 & 0.36 \\ 
  3a &  &  &  &  & 1 & 1 & 1 & 1 & 1 & 1 & 1 & 1 & 1 & 1 & 1 & 1 & 1 & 1 & 1 & 1 & 1 & 1 & 1 & 1 & 1 & 1 & 1 & 1 & 1 & 1 & 1 \\ 
  3b &  &  &  &  &  & 1 & 0.90 & 1 & 0.87 & 1 & 0.95 & 1 & 1.01 & 1 & 1 & 1 & 1.02 & 1 & 1 & 1 & 0.99 & 1 & 0.94 & 1 & 0.94 & 1 & 0.95 & 1 & 1.02 & 1.05 & 1.06 \\ 
  4a &  &  &  &  &  &  & 1 & 1 & 1 & 1 & 1 & 1 & 1 & 1 & 1 & 1 & 1 & 1 & 1 & 1 & 1 & 1 & 1 & 1 & 1 & 1 & 1 & 1 & 1 & 1 & 1 \\ 
  4b &  &  &  &  &  &  &  & 0.98 & 3.34 & 1 & 1.92 & 0.94 & 0.73 & 1 & 1 & 0.93 & 0.56 & 0.99 & 0.96 & 0.99 & 1.25 & 0.99 & 2.09 & 0.96 & 2.19 & 0.97 & 2.00 & 0.97 & 0.64 & 0.09 & 0.06 \\ 
  5a &  &  &  &  &  &  &  &  & 0.98 & 1 & 0.99 & 1 & 1 & 1 & 1 & 1 & 1.01 & 1 & 1 & 1 & 1 & 1 & 0.99 & 1 & 0.99 & 1 & 0.99 & 1 & 1 & 1.02 & 1.02 \\ 
  5b &  &  &  &  &  &  &  &  &  & 1 & 2.13 & 0.92 & 0.65 & 1 & 1 & 0.91 & 0.44 & 0.99 & 0.89 & 0.98 & 1.32 & 0.98 & 2.30 & 0.95 & 2.48 & 0.96 & 2.18 & 0.96 & 0.55 & 0.03 & 0.01 \\ 
  6a &  &  &  &  &  &  &  &  &  &  & 1 & 1 & 1 & 1 & 1 & 1 & 1 & 1 & 1 & 1 & 1 & 1 & 1 & 1 & 1 & 1 & 1 & 1 & 1 & 1 & 1 \\ 
  6b &  &  &  &  &  &  &  &  &  &  &  & 0.97 & 0.87 & 1 & 1 & 0.96 & 0.73 & 0.99 & 1.02 & 0.99 & 1.16 & 0.99 & 1.67 & 0.98 & 1.69 & 0.99 & 1.67 & 0.98 & 0.75 & 0.11 & 0.06 \\ 
  7a &  &  &  &  &  &  &  &  &  &  &  &  & 1.01 & 1 & 1 & 1 & 1.02 & 1 & 1 & 1 & 0.99 & 1 & 0.97 & 1 & 0.96 & 1 & 0.97 & 1 & 1.01 & 1.04 & 1.04 \\ 
  7b &  &  &  &  &  &  &  &  &  &  &  &  &  & 1 & 1 & 1.01 & 1.07 & 1 & 1.01 & 1 & 0.96 & 1 & 0.85 & 1.01 & 0.83 & 1 & 0.86 & 1.01 & 1.05 & 1.12 & 1.12 \\ 
  8a &  &  &  &  &  &  &  &  &  &  &  &  &  &  & 1 & 1 & 1 & 1 & 1 & 1 & 1 & 1 & 1 & 1 & 1 & 1 & 1 & 1 & 1 & 1 & 1 \\ 
  8b  &  &  &  &  &  &  &  &  &  &  &  &  &  &  &  & 1 & 1 & 1 & 1 & 1 & 1 & 1 & 1 & 1 & 1 & 1 & 1 & 1 & 1 & 1 & 1 \\ 
  9a &  &  &  &  &  &  &  &  &  &  &  &  &  &  &  &  & 1.02 & 1 & 1.01 & 1 & 0.99 & 1 & 0.96 & 1 & 0.95 & 1 & 0.97 & 1 & 1.02 & 1.07 & 1.07 \\ 
  9b &  &  &  &  &  &  &  &  &  &  &  &  &  &  &  &  &  & 1 & 1.04 & 1 & 0.92 & 1 & 0.74 & 1.01 & 0.70 & 1.01 & 0.76 & 1.01 & 1.11 & 1.34 & 1.35 \\ 
  10a &  &  &  &  &  &  &  &  &  &  &  &  &  &  &  &  &  &  & 1 & 1 & 1 & 1 & 1 & 1 & 0.99 & 1 & 1 & 1 & 1 & 1.01 & 1.01 \\ 
  10b &  &  &  &  &  &  &  &  &  &  &  &  &  &  &  &  &  &  &  & 1 & 0.94 & 1 & 1.09 & 1.01 & 1 & 1 & 1.11 & 1.01 & 1.01 & 1.12 & 1.12 \\ 
  11a &  &  &  &  &  &  &  &  &  &  &  &  &  &  &  &  &  &  &  &  & 1 & 1 & 0.99 & 1 & 0.99 & 1 & 0.99 & 1 & 1 & 1.02 & 1.02 \\ 
  11b &  &  &  &  &  &  &  &  &  &  &  &  &  &  &  &  &  &  &  &  &  & 1 & 1.13 & 0.99 & 1.17 & 0.99 & 1.12 & 0.99 & 0.93 & 0.64 & 0.61 \\ 
  12a &  &  &  &  &  &  &  &  &  &  &  &  &  &  &  &  &  &  &  &  &  &  & 0.99 & 1 & 0.99 & 1 & 1 & 1 & 1 & 1.01 & 1.02 \\ 
  12b &  &  &  &  &  &  &  &  &  &  &  &  &  &  &  &  &  &  &  &  &  &  &  & 0.98 & 1.76 & 0.99 & 1.77 & 0.98 & 0.76 & 0.22 & 0.17 \\ 
  13a &  &  &  &  &  &  &  &  &  &  &  &  &  &  &  &  &  &  &  &  &  &  &  &  & 0.97 & 1 & 0.98 & 1 & 1.01 & 1.04 & 1.04 \\ 
  13b &  &  &  &  &  &  &  &  &  &  &  &  &  &  &  &  &  &  &  &  &  &  &  &  &  & 0.98 & 1.73 & 0.98 & 0.74 & 0.18 & 0.13 \\ 
  14a &  &  &  &  &  &  &  &  &  &  &  &  &  &  &  &  &  &  &  &  &  &  &  &  &  &  & 0.99 & 1 & 1.01 & 1.02 & 1.02 \\ 
  14b &  &  &  &  &  &  &  &  &  &  &  &  &  &  &  &  &  &  &  &  &  &  &  &  &  &  &  & 0.98 & 0.77 & 0.20 & 0.15 \\ 
  15a &  &  &  &  &  &  &  &  &  &  &  &  &  &  &  &  &  &  &  &  &  &  &  &  &  &  &  &  & 1.01 & 1.04 & 1.04 \\ 
  15b &  &  &  &  &  &  &  &  &  &  &  &  &  &  &  &  &  &  &  &  &  &  &  &  &  &  &  &  &  & 1.38 & 1.41 \\ 
  16a &  &  &  &  &  &  &  &  &  &  &  &  &  &  &  &  &  &  &  &  &  &  &  &  &  &  &  &  &  &  & 8.99 \\ 

\end{tabular}
\end{sidewaystable}

\begin{sidewaystable}
\caption{Estimate of the $\mathit{lift}$ for $g=2$ in the selected model} \label{tab:cvot.lift2}
\smallskip
\scriptsize
\begin{tabular}{@{\extracolsep{-6pt}}r|cccccccccccccccccccccccccccccccc}
  \multicolumn{16}{c}{}
      \vspace{-11pt}\\
  \hline
$m$ & 1b & 2a & 2b & 3a & 3b & 4a & 4b & 5a & 5b & 6a & 6b & 7a & 7b & 8a & 8b & 9a & 9b & 10a & 10b & 11a & 11b & 12a & 12b & 13a & 13b & 14a & 14b & 15a & 15b & 16a & 16b \\ 
  \hline
1a & 1.12 & 0.98 & 0.93 & 1.04 & 1.11 & 1.04 & 0.80 & 1.02 & 0.73 & 0.99 & 0.82 & 1.04 & 1.09 & 1.03 & 1.07 & 1.02 & 1.06 & 1.01 & 0.97 & 1.04 & 0.96 & 1.02 & 0.86 & 1.03 & 0.88 & 1.03 & 0.86 & 1.03 & 1.14 & 1.21 & 1.24 \\ 
  1b &  & 0.96 & 0.89 & 1.10 & 1.30 & 1.09 & 0.53 & 1.06 & 0.41 & 0.96 & 0.53 & 1.12 & 1.23 & 1.08 & 1.17 & 1.04 & 1.15 & 1.03 & 0.86 & 1.09 & 0.90 & 1.03 & 0.62 & 1.06 & 0.68 & 1.07 & 0.60 & 1.06 & 1.41 & 1.66 & 1.76 \\ 
  2a &  &  & 1.55 & 0.94 & 0.83 & 0.92 & 1.37 & 0.94 & 1.64 & 0.97 & 1.25 & 0.90 & 0.83 & 0.92 & 0.86 & 0.95 & 0.86 & 0.95 & 0.85 & 0.89 & 1.17 & 0.90 & 1.12 & 0.93 & 1.18 & 0.91 & 1.10 & 0.92 & 0.80 & 0.67 & 0.64 \\ 
  2b &  &  &  & 0.84 & 0.56 & 0.78 & 2.26 & 0.84 & 3.58 & 0.92 & 1.76 & 0.71 & 0.53 & 0.74 & 0.58 & 0.83 & 0.59 & 0.87 & 0.61 & 0.71 & 1.53 & 0.73 & 1.34 & 0.80 & 1.53 & 0.74 & 1.28 & 0.79 & 0.51 & 0.29 & 0.25 \\ 
  3a &  &  &  &  & 1.12 & 1.05 & 0.76 & 1.03 & 0.66 & 1 & 0.80 & 1.06 & 1.11 & 1.04 & 1.08 & 1.03 & 1.08 & 1.02 & 1.01 & 1.06 & 0.93 & 1.04 & 0.86 & 1.04 & 0.87 & 1.04 & 0.86 & 1.04 & 1.16 & 1.25 & 1.28 \\ 
  3b &  &  &  &  &  & 1.14 & 0.39 & 1.09 & 0.20 & 0.99 & 0.44 & 1.17 & 1.32 & 1.12 & 1.24 & 1.08 & 1.23 & 1.06 & 1.04 & 1.17 & 0.79 & 1.11 & 0.61 & 1.11 & 0.62 & 1.13 & 0.61 & 1.11 & 1.49 & 1.80 & 1.89 \\ 
  4a &  &  &  &  &  &  & 0.73 & 1.04 & 0.59 & 1 & 0.78 & 1.07 & 1.12 & 1.05 & 1.10 & 1.03 & 1.09 & 1.03 & 1.04 & 1.07 & 0.91 & 1.05 & 0.85 & 1.04 & 0.85 & 1.05 & 0.86 & 1.05 & 1.17 & 1.28 & 1.30 \\ 
  4b &  &  &  &  &  &  &  & 0.82 & 5.17 & 1.02 & 2.51 & 0.63 & 0.37 & 0.70 & 0.43 & 0.83 & 0.48 & 0.88 & 0.95 & 0.70 & 1.48 & 0.79 & 2.05 & 0.78 & 1.95 & 0.74 & 1.97 & 0.78 & 0.34 & 0.15 & 0.11 \\ 
  5a &  &  &  &  &  &  &  &  & 0.73 & 1 & 0.86 & 1.05 & 1.08 & 1.04 & 1.07 & 1.02 & 1.06 & 1.02 & 1.03 & 1.05 & 0.94 & 1.04 & 0.91 & 1.03 & 0.90 & 1.04 & 0.91 & 1.03 & 1.12 & 1.19 & 1.21 \\ 
  5b &  &  &  &  &  &  &  &  &  & 1 & 3.29 & 0.44 & 0.14 & 0.53 & 0.17 & 0.73 & 0.23 & 0.82 & 0.84 & 0.57 & 1.80 & 0.67 & 2.55 & 0.66 & 2.43 & 0.59 & 2.40 & 0.67 & 0.17 & 0.03 & 0.02 \\ 
  6a &  &  &  &  &  &  &  &  &  &  & 1.03 & 1 & 1 & 1 & 1 & 1 & 1 & 1.01 & 1.05 & 1.01 & 0.98 & 1.01 & 1.04 & 1 & 1.01 & 1.01 & 1.04 & 1.01 & 0.99 & 0.99 & 0.99 \\ 
  6b &  &  &  &  &  &  &  &  &  &  &  & 0.70 & 0.45 & 0.78 & 0.53 & 0.87 & 0.58 & 0.91 & 1.03 & 0.76 & 1.34 & 0.84 & 1.84 & 0.83 & 1.76 & 0.80 & 1.83 & 0.83 & 0.37 & 0.14 & 0.10 \\ 
  7a &  &  &  &  &  &  &  &  &  &  &  &  & 1.16 & 1.07 & 1.13 & 1.04 & 1.12 & 1.03 & 1.04 & 1.08 & 0.88 & 1.06 & 0.81 & 1.05 & 0.80 & 1.07 & 0.81 & 1.06 & 1.20 & 1.31 & 1.34 \\ 
  7b &  &  &  &  &  &  &  &  &  &  &  &  &  & 1.12 & 1.24 & 1.07 & 1.23 & 1.05 & 1.06 & 1.15 & 0.80 & 1.11 & 0.64 & 1.10 & 0.64 & 1.12 & 0.64 & 1.10 & 1.37 & 1.56 & 1.60 \\ 
  8a &  &  &  &  &  &  &  &  &  &  &  &  &  &  & 1.10 & 1.03 & 1.10 & 1.02 & 1.04 & 1.06 & 0.91 & 1.05 & 0.86 & 1.04 & 0.85 & 1.05 & 0.87 & 1.05 & 1.15 & 1.22 & 1.23 \\ 
  8b &  &  &  &  &  &  &  &  &  &  &  &  &  &  &  & 1.06 & 1.19 & 1.04 & 1.05 & 1.11 & 0.83 & 1.08 & 0.70 & 1.08 & 0.70 & 1.10 & 0.71 & 1.08 & 1.27 & 1.38 & 1.40 \\ 
  9a &  &  &  &  &  &  &  &  &  &  &  &  &  &  &  &  & 1.06 & 1.02 & 1.03 & 1.04 & 0.94 & 1.03 & 0.92 & 1.03 & 0.91 & 1.03 & 0.93 & 1.03 & 1.09 & 1.15 & 1.16 \\ 
  9b &  &  &  &  &  &  &  &  &  &  &  &  &  &  &  &  &  & 1.04 & 1.06 & 1.11 & 0.84 & 1.08 & 0.74 & 1.08 & 0.73 & 1.09 & 0.75 & 1.08 & 1.26 & 1.38 & 1.41 \\ 
  10a &  &  &  &  &  &  &  &  &  &  &  &  &  &  &  &  &  &  & 1.04 & 1.03 & 0.95 & 1.03 & 0.95 & 1.02 & 0.94 & 1.03 & 0.95 & 1.02 & 1.07 & 1.12 & 1.13 \\ 
  10b &  &  &  &  &  &  &  &  &  &  &  &  &  &  &  &  &  &  &  & 1.08 & 0.88 & 1.09 & 1.10 & 1.04 & 0.99 & 1.06 & 1.13 & 1.05 & 1.04 & 1.11 & 1.11 \\ 
  11a &  &  &  &  &  &  &  &  &  &  &  &  &  &  &  &  &  &  &  &  & 0.88 & 1.07 & 0.85 & 1.05 & 0.83 & 1.07 & 0.86 & 1.06 & 1.22 & 1.37 & 1.41 \\ 
  11b &  &  &  &  &  &  &  &  &  &  &  &  &  &  &  &  &  &  &  &  &  & 0.90 & 1.20 & 0.92 & 1.24 & 0.90 & 1.18 & 0.92 & 0.75 & 0.61 & 0.58 \\ 
  12a &  &  &  &  &  &  &  &  &  &  &  &  &  &  &  &  &  &  &  &  &  &  & 0.92 & 1.04 & 0.89 & 1.05 & 0.93 & 1.05 & 1.14 & 1.24 & 1.26 \\ 
  12b &  &  &  &  &  &  &  &  &  &  &  &  &  &  &  &  &  &  &  &  &  &  &  & 0.89 & 1.52 & 0.88 & 1.62 & 0.89 & 0.54 & 0.35 & 0.30 \\ 
  13a &  &  &  &  &  &  &  &  &  &  &  &  &  &  &  &  &  &  &  &  &  &  &  &  & 0.88 & 1.04 & 0.90 & 1.04 & 1.13 & 1.21 & 1.23 \\ 
  13b &  &  &  &  &  &  &  &  &  &  &  &  &  &  &  &  &  &  &  &  &  &  &  &  &  & 0.86 & 1.52 & 0.88 & 0.55 & 0.34 & 0.29 \\ 
  14a &  &  &  &  &  &  &  &  &  &  &  &  &  &  &  &  &  &  &  &  &  &  &  &  &  &  & 0.89 & 1.05 & 1.15 & 1.24 & 1.26 \\ 
  14b &  &  &  &  &  &  &  &  &  &  &  &  &  &  &  &  &  &  &  &  &  &  &  &  &  &  &  & 0.90 & 0.53 & 0.34 & 0.28 \\ 
  15a &  &  &  &  &  &  &  &  &  &  &  &  &  &  &  &  &  &  &  &  &  &  &  &  &  &  &  &  & 1.14 & 1.24 & 1.26 \\ 
  15b &  &  &  &  &  &  &  &  &  &  &  &  &  &  &  &  &  &  &  &  &  &  &  &  &  &  &  &  &  & 2.17 & 2.33 \\ 
  16a &  &  &  &  &  &  &  &  &  &  &  &  &  &  &  &  &  &  &  &  &  &  &  &  &  &  &  &  &  &  & 3.48 \\ 
  \end{tabular}
\end{sidewaystable}

\begin{sidewaystable}
\caption{Estimate of the $\mathit{lift}$ for $g=3$ in the selected model} \label{tab:cvot.lift3}
\smallskip
\scriptsize
\begin{tabular}{@{\extracolsep{-5pt}}r|cccccccccccccccccccccccccccccccc}
  \multicolumn{16}{c}{}
      \vspace{-11pt}\\
  \hline
$m$ & 1b & 2a & 2b & 3a & 3b & 4a & 4b & 5a & 5b & 6a & 6b & 7a & 7b & 8a & 8b & 9a & 9b & 10a & 10b & 11a & 11b & 12a & 12b & 13a & 13b & 14a & 14b & 15a & 15b & 16a & 16b \\ 
  \hline
1a & 1 & 1 & 1 & 1 & 1 & 1 & 1 & 1 & 1 & 1 & 1 & 1 & 1 & 1 & 1 & 1 & 1 & 1 & 1 & 1 & 1 & 1 & 1 & 1 & 1 & 1 & 1 & 1 & 1 & 1 & 1 \\ 
  1b &  & 1 & 0.98 & 1 & 1.04 & 1 & 0.78 & 1 & 0.78 & 1 & 0.86 & 1 & 1.08 & 1.01 & 1.07 & 1 & 1.08 & 1 & 0.96 & 1 & 0.97 & 1 & 0.78 & 1.01 & 0.85 & 1 & 0.83 & 1 & 1.09 & 1.02 & 1.04 \\ 
  2a &  &  & 1.01 & 1 & 1 & 1 & 1.01 & 1 & 1.02 & 1 & 1.01 & 1 & 0.99 & 1 & 0.99 & 1 & 0.99 & 1 & 0.99 & 1 & 1 & 1 & 1.01 & 1 & 1.01 & 1 & 1 & 1 & 0.99 & 1 & 1 \\ 
  2b &  &  &  & 1 & 0.95 & 0.99 & 1.30 & 1 & 1.36 & 1 & 1.15 & 0.99 & 0.86 & 0.99 & 0.88 & 0.99 & 0.84 & 1 & 0.88 & 0.99 & 1.10 & 0.98 & 1.14 & 0.98 & 1.18 & 0.98 & 1.09 & 0.99 & 0.88 & 0.97 & 0.94 \\ 
  3a &  &  &  &  & 1 & 1 & 0.99 & 1 & 0.99 & 1 & 1 & 1 & 1 & 1 & 1 & 1 & 1 & 1 & 1 & 1 & 1 & 1 & 1 & 1 & 1 & 1 & 1 & 1 & 1 & 1 & 1 \\ 
  3b &  &  &  &  &  & 1 & 0.81 & 1 & 0.81 & 1 & 0.91 & 1 & 1.07 & 1.01 & 1.06 & 1 & 1.07 & 1 & 1.01 & 1 & 0.97 & 1.01 & 0.86 & 1.01 & 0.89 & 1.01 & 0.91 & 1 & 1.06 & 1.02 & 1.04 \\ 
  4a &  &  &  &  &  &  & 0.98 & 1 & 0.98 & 1 & 0.99 & 1 & 1.01 & 1 & 1.01 & 1 & 1.01 & 1 & 1 & 1 & 1 & 1 & 0.99 & 1 & 0.99 & 1 & 0.99 & 1 & 1.01 & 1 & 1 \\ 
  4b &  &  &  &  &  &  &  & 0.99 & 2.16 & 1 & 1.50 & 0.97 & 0.57 & 0.96 & 0.61 & 0.97 & 0.57 & 1 & 0.97 & 0.99 & 1.18 & 0.97 & 1.88 & 0.96 & 1.65 & 0.96 & 1.58 & 0.98 & 0.63 & 0.85 & 0.74 \\ 
  5a &  &  &  &  &  &  &  &  & 0.99 & 1 & 0.99 & 1 & 1.01 & 1 & 1 & 1 & 1.01 & 1 & 1 & 1 & 1 & 1 & 0.99 & 1 & 0.99 & 1 & 0.99 & 1 & 1 & 1 & 1 \\ 
  5b &  &  &  &  &  &  &  &  &  & 1 & 1.56 & 0.97 & 0.53 & 0.97 & 0.58 & 0.97 & 0.52 & 1 & 0.94 & 0.98 & 1.21 & 0.96 & 1.84 & 0.96 & 1.67 & 0.96 & 1.58 & 0.98 & 0.60 & 0.87 & 0.75 \\ 
  6a &  &  &  &  &  &  &  &  &  &  & 1 & 1 & 1 & 1 & 1 & 1 & 1 & 1 & 1 & 1 & 1 & 1 & 1 & 1 & 1 & 1 & 1 & 1 & 1 & 1 & 1 \\ 
  6b &  &  &  &  &  &  &  &  &  &  &  & 0.99 & 0.78 & 0.99 & 0.82 & 0.99 & 0.78 & 1 & 1.01 & 0.99 & 1.10 & 0.99 & 1.42 & 0.98 & 1.35 & 0.98 & 1.35 & 0.99 & 0.78 & 0.95 & 0.91 \\ 
  7a &  &  &  &  &  &  &  &  &  &  &  &  & 1.01 & 1 & 1.01 & 1 & 1.01 & 1 & 1 & 1 & 1 & 1 & 0.98 & 1 & 0.99 & 1 & 0.99 & 1 & 1.01 & 1 & 1.01 \\ 
  7b &  &  &  &  &  &  &  &  &  &  &  &  &  & 1.01 & 1.16 & 1.01 & 1.19 & 1 & 1.03 & 1.01 & 0.92 & 1.01 & 0.69 & 1.02 & 0.74 & 1.01 & 0.78 & 1.01 & 1.16 & 1.05 & 1.09 \\ 
  8a &  &  &  &  &  &  &  &  &  &  &  &  &  &  & 1.01 & 1 & 1.01 & 1 & 1 & 1 & 0.99 & 1 & 0.98 & 1 & 0.98 & 1 & 0.99 & 1 & 1.01 & 1 & 1.01 \\ 
  8b &  &  &  &  &  &  &  &  &  &  &  &  &  &  &  & 1.01 & 1.16 & 1 & 1.02 & 1.01 & 0.93 & 1.01 & 0.72 & 1.01 & 0.78 & 1.01 & 0.81 & 1.01 & 1.13 & 1.05 & 1.08 \\ 
  9a &  &  &  &  &  &  &  &  &  &  &  &  &  &  &  &  & 1.01 & 1 & 1 & 1 & 0.99 & 1 & 0.98 & 1 & 0.98 & 1 & 0.99 & 1 & 1.01 & 1 & 1.01 \\ 
  9b &  &  &  &  &  &  &  &  &  &  &  &  &  &  &  &  &  & 1 & 1.04 & 1.01 & 0.91 & 1.02 & 0.70 & 1.02 & 0.74 & 1.02 & 0.78 & 1.01 & 1.16 & 1.05 & 1.09 \\ 
  10a &  &  &  &  &  &  &  &  &  &  &  &  &  &  &  &  &  &  & 1 & 1 & 1 & 1 & 1 & 1 & 1 & 1 & 1 & 1 & 1 & 1 & 1 \\ 
  10b &  &  &  &  &  &  &  &  &  &  &  &  &  &  &  &  &  &  &  & 1 & 0.96 & 1.01 & 1.07 & 1.01 & 1 & 1.01 & 1.06 & 1 & 1.01 & 1.01 & 1.01 \\ 
  11a &  &  &  &  &  &  &  &  &  &  &  &  &  &  &  &  &  &  &  &  & 1 & 1 & 0.99 & 1 & 0.99 & 1 & 0.99 & 1 & 1.01 & 1 & 1 \\ 
  11b &  &  &  &  &  &  &  &  &  &  &  &  &  &  &  &  &  &  &  &  &  & 0.99 & 1.11 & 0.99 & 1.11 & 0.99 & 1.08 & 1 & 0.93 & 0.98 & 0.96 \\ 
  12a &  &  &  &  &  &  &  &  &  &  &  &  &  &  &  &  &  &  &  &  &  &  & 0.98 & 1 & 0.98 & 1 & 0.99 & 1 & 1.01 & 1 & 1.01 \\ 
  12b &  &  &  &  &  &  &  &  &  &  &  &  &  &  &  &  &  &  &  &  &  &  &  & 0.97 & 1.51 & 0.98 & 1.53 & 0.99 & 0.71 & 0.90 & 0.82 \\ 
  13a &  &  &  &  &  &  &  &  &  &  &  &  &  &  &  &  &  &  &  &  &  &  &  &  & 0.98 & 1 & 0.98 & 1 & 1.01 & 1 & 1.01 \\ 
  13b &  &  &  &  &  &  &  &  &  &  &  &  &  &  &  &  &  &  &  &  &  &  &  &  &  & 0.98 & 1.37 & 0.99 & 0.76 & 0.93 & 0.87 \\ 
  14a &  &  &  &  &  &  &  &  &  &  &  &  &  &  &  &  &  &  &  &  &  &  &  &  &  &  & 0.99 & 1 & 1.01 & 1 & 1.01 \\ 
  14b &  &  &  &  &  &  &  &  &  &  &  &  &  &  &  &  &  &  &  &  &  &  &  &  &  &  &  & 0.99 & 0.78 & 0.94 & 0.89 \\ 
  15a &  &  &  &  &  &  &  &  &  &  &  &  &  &  &  &  &  &  &  &  &  &  &  &  &  &  &  &  & 1.01 & 1 & 1 \\ 
  15b &  &  &  &  &  &  &  &  &  &  &  &  &  &  &  &  &  &  &  &  &  &  &  &  &  &  &  &  &  & 1.04 & 1.07 \\ 
  16a &  &  &  &  &  &  &  &  &  &  &  &  &  &  &  &  &  &  &  &  &  &  &  &  &  &  &  &  &  &  & 1.03 \\ 
\end{tabular}
\end{sidewaystable}

\begin{sidewaystable}
\caption{Estimate of the $\mathit{lift}$ for $g=4$ in the selected model}\label{tab:cvot.lift4}
\smallskip
\scriptsize
\begin{tabular}{@{\extracolsep{-5pt}}r|cccccccccccccccccccccccccccccccc}
  \multicolumn{16}{c}{}
      \vspace{-11pt}\\
  \hline
$m$ & 1b & 2a & 2b & 3a & 3b & 4a & 4b & 5a & 5b & 6a & 6b & 7a & 7b & 8a & 8b & 9a & 9b & 10a & 10b & 11a & 11b & 12a & 12b & 13a & 13b & 14a & 14b & 15a & 15b & 16a & 16b \\ 
  \hline
1a & 1 & 1 & 1 & 1 & 1 & 1 & 1 & 1 & 1 & 1 & 1 & 1 & 1 & 1 & 1 & 1 & 1 & 1 & 1 & 1 & 1 & 1 & 1 & 1 & 1 & 1 & 1 & 1 & 1 & 1 & 1 \\ 
  1b &  & 1 & 0.96 & 1 & 1.43 & 1 & 0.97 & 1 & 0.98 & 1 & 0.91 & 1 & 1.45 & 1.01 & 1.62 & 1 & 1.50 & 1 & 0.92 & 1 & 0.92 & 1 & 0.93 & 1 & 0.93 & 1 & 1 & 1 & 1.42 & 1.05 & 1.21 \\ 
  2a &  &  & 1.02 & 1 & 0.97 & 1 & 1 & 1 & 1 & 1 & 1 & 1 & 0.96 & 1 & 0.94 & 1 & 0.95 & 1 & 0.99 & 1 & 1.02 & 1 & 1 & 1 & 1 & 1 & 1 & 1 & 0.97 & 1 & 0.98 \\ 
  2b &  &  &  & 1 & 0.67 & 1 & 1.02 & 1 & 1.01 & 1 & 1.04 & 1 & 0.59 & 0.98 & 0.47 & 1 & 0.48 & 1 & 0.87 & 0.99 & 1.20 & 0.99 & 1.03 & 1 & 1.04 & 1 & 1 & 0.99 & 0.69 & 0.95 & 0.79 \\ 
  3a &  &  &  &  & 1 & 1 & 1 & 1 & 1 & 1 & 1 & 1 & 1 & 1 & 1 & 1 & 1 & 1 & 1 & 1 & 1 & 1 & 1 & 1 & 1 & 1 & 1 & 1 & 1 & 1 & 1 \\ 
  3b &  &  &  &  &  & 1 & 0.93 & 1 & 0.95 & 1 & 0.82 & 1.01 & 2.11 & 1.02 & 2.69 & 1 & 2.46 & 1 & 1.03 & 1.01 & 0.73 & 1.01 & 0.88 & 1.01 & 0.86 & 1 & 1 & 1.01 & 1.97 & 1.08 & 1.40 \\ 
  4a &  &  &  &  &  &  & 1 & 1 & 1 & 1 & 1 & 1 & 1 & 1 & 1 & 1 & 1 & 1 & 1 & 1 & 1 & 1 & 1 & 1 & 1 & 1 & 1 & 1 & 1 & 1 & 1 \\ 
  4b &  &  &  &  &  &  &  & 1 & 1 & 1 & 1.01 & 1 & 0.92 & 1 & 0.87 & 1 & 0.89 & 1 & 1 & 1 & 1.02 & 1 & 1.01 & 1 & 1.01 & 1 & 1 & 1 & 0.93 & 1 & 0.97 \\ 
  5a &  &  &  &  &  &  &  &  & 1 & 1 & 1 & 1 & 1 & 1 & 1 & 1 & 1 & 1 & 1 & 1 & 1 & 1 & 1 & 1 & 1 & 1 & 1 & 1 & 1 & 1 & 1 \\ 
  5b &  &  &  &  &  &  &  &  &  & 1 & 1.01 & 1 & 0.95 & 1 & 0.90 & 1 & 0.91 & 1 & 1 & 1 & 1.01 & 1 & 1.01 & 1 & 1.01 & 1 & 1 & 1 & 0.95 & 1 & 0.99 \\ 
  6a &  &  &  &  &  &  &  &  &  &  & 1 & 1 & 1 & 1 & 1 & 1 & 1 & 1 & 1 & 1 & 1 & 1 & 1 & 1 & 1 & 1 & 1 & 1 & 1 & 1 & 1 \\ 
  6b &  &  &  &  &  &  &  &  &  &  &  & 1 & 0.81 & 1 & 0.69 & 1 & 0.74 & 1 & 1 & 1 & 1.04 & 1 & 1.02 & 1 & 1.03 & 1 & 1 & 1 & 0.82 & 0.99 & 0.94 \\ 
  7a &  &  &  &  &  &  &  &  &  &  &  &  & 1.01 & 1 & 1.01 & 1 & 1.01 & 1 & 1 & 1 & 1 & 1 & 1 & 1 & 1 & 1 & 1 & 1 & 1.01 & 1 & 1.01 \\ 
  7b &  &  &  &  &  &  &  &  &  &  &  &  &  & 1.02 & 2.85 & 1.01 & 2.63 & 1 & 1.07 & 1.01 & 0.67 & 1.01 & 0.87 & 1.01 & 0.85 & 1 & 1 & 1.01 & 2.08 & 1.09 & 1.48 \\ 
  8a &  &  &  &  &  &  &  &  &  &  &  &  &  &  & 1.03 & 1 & 1.03 & 1 & 1 & 1 & 0.99 & 1 & 1 & 1 & 1 & 1 & 1 & 1 & 1.02 & 1.01 & 1.02 \\ 
  8b &  &  &  &  &  &  &  &  &  &  &  &  &  &  &  & 1.01 & 3.65 & 1 & 1.10 & 1.01 & 0.59 & 1.01 & 0.81 & 1.01 & 0.77 & 1 & 1 & 1.02 & 2.66 & 1.09 & 1.55 \\ 
  9a &  &  &  &  &  &  &  &  &  &  &  &  &  &  &  &  & 1.01 & 1 & 1 & 1 & 1 & 1 & 1 & 1 & 1 & 1 & 1 & 1 & 1 & 1 & 1 \\ 
  9b &  &  &  &  &  &  &  &  &  &  &  &  &  &  &  &  &  & 1 & 1.12 & 1.01 & 0.61 & 1.01 & 0.84 & 1.01 & 0.80 & 1 & 1 & 1.02 & 2.43 & 1.09 & 1.51 \\ 
  10a &  &  &  &  &  &  &  &  &  &  &  &  &  &  &  &  &  &  & 1 & 1 & 1 & 1 & 1 & 1 & 1 & 1 & 1 & 1 & 1 & 1 & 1 \\ 
  10b &  &  &  &  &  &  &  &  &  &  &  &  &  &  &  &  &  &  &  & 1 & 0.93 & 1 & 1.01 & 1 & 1 & 1 & 1 & 1 & 1.03 & 1.01 & 1.03 \\ 
  11a &  &  &  &  &  &  &  &  &  &  &  &  &  &  &  &  &  &  &  &  & 1 & 1 & 1 & 1 & 1 & 1 & 1 & 1 & 1.01 & 1 & 1.01 \\ 
  11b &  &  &  &  &  &  &  &  &  &  &  &  &  &  &  &  &  &  &  &  &  & 0.99 & 1.03 & 1 & 1.04 & 1 & 1 & 0.99 & 0.74 & 0.95 & 0.82 \\ 
  12a &  &  &  &  &  &  &  &  &  &  &  &  &  &  &  &  &  &  &  &  &  &  & 1 & 1 & 1 & 1 & 1 & 1 & 1.01 & 1 & 1.01 \\ 
  12b &  &  &  &  &  &  &  &  &  &  &  &  &  &  &  &  &  &  &  &  &  &  &  & 1 & 1.02 & 1 & 1 & 1 & 0.88 & 0.99 & 0.95 \\ 
  13a &  &  &  &  &  &  &  &  &  &  &  &  &  &  &  &  &  &  &  &  &  &  &  &  & 1 & 1 & 1 & 1 & 1.01 & 1 & 1 \\ 
  13b &  &  &  &  &  &  &  &  &  &  &  &  &  &  &  &  &  &  &  &  &  &  &  &  &  & 1 & 1 & 1 & 0.86 & 0.99 & 0.94 \\ 
  14a &  &  &  &  &  &  &  &  &  &  &  &  &  &  &  &  &  &  &  &  &  &  &  &  &  &  & 1 & 1 & 1 & 1 & 1 \\ 
  14b &  &  &  &  &  &  &  &  &  &  &  &  &  &  &  &  &  &  &  &  &  &  &  &  &  &  &  & 1 & 1 & 1 & 1 \\ 
  15a &  &  &  &  &  &  &  &  &  &  &  &  &  &  &  &  &  &  &  &  &  &  &  &  &  &  &  &  & 1.01 & 1 & 1.01 \\ 
  15b &  &  &  &  &  &  &  &  &  &  &  &  &  &  &  &  &  &  &  &  &  &  &  &  &  &  &  &  &  & 1.07 & 1.39 \\ 
  16a &  &  &  &  &  &  &  &  &  &  &  &  &  &  &  &  &  &  &  &  &  &  &  &  &  &  &  &  &  &  & 1.07 \\ 
\end{tabular}
\end{sidewaystable}

\end{document}